\documentclass[aps,prd,superscriptaddress,nofootinbib,amsmath, showkeys, showpacs,amsfonts,preprintnumbers,notitlepage,10pt,english]{revtex4-1}
\usepackage{amsmath}
\usepackage{amssymb}
\usepackage{babel}

\usepackage{bm}
\usepackage{booktabs}
\usepackage{multirow}
\usepackage{array}
\usepackage{enumitem}
\usepackage{mathtools}
\makeatletter
\usepackage{array,multirow,graphicx}
\usepackage{dcolumn}
\usepackage{newlfont}
\usepackage{bm}
\usepackage[colorlinks,citecolor=blue,urlcolor=blue,linkcolor=blue]{hyperref}
\usepackage[figtopcap]{subfigure}
\usepackage{color}

\begin{document}

\date{\today}

\title{Anisotropic compact stars in $D\rightarrow 4$ limit of Gauss-Bonnet gravity}
\author{G.G.L. Nashed}
\email{nashed@bue.edu.eg}
\affiliation {Centre for Theoretical Physics, The British University, P.O. Box
43, El Sherouk City, Cairo 11837, Egypt}
\affiliation{Laboratory for Theoretical Cosmology, Tomsk State University of Control Systems and Radioelectronics (TUSUR), 634050 Tomsk, Russia}
\author{S.D. Odintsov}
\email{odintsov@ieec.uab.es}
\affiliation{Institut de Ci\`encies de l'Espai (ICE-CSIC/IEEC),\\
Campus, c. Can Magrans s/n, 08193, Barcelona, Spain}
\affiliation{Instituci\'o Catalana de Recerca i Estudis
Avan\c{c}ats (ICREA), Barcelona, Spain}
\author{V.K. Oikonomou}
\email{v.k.oikonomou1979@gmail.com}
\affiliation{Department of Physics, Aristotle University of Thessaloniki, Thessaloniki 54124, Greece}
\affiliation{Laboratory for Theoretical Cosmology, Tomsk State University of Control Systems and Radioelectronics (TUSUR), 634050 Tomsk, Russia}

\begin{abstract}
{In the frame of Gauss-Bonnet gravity and in the limit of $D\to 4$,  based on the fact that  spherically symmetric solution derived using any of regularization schemes will be the same form as the original
theory  \cite{Banerjee:2020yhu,Hennigar:2020lsl,Casalino:2020kbt,Aoki:2020lig},}
we derive a new interior spherically symmetric solution assuming specific forms of the metric potentials that have two constants . Using the
junction condition we determine these two constants. By using the data of the star EXO
$1785-248 $, whose mass is $ { M=1.3\pm0.2 \,{\textrm M}_\odot}$ and radius
$ { l=8.849\pm0.4\,{\textrm km}}$, we calculate the numerical values of
these constants, in terms of the dimensionful coupling parameter
of the Gauss-Bonnet term, and eventually, we get real values for
these constants. In this regard, we show that the components of
the energy-momentum tensor has a finite value at the center of
the star as well as a smaller value to the surface of the star.
Moreover, we show that the equations of the state behave in a
non-linear way due to the impact of the Gauss-Bonnet term. Using
the Tolman-Oppenheimer-Volkoff equation, the adiabatic index, and
stability in the static state we show that the model under
consideration is always stable.  Finally, the solution of this
study is matched with observational data of other pulsars showing
satisfactory results.
\end{abstract}


\maketitle
\section{Introduction}
To date a considerable amount of observational data of compact
stellar objects data, such as neutron stars and spherically
symmetric black holes, are available from gravitational-wave (GW)
detectors, like X-ray observations, advanced LIGO
\cite{LIGOScientific:2014pky},  Kagra
\cite{Somiya:2011np,Aso:2013eba},  Advanced Virgo
\cite{VIRGO:2014yos}, { and  from the networks of radio telescopes,
like the event horizon telescope
\cite{EventHorizonTelescope:2020qrl}, from  the NICER mission
\cite{Raaijmakers:2019dks} and there is much activity in the study of such objects also, i.e., neutron stars in $F(R)$ \cite{Astashenok:2013vza,Astashenok:2014pua}.} In the frame of current observational
uncertainties, all data are consistent with Einstein's general
relativity (GR) and despite this. However, there is a venue for a
extended gravity theories, which might be useful to future
observations which require alternative descriptions than that of
Einstein's  GR theory, like for example the curious GW190814 event
\cite{LIGOScientific:2020zkf}. Regardless of the direct tests of
Einstein GR, there are also pending questions to be clarified,
like the maximum mass of neutron stars, the high-density equation
of state (EoS), and so on. The observations obtained recently by
the compact object merger GW190814 \cite{LIGOScientific:2020zkf}
shows that the low-mass component of the binary, possesses a mass
of $2.59_{-0.09}^{+0.08}~M_\odot$, which directly puts it in the
present observational mass gap between neutron stars and black
holes. { Moreover, among
the gravitational waves, there are three events/candidates involving the
merger of one neutron star   or light mass black
hole   with another compact object.} If this event is described solely by GR, the lower mass
component of the binary can either be a neutron star with an
unexpectedly stiff (or exotic) EoS,  a black hole with an
unexpectedly small mass, or a neutron star with an unexpectedly
rapid rotation
 \cite{Huang:2020cab,Bombaci:2020vgw,Roupas:2020nua,Zhou:2020xan,Awad:2017sau,Most:2020bba,Nashed:2014sea,Nashed:2005kn,Tan:2020ics,Vattis:2020iuz,Nashed:2005kn,Zhang:2020zsc,Fattoyev:2020cws,Shirafuji:1996im,Tsokaros:2020hli,Tews:2020ylw,Dexheimer:2020rlp,Godzieba:2020tjn,Kanakis-Pegios:2020kzp,Nathanail:2021tay,Roupas:2020jyv,2021MNRAS.505.1600B}.
On the other hand, an natural and non-exotic description of the
lower mass component of the event GW190814 can be given by
extended gravity
 \cite{Nunes:2020cuz,Astashenok:2020qds,Astashenok:2021peo}. { In the frame of modified gravitational theories  there are many studied tackled the problem of anisotropic compact stars for example in the frame of $f(R,G)$, where $R$ is the Ricci scalar and $G$ is the Gauss-Bonnet term   \cite{Mustafa:2020jln}, in the frame of mimetic theory \cite{Nashed:2021pkc}, in the frame of $f(R)$ \cite{Nashed:2021gkp,Nashed:2021sji,Astashenok:2021btj,Astashenok:2021xpm}, in the frame of teleparallel gravity \cite{Nashed:2020buf,Nashed:2020kjh}, and in the frame of scalar Gauss-Bonnet theory \cite{Ovgun:2021ttv}.}

To date, two GW outcomes have been verified as binary neutron star
(BNS) mergers, GW170817 \citep{LIGOScientific:2017vwq} and
GW190425 \citep{LIGOScientific:2020aai}, and extra outcomes are
anticipated in the near future \citep{KAGRA:2013rdx}. Revealing of
the GW from the inspiral phase of GW170817, in connection with
observations of its kilonova electromagnetic counterpart
\citep{LIGOScientific:2017ync,LIGOScientific:2017zic,Goldstein:2017mmi},
have put new constraints on the dimensionless tidal deformability
of neutron stars and hence on their EoS
\citep{Bauswein:2017vtn,LIGOScientific:2018cki,Capano:2019eae,Thapa:2021ifv,Dietrich:2020efo,Breschi:2021tbm,Chatziioannou:2020pqz,Chatziioannou:2020pqz}.
Such EoS constraints are expected to be more precise in the next
years, via the combination of a larger number of observations
\cite{DelPozzo:2013ala,Chatziioannou:2015uea,Lackey:2014fwa,HernandezVivanco:2019vvk,Chatziioannou:2019yko}.
Despite the sensitivity of the Advanced LIGO and Advanced Virgo
detectors, these were not sensitive enough in order to determine
the post-merger phase in GW170817
\citep{LIGOScientific:2017vwq,LIGOScientific:2017fdd}. However,
this post-merger phase may be revealed in the near future, by
several experiments and collaborations, like
\citep{KAGRA:2013rdx}, or by third generation detectors
\citep{LIGOScientific:2016wof,Maggiore:2019uih}, or even with
high-frequency detectors \citep{Ganapathy:2020thy,Page:2020zbr}.
The observation of GW in the post-merger phase of a BNS merger
would then offer another opportunity to probe the high-density EoS
of NSs, see
\citep{1992ApJ...401..226R,Shibata:2005xz,Elizalde:2020icc,Bauswein:2011tp,Clark:2014wua,Rezzolla:2016nxn,Bauswein:2019ybt,Breschi:2019srl,Tsang:2019esi,Vretinaris:2019spn,Easter:2020ifj,Friedman:2020xac}.


In view of the future observational data which might indicate
deviations from GR, it is important to study compact objects in
the context of extended theories of gravity. In the present work,
we will study compact objects in the context of
Einstein-Gauss-Bonnet gravity (EGB)
\cite{zwiebach1985curvature,gross1987quartic} which provides a
number of attractive analytic solutions. The action of this theory
is given as:
\begin{equation}
S=\dfrac{1}{2\kappa}\displaystyle\int{\mathrm{d}^4x\sqrt{-g}\left\{R+\beta \mathrm{G}\right\}}+S_\mathrm{m},
\label{eq:action}
\end{equation}
with $\kappa=8\pi {\textit G}/c^4$, and
$\mathrm{G}=R^2-4R_{\mu\nu}R^{\mu\nu}+R_{\mu\nu\rho\sigma}R^{\mu\nu\rho\sigma}$
is the Gauss-Bonnet scalar while $S_\mathrm{m}$ is the matter
Lagrangian. Here  $\beta$ has a dimension of length squared. We
will then formulate models of neutron stars, and by using
numerical methods we will study several solutions for neutron
stars.

Lately, Glavan and Lin \cite{Glavan:2019inb} provided a new
general covariant modified model of the EGB theory that elucidates
multiple non-trivial issues taking place in $4$-dimensional
spacetime. Note that such scenario to get consistent 2-dimensional GR from D-dimensional GR in the limit of $D \rightarrow 2$ was proposed in \cite{Mann:1992ar,Nojiri:2020tph}. The GB expression ($\mathrm{G}$) in $4$-dimensions   is a total derivative term and hence does not impact the
gravitational dynamics. However, on a higher dimension, i.e.,
$D>4$, it involves an exterior result of the total derivative.
When $D>4$, Torii et al. \cite{Torii:2008ru}  and, Mardones et al.
\cite{mardones1991lovelock} showed that the support of the GB term
is proportional to a vanishing factor in $4$-dimensions. Moreover,
Glavan and Lin \cite{Glavan:2019inb} rescaled the dimensional
parameter $\beta$,  i.e., $\beta \to \frac{\beta}{(D-4)}$ showing
a very much identical to GR behavior, as it preserves equal
degrees of freedom in all dimensions and sets aside the
Ostrogradsky instability  \cite{Woodard:2015zca}. Using the
regularization of the GB expression at $D=4$, one can derive a
non-trivial impact in the four-dimensional dynamics
\cite{Tomozawa:2011gp}.

Here, we stress the fact  that the novel 4D EGB theory \cite{Glavan:2019inb} has got
several quibbles. The  criticisms
on the correctness  by taking the limit $D \to 4$ have been discussed in
\cite{Aoki:2020ila, Bonifacio:2020vbk}. Nevertheless, it was shown  in Refs. \cite{Ai:2020peo,Mahapatra:2020rds,Gurses:2020ofy}   that in 4-dimensions spacetime the output field  equations are not regular   and  no regular
action which creates the proposed regularized field equations can be exist \cite{hohmann2021canonical}.  Kaluza-Klein-reduction method of  the limit
$D \to 4$ yields a special class of scalar-tensor theories  in the Horndeski family \cite{Ma:2020ufk,Lu:2020iav}. Similar
method was used in \cite{Hennigar:2020lsl,Fernandes:2020nbq,Banerjee:2021bmv} by adding  a counter
term in D-dimensions and then taking  $D \to 4$ limit.
Furthermore, many projects have been analyzed  the previous issues coming from the new 4-dimensions EGB gravity.  Depending on
the selection of the ``regularization method",  several  theories have been presented with various degrees of
freedom and various characteristic.  Using the procedure of  \cite{Aoki:2020ila,Aoki:2020lig}, the
handling was shown to be symmetrical by ignoring the 4-dimensions diffeomorphism invariance. Nevertheless, it is interesting to note that
 spherically symmetric 4-dimensions solutions are valid in
these regularized theories \cite{Banerjee:2020yhu}. Actually, black hole solutions through the rescaling method \cite{Glavan:2019inb}  still  correct in these
regularized theories \cite{Hennigar:2020lsl,Casalino:2020kbt}. Therefore, it becomes clear that the spherically symmetric solution derived using any of these regularization schemes  will be identical to the authentic
theory. In this frame, the interior spherically symmetric   star solution  is significant to study. Therefore, we
decide to obtain the field equations  starting with outlines depending on the new 4-dimensions EGB gravity. { Moreover, one of the motivation of the present study is to constrains the numerical value of Gauss-Bonnet parameter in the frame of anisotropic model so that we get a realizable astrophysical model. Additionally, we want to test the behavior of the equation of state in this model and compared it with the Einstein general relativity.}

The GB  theory has various applications in different topics of
gravity and therefore, it appeared to be quite unconventional to
scientists in the field of cosmology and astrophysics. Glavan and
Lin \cite{Glavan:2019inb} in the frame of GB theory derived a
static spherically symmetric vacuum black hole which was proved to
be quite different from  Schwarzschild black holes. Gurses et. al
\cite{Gurses:2020ofy} showed that EGB theory does not support any
intrinsic 4-dimensional solution in terms of the metric, thus its
viability of the black hole solutions is appropriate for the case
$D>4$. More applications of EGB theory were used to calculate the
thermodynamics of non-rotating and rotating black holes and
evaluate the thermodynamical quantities as entropy, mass,
temperature
\cite{Ghosh:2020syx,Kumar:2020xvu,Kumar:2020uyz,Zhang:2020sjh}.
Moreover, EGB theory is used for the study of various black hole
related issues, like thermodynamics of AdS black holes
\cite{HosseiniMansoori:2020yfj}, quasi-normal modes
\cite{Mishra:2020gce}, black gravitational Lensing of strong and
weak types \cite{Ghosh:2020syx}, Hawking radiation
\cite{Zhang:2020qam}, geodesic motions  for spinning test
particles \cite{Zhang:2020qew}, and wormhole solutions
\cite{Jusufi:2020yus}. Aside from the issues of the black hole,
EGB theory is applied for the study of neutron stars, which is a
subject of continuous research interest. Recent observations of
neutron stars have further constrained the equation of state of
nuclear matter, but the inner structure of neutron stars still
remains a mystery. { An investigation of the photon sphere and the shadow observed by a distant observer and the exploration of the effects of
black hole parameters have been studied in the frame of EGB \cite{EslamPanah:2020hoj}}

The determination of the inner core matter is quite non-trivial as
the composition varies with the nature of interaction and quark
matter, hyperon matter, Bose-Einstein condensate, strange mesons
are all considered to be the core matter. But the invention of
2$M_\odot$ Neutron stars like PSR J1614-2230 [$1.97 \pm
0.04M_\odot$]  \cite{Huang:2020cab} and J0348+0432 [$2.01 \pm
0.04M_\odot$]
 \cite{Antoniadis:2013pzd} brought criterion on the nature of the core matter. Quantum
chromodynamics [QCD], supports the conversion of hadronic matter
into deconfined quarks inside the neutron stars. This invention
makes the Bodmer-Witten hypothesis quite important to the context
as it already predicted for a quark-matter component in the cosmic
rays detected from neutron stars
\cite{bodmer1971collapsed,PhysRevD.30.272}.  This strengthens the
presence of quark stars.

The present study aims to derive an interior
solutions for EGB theory,  using the fact that  spherically symmetric solution obtained from any  regularization methods will be identical with the original
theory  \cite{Banerjee:2020yhu,Hennigar:2020lsl,Casalino:2020kbt} , and then check the physical viability of
such a solution. According to the procedure,  it is possible to
construct self-consistent compact stellar models capable of
describing anisotropic neutron stars in the context of EGB
gravity. The layout of the present paper is as follows: In Sec.
\ref{S2}, a summary of EGB theory is reported. Also in Sec.
\ref{S2}, we apply the field equation of EGB to $D$-dimensional
spherically symmetric spacetime and write them in the limit  $D\to
4$ and solve the resulting differential equations assuming
specific forms of the metric potentials. In Sec. \ref{S3}, we use
the junction condition, i.e. match the interior solution with the
exterior solution derived \cite{Glavan:2019inb}, and fix the two
constants that characterize our solution.  In Sec. \ref{S4}, we
discuss the physical conditions that any real compact star must
satisfy and show that the energy components of our model have
finite values at the center of the star and decrease towards the
star of the surface. Moreover, we show that the model under
consideration has a positive anisotropy which means that the
tangential pressure is greater than the radial one. Also, we prove
that our model satisfies the causality condition and the
energy-momentum conditions. Finally, we study the equation of
state, (EoS),  and show that the impact of Gauss-Bonnet expression
makes its pattern has a non-linear form.  In Sec. \ref{S5} we show
that our model is stable under various conditions, like TOV
equation, adiabatic index, and the static state.  We confront our
results with observational data using and we tabulated our results
in Tables \ref{Table1} and \ref{Table2}.  Finally, the conclusions
follow in the end of the paper\footnote{ In this study, we are going to use the geometrized units units.}.

\section{Compact star  in  Gauss-Bonnet theory} \label{S2}

The Lagrangian of Einstein-Gauss-Bonnet, in $D-$dimensional
spacetime has the form:
\begin{equation}\label{e1}
\mathcal{L}_{EGB}=\mathcal{R}+{\beta \over
D-4}~\mathrm{G}+\mathcal{L}_{m}\,,
\end{equation}
whose corresponding action takes the form:
\begin{eqnarray}  \label{e2}
\mathrm{S}_{EGB}={1 \over 16\pi} \int d^Dx \sqrt{-g}
\Big(\mathcal{R}+{\beta \over
D-4}~\mathrm{G}\Big)+\mathcal{S}_{matter}\,,
\end{eqnarray}
where ${\cal R}$ is the Ricci scalar curvature and $\mathrm{G}$ is
the Gauss-Bonnet invariant  associated with the $D$-dimensional
spacetime. The dimensionful parameter $\beta$ is the Gauss-Bonnet
coupling parameter and  $\mathcal{S}_{\text{matter}}$ is the
action  contributed from the matter perfect fluid. The
Gauss-Bonnet term, $\mathrm{G}$, takes the form:
\begin{eqnarray}
\mathrm{G} = \mathcal{R}^2-4\mathcal{R}_{\mu \nu} \mathcal{R}^{\mu \nu}+\mathcal{R}_{\mu \nu \kappa \lambda} \mathcal{R}^{\mu \nu \kappa \lambda}\,,
\end{eqnarray}
where ${R}_{\mu \nu}$ and $\mathcal{R}_{\mu \nu \kappa \lambda}$
are the Ricci curvature and Riemann curvature tensors,
respectively. Variation of the action  \eqref{e2} with respect to
the metric tensor yields:
\begin{eqnarray}
\mathcal{G}_{\mu \nu} + {\beta \over N-4}~\mathcal{H}_{\mu \nu}=8\pi \mathcal{T}_{\mu \nu}\,, \label{e4}
\end{eqnarray}
where the Einstein tensor ${\cal G}_{\mu \nu}$, the Lancoz tensor ${\cal H}_{\mu \nu}$ and the stress tensor ${\cal T}_{\mu \nu}$ are defined respectively as:
\begin{eqnarray}
\mathcal{G}_{\mu \nu} &=& \mathcal{R}_{\mu \nu} -{1 \over 2}~\mathcal{R}~g_{\mu \nu}~,\\
\mathcal{H}_{\mu \nu} &=& 2\Big(\mathcal{R} \mathcal{R}_{\mu \nu}-2\mathcal{R}_{\mu \lambda} \mathcal{R}^\lambda_\nu-2\mathcal{R}_{\mu \lambda \nu \rho} \mathcal{R}^{\lambda \rho}
- \mathcal{R}_{\mu \alpha \beta \gamma} \mathcal{R}^{\alpha \beta \gamma}_\nu \Big)-{1 \over 2}~g_{\mu \nu} \mathrm{G}\,,\\
\mathcal{T}_{\mu \nu} &=& -{2 \over \sqrt{-g}}~{\delta (\sqrt{-g} ~\mathcal{S}_m) \over \delta g^{\mu \nu}}\,.
\end{eqnarray}
The field equation \eqref{e4} has no meaning  when $D=4$. In spite
of this, Glavan  et. al., \cite{Glavan:2019inb,Ghosh:2020vpc} show
that using a re-scale of the parameter $\beta$ as   $\beta
\rightarrow \beta / (D-4)$ and by specifying spacetimes of
curvature scale $\mathrm{K}$ which are maximally symmetric that
set the variation of  GB contribution as:
\begin{eqnarray}
{g_{\mu \lambda} \over \sqrt{-g}}~{\delta \mathrm{G} \over \delta
g_{\nu \lambda}}={\beta(D-2)(D-3) \over
2(D-1)}~\mathrm{K}^2~\delta^\nu_\mu\,,
\end{eqnarray}
that shows in  a clear way the non-vanishing of $\mathrm{G}$ when
$D=4$.

Now we consider a spacetime of  $D-$dimensions that has the form:
\begin{eqnarray}
ds^2 &=&- e^{2h(r)} dt^2+e^{2h_1(r)} dr^2+r^2 d\Omega^2_{D-2}\,,
\label{e9} \label{e9}
\end{eqnarray}
where, $d\Omega^2_{D-2}$ refers to the $(D-2)$-dimensional surface
of a unit sphere.

As shown in \cite{Glavan:2019inb}, the effective 4-dimensional
theory is obtained in the singular limit $D\to 4$ with
$\beta=\beta/(D-4)$.  Using the limit $D\to 4$ we derive the
dimensionally reduced field equations as:
\begin{eqnarray}\label{DRE1}
&&8 \pi \rho=\frac{\beta (1-e^{-2h_1 })}{r^3}\left[ 4h'_1 e^{-2h_1 }-\frac{(1-e^{-2h_1 })}{r}\right]+e^{-2h_1}\left( {2h'_1 \over r}-{1\over r^2}\right) +{1\over r^2}\,, \\ \notag \\
&&8\pi p_r =\frac{\beta(1-e^{-2h_1 })}{r^3}\left[ 4h' e^{-2h_1 }+\frac{(1-e^{-2h_1})}{r}\right] +e^{-2h_1}\left( {2h' \over r}+{1\over r^2}\right) -{1\over r^2}\,, \\ \notag \\
 &&8 \pi p_\theta=e^{-2h_1 }\left[ h''+h'^2+\frac{1}{r}\left( h'-h'_1 \right) \right. \left. +h'h'_1 \left( \frac{8\beta e^{-2h_1} }{r^2}-1\right) - \frac{2\beta (1-e^{-2h_1})}{r^2} \right. \nonumber \\
 &&\left. \times \Bigg \{ \frac{1}{r}\left( h'-h'_1 \right) \right. \left. -2\left( h''+h'^2 -h' h'_1 \right) + \frac{1}{r^2}(e^{2h_1 }-1) \Bigg \} \right]
 \,,  \label{DRE3}
\end{eqnarray}
where the ``prime'' indicates differentiation with respect to the
radial coordinate.

Moreover, we suppose that the stress energy-momentum tensor of
anisotropic fluid has the form:
\begin{eqnarray}
\mathcal{T}_{\mu \nu} = (\rho+p_\theta)u_\mu u_\nu+p_\theta g_{\mu \nu}+(p_r-p_\theta)\xi_\nu\xi^\mu,
\end{eqnarray}
where $\rho$ is the energy-density, $p_r$ and $p_\theta$ are the
radial and tangential pressures, where $u^\mu$ is the time-like
vector defined as $u^\mu = [1, 0, 0, 0]$ and $\xi_\mu$ is the unit
radial vector with its spacelike property, defined by $\xi_\mu =
[0, 1, 0, 0]$ such that $u^\mu u_\mu=-1$ and $\xi^\mu \xi_\mu=1$.
In the present study, we shall consider the matter content
described by the energy density $\rho$, radial and tangential
pressures, $p_r$ and $p_\theta$ respectively\footnote{Throughout this study we shall use geometrized units in which $G=c=1$.}.
quantities

The above system of differential equations,
(\ref{DRE1})--(\ref{DRE3}),  constitutes three differential
equations with five variables, thus we need two extra conditions.
In this study we are going to assume the metric potentials,
 $g_{tt}$ and $g_{rr}$ to have special forms so that the resulting model can be consistent with a real compact star. Here we assume
 \begin{eqnarray}\label{metpot}
h(r) =-4\ln(c_1-c_0r^2)\,, \qquad \qquad h_1(r) =-4\ln(1+c_0r^2)\,,
\end{eqnarray}
where $c_0$ and $c_1$ are constants that will be fixed from the
junction conditions.  { We assume the ansatz given by Eq. (\ref{metpot}) so that we can get non-singular values of energy-density, radial and tangential pressures and also finite values of these quantities at the center of the star as we will show in the following sections}.
Using Eq. (\ref{metpot}) in Eqs.
(\ref{DRE1})--(\ref{DRE3}) we get the form of energy-density,
radial and tangential pressures which are quite lengthy to quote
them here and therefore, we present these in Appendix A. In the
next section we are going to fix the two constants using the
junction conditions, i.e., we will match the interior solution
with exact exterior solution.


\section{Junction conditions }\label{S3}

The vacuum spherically symmetric  solution in Gauss-Bonnet
gravitational theory is  derived in \cite{Glavan:2019inb} and has
the following form in the limit $D\rightarrow 4$:
\begin{eqnarray}
ds^2 &=&- A(r) dt^2 + {dr^2 \over A(r)} + r^2 d\Omega_2^2,
\end{eqnarray}
where, the unknown function $A(r)$ has the form
\begin{eqnarray}
A(r) &=& 1+{r^2 \over 32\pi \beta} \left[1 \pm \left\{1+{128\pi
\beta m \over r^3} \right\}^{1/2} \right]\,.
\label{e27}\end{eqnarray}
 In this study, we consider the negative
branch in Eq. \eqref{e27} that yields\footnote{ The matching
condition of the  positive branch of Eq. \eqref{e27} yields
unphysical quantities, that is, negative values of density, radial
and tangential pressures. So we will not consider this case in
this study.}:
{
\begin{eqnarray}\label{ref}
A(r) &=& 1+{r^2 \over 32\pi { \beta}} \left[1 - \left\{1+{128\pi { \beta} M \over r^3} \right\}^{1/2} \right]\approx 1-\frac{2M}{r}+\frac{64\pi \beta M^2}{r^4}. \label{e29}
\end{eqnarray}
Equation (\ref{ref}) shows that the direct effect of the GB parameter will be of order $\mathcal{O}\left( r^{-4} \right)$ and in case we neglect it we will return to the  usual Schwarzschild solution.}
Now matching the interior and exterior spacetimes at the boundary
$r=l$, i.e., $e^{2h(l)} = e^{-2h_1(l)}=A(l)$ yields the value of
the two constants $c_0$ and $c_1$ which are:
{
\begingroup
\small
\begin{eqnarray}
c_0 &=&\frac{\sqrt {2}\sqrt [4]{\frac{2\, \left( 32\,\pi\,\beta+{l}^{2}-{l}^{2}\sqrt {{\frac {{l}^{3}+128\,\pi\,\beta\,M}{{l}^{3}}}} \right)}{ {
\pi}{\alpha}}}-4}{ 4{l}^{2}}
\,,\label{const}\\
c_1 &=& \frac {1}{2\sqrt [4]{\pi\beta} \parallel\mathbb{B}\parallel \left[2\sqrt [4]{2}\parallel\mathbb{B}\parallel-4\,\sqrt [4]{\pi}\sqrt [4]{\beta}\right] }\Bigg\{\sqrt {2}\parallel\mathbb{B}^{3}\parallel-8\sqrt [4]{2\pi\beta}\mathbb{B}^2+8\,\sqrt {\pi\beta}
\mathbb{B}\nonumber\\
&&-8\,\sqrt {
\pi\beta}\sqrt [4]{[8\,\sqrt {\pi\beta}-8\sqrt [4]
{2\pi\beta}\parallel\mathbb{B}\parallel+\sqrt {2}\mathbb{B}^2][2\sqrt [4]{2}\parallel\mathbb{B}\parallel-4\,\sqrt [4]{\pi\beta}]}\Bigg\}\,,\label{const1}
\end{eqnarray}
\endgroup
where $\mathbb{B}=\sqrt [4]{\sqrt {l}\sqrt
{{l}^{3}+128\,\pi\, \beta\,M} -32\,\pi\,\beta-{l}^{2}}$.}  Now we are going to calculate the derivative of the metric potentials (\ref{metpot}) and the exterior solution given by Eq. (\ref{ref}) and get:
\begin{eqnarray}\label{ref1}
&&g'_{tt}(r)=\frac{dg_{tt}(r)}{dr}=\frac{8rc_0}{(c_1-c_0 r^2)^5}\,, \qquad  g'_{rr}(r)=\frac{dg_{rr}(r)}{dr}=-\frac{8rc_0}{(1+c_0 r^2)^5}\,, \nonumber\\
&& A'(r)=\frac{dA(r)}{dr}=\frac{r^{3/2}\sqrt{r^3+128\pi \beta M}+r^3+32\pi \beta M}{\pi \beta\sqrt{r^4+128r \pi \beta M}}\,.
\end{eqnarray}

{ The second junction condition should be checked as well which deals with the first derivative of the metric with respect to the radial coordinate. Having the results in Eq. (\ref{ref1}), one can check the second junction condition to see whether it trivially satisfies or one needs to consider a shell with an appropriate energy-momentum tensor.}
\section{Physics of the solution given in Appendix A}\label{S4}

Now, we are ready to examine if the interior solution which is
given in Appendix A can describe a realistic star. For such a
purpose, we are going to derive some necessary conditions which
must be satisfied in order to have a real star.

\subsection{Energy--momentum tensor}

\begin{figure}
\centering \subfigure[~The energy-density given by the first
equation of Appendix
A]{\label{fig:1a}\includegraphics[scale=0.3]{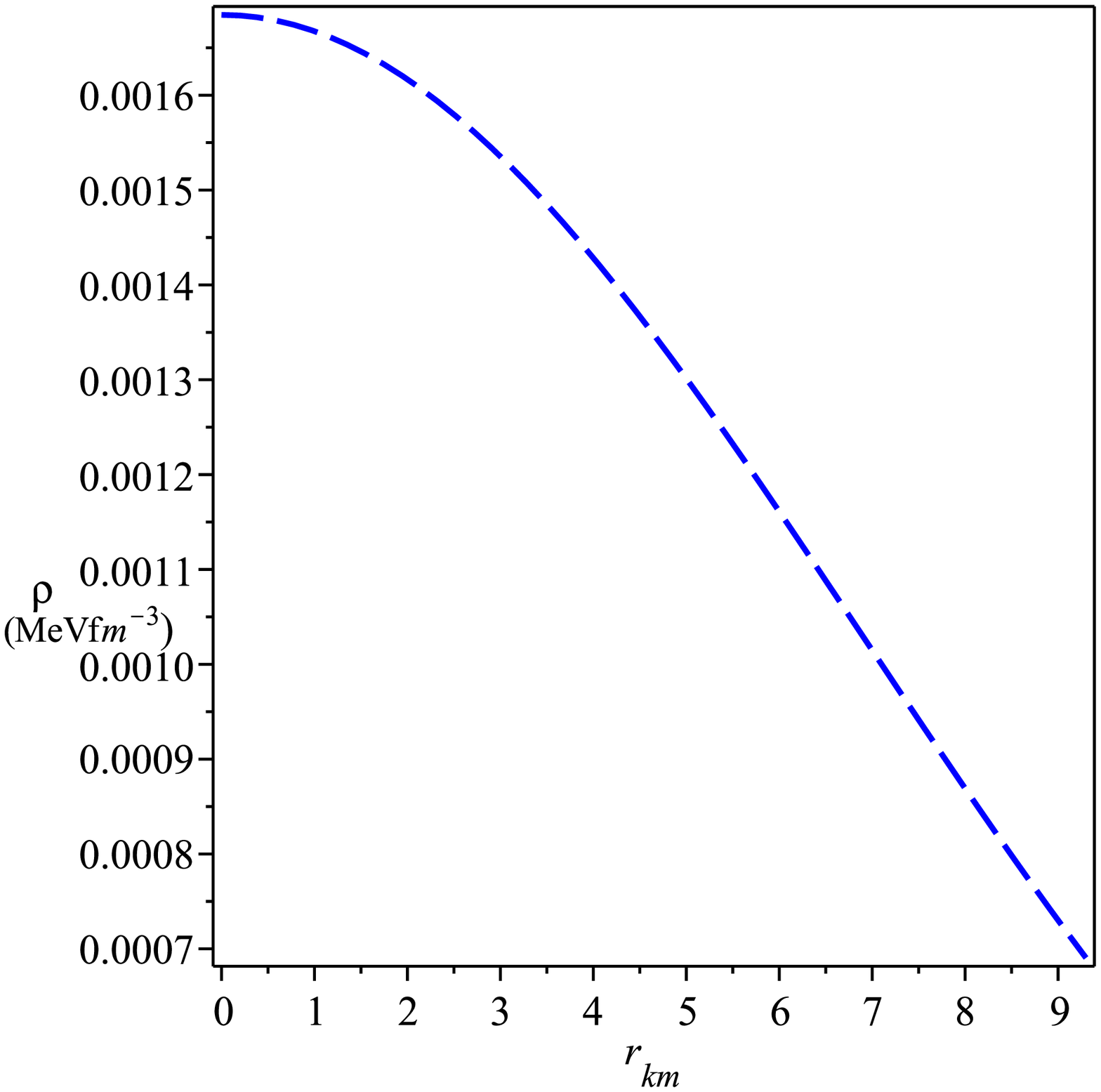}}
\subfigure[The radial pressure given  by the second equation of
Appendix
A]{\label{fig:1b}\includegraphics[scale=0.3]{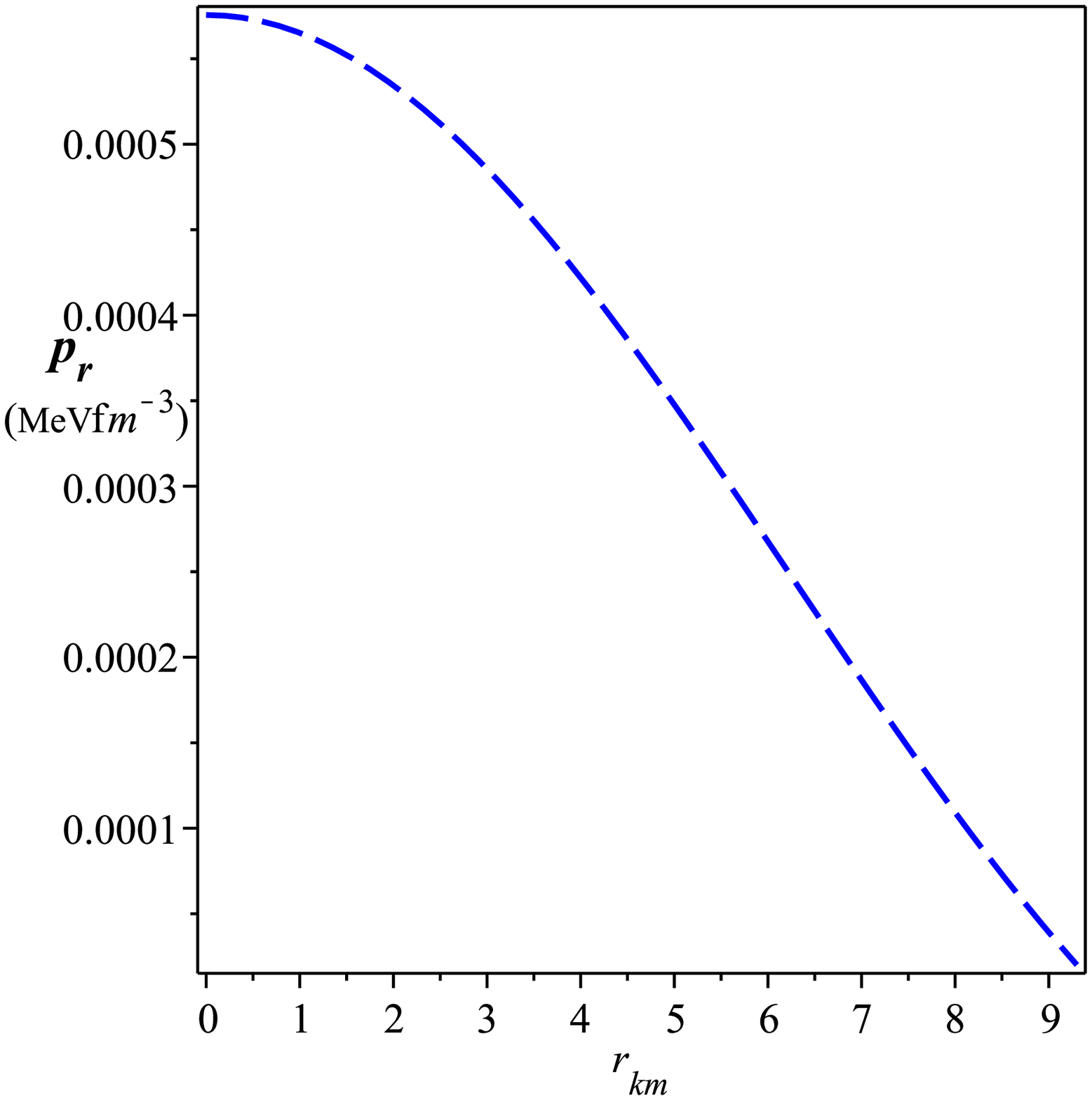}}
\subfigure[~The tangential pressure given  by the third equation
of Appendix
A]{\label{fig:1c}\includegraphics[scale=0.3]{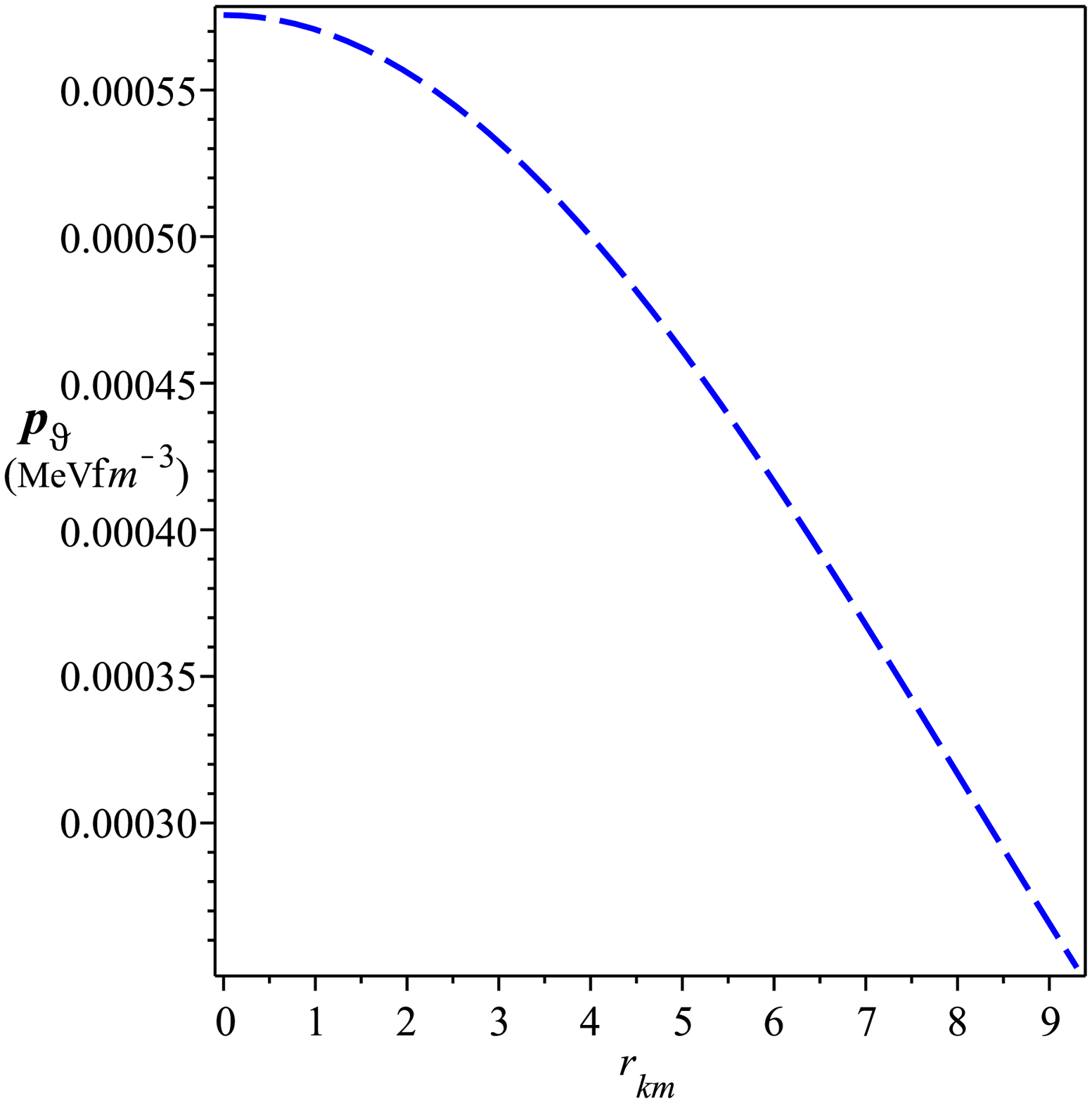}}
\subfigure[~The anisotropy $\Delta$ and anisotropic force of the
model given by the fourth equation of Appendix
A]{\label{fig:1d}\includegraphics[scale=0.3]{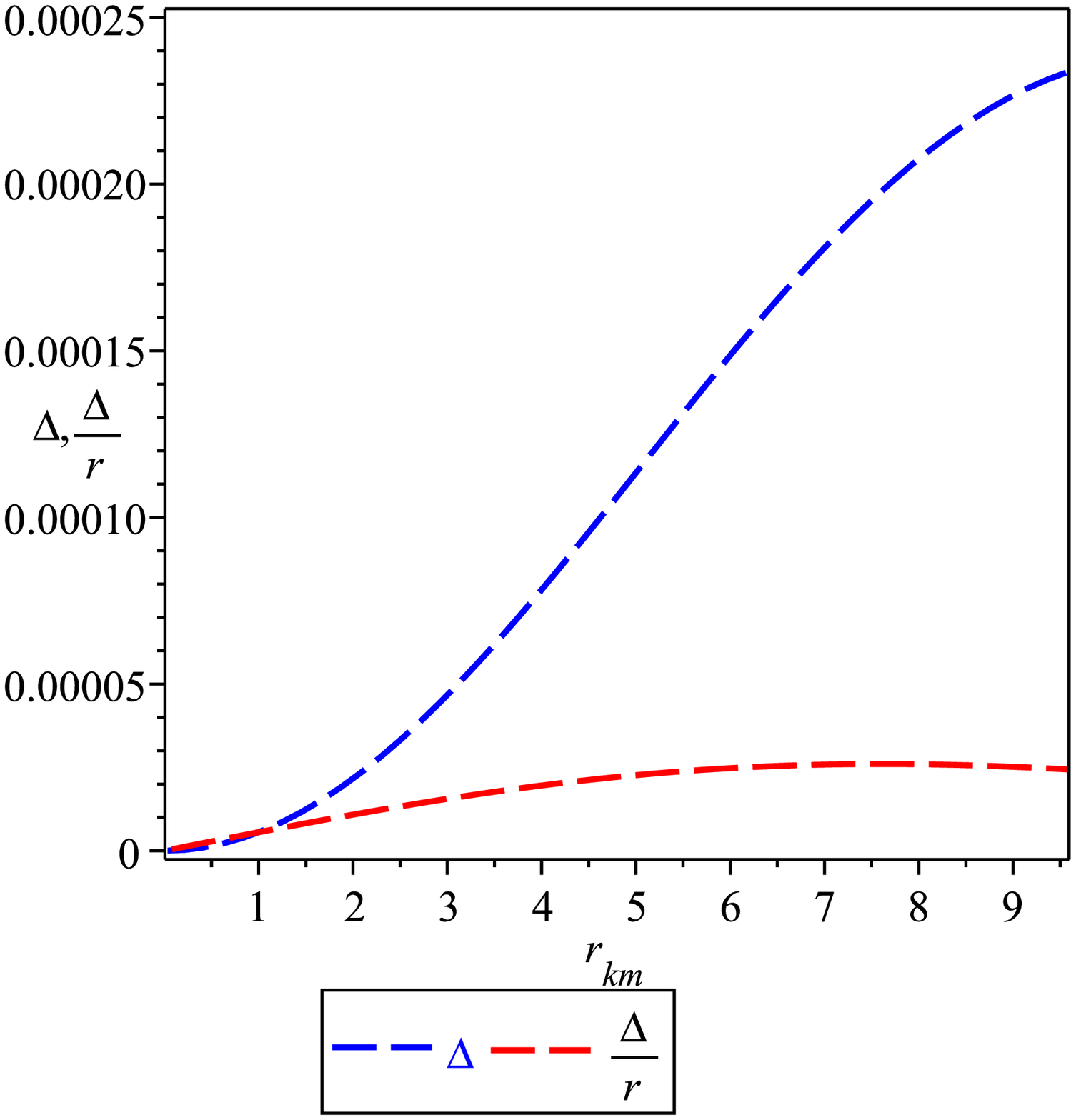}}
\caption[figtopcap]{\small{{Plot of the components of the
energy-momentum tensor and anisotropic force where we put
$\beta=-0.0001$ and the constants $c_0$ and $c_1$ take the
numerical values, $-0.001763358117$ and $-0.9329227851$
respectively.}}} \label{Fig:1}
\end{figure}

For any real interior solution, we must have a positive value of
all the components of the energy-momentum tensor, i.e., the
energy-density, and the radial, and transverse pressures should
have positive well-defined values. Additionally, all these
components must be finite at the center of the star and should
decrease in the radial direction towards the surface of the star.
{ Finally, the tangential pressure should exceed the radial  one at the center of the star}. { In
the present study we shall consider the stellar model of the
observed pulsar EXO $1785-248$ whose mass is $M=1.3\pm0.2 \, {\textrm M}_\odot$ and its
radius $l=8.849\pm0.4 \,{\textrm km}$ for which the constants $c_0$ and $c_1$ can
be calculated numerically. We must stress that the model solution
given in Appendix A cannot be reduced to the GR solution because
the parameter $\beta$ is not allowed to take zero values. We
depict the components of energy-momentum in Fig. \ref{Fig:1}. The
numerical values of the parameters $c_0$ and $c_1$ used in Fig.
\ref{Fig:1} are $c_0=-0.001763358117,\qquad c_1=-0.9329227851$.}
Moreover, $\rho(r\to 0)=0.00168473468$,\,\, ${p_r}(r\to
0)=0.0005755839216$,\, and ${p}_\theta(r\to 0)=0.0005755799012$.
Figure \ref{Fig:1} ensures that the energy-density, radial and
tangential pressures are decreasing towards the surface stellar.
Also, in  Fig.\ref{Fig:1} \subref{fig:1d}, we present  the
behavior of the anisotropy parameter  which is defined as
$\Delta(r)={p}_\theta-{p_r}$. Moreover, in Fig. \ref{Fig:1}
\subref{fig:1d}  we present the anisotropic force that has a
positive behavior which means that we have a repulsive force, due
to the fact $p_\theta\geq p_r$.

\subsection{Causality}

\begin{figure}
\centering \subfigure[~The gradient of the energy-momentum
components  given by the  three equations presented Appendix B
]{\label{fig:2a}\includegraphics[scale=0.3]{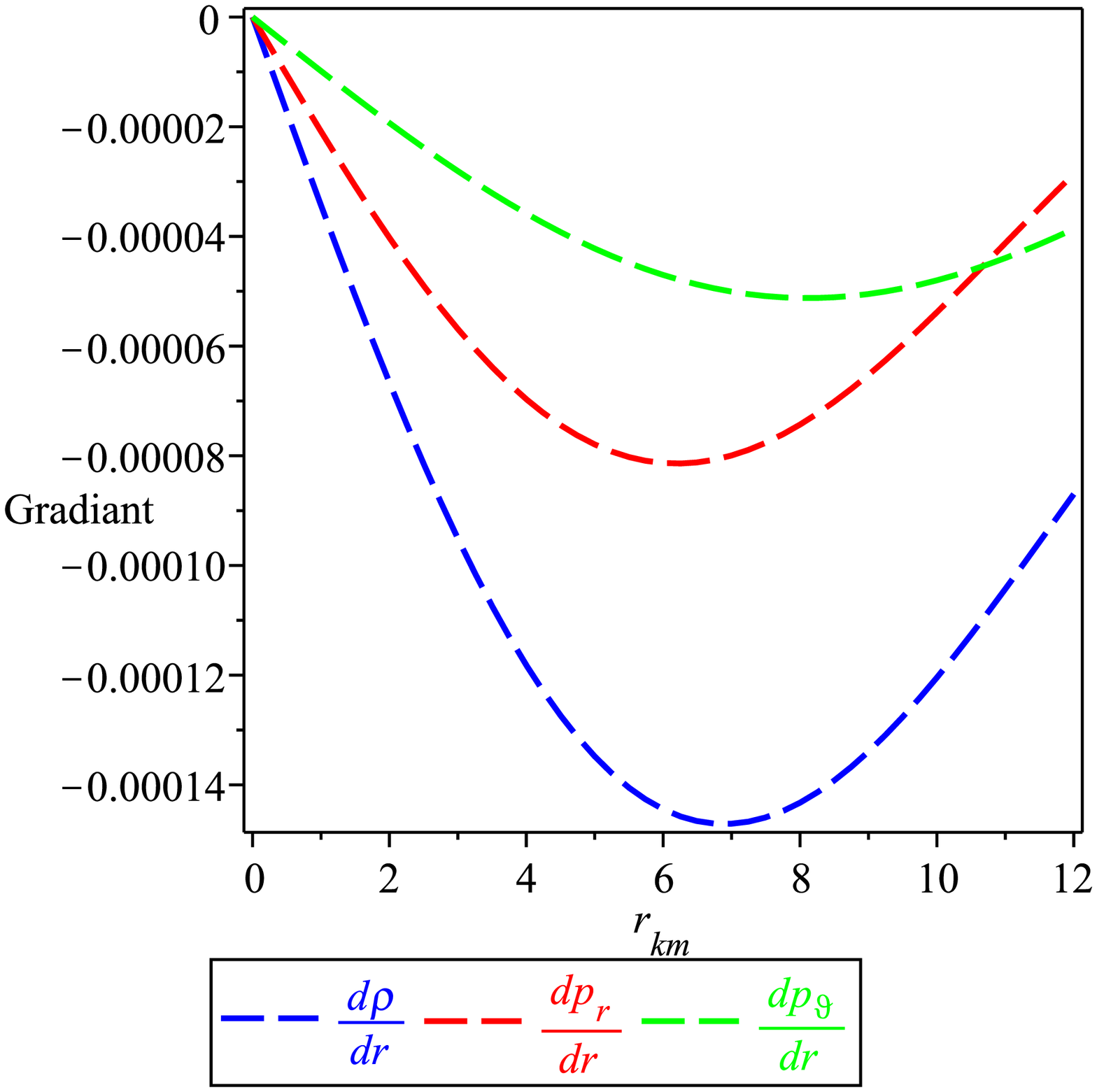}}
\subfigure[~The  speed of sound given by the  two equations
presented Appendix C]
{\label{fig:2b}\includegraphics[scale=0.3]{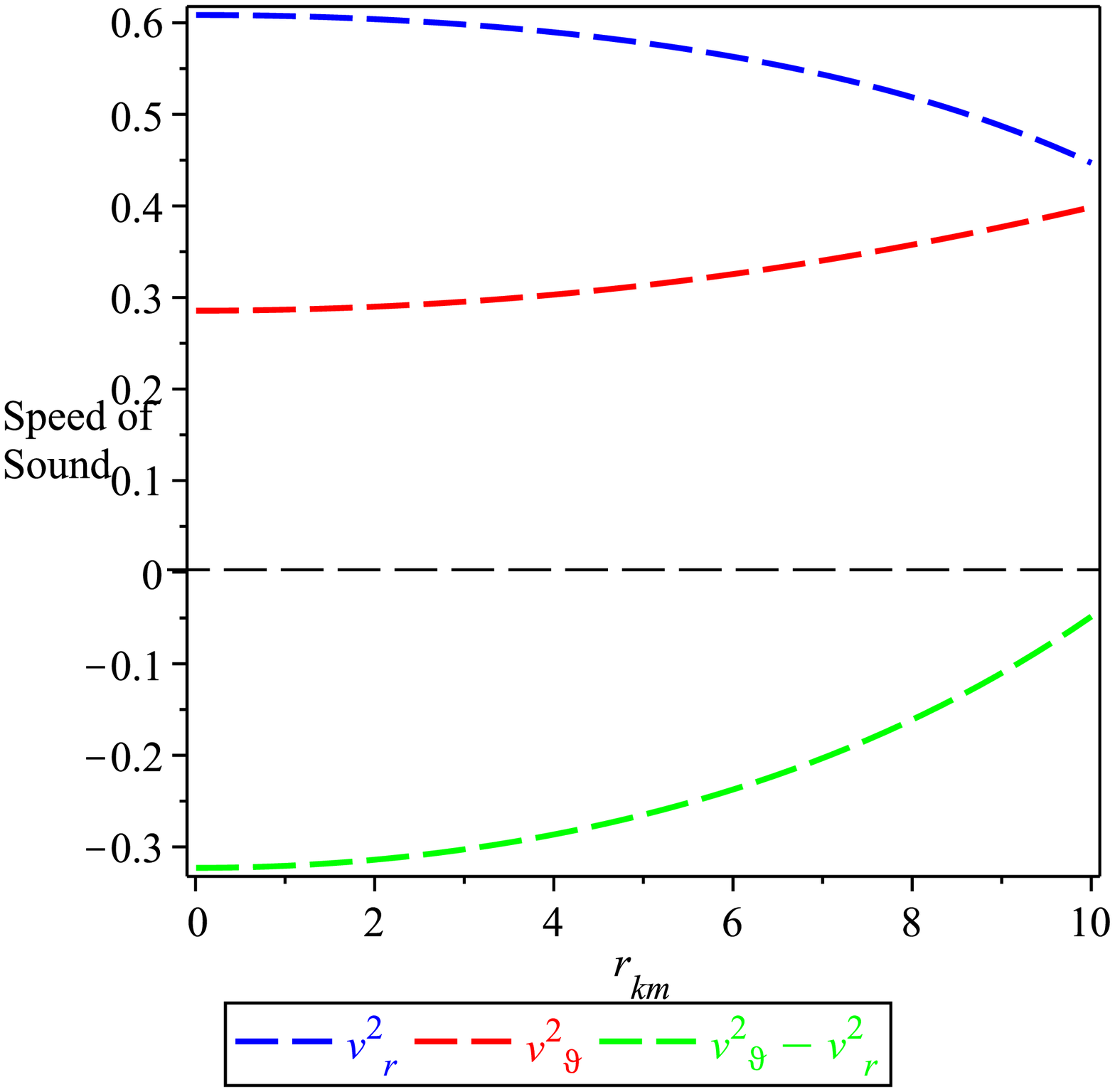}}
%
\caption[figtopcap]{\small{{The plot of the gradients of the
energy-momentum components  where we put $\beta=-0.0001$ and the
constants $c_0$ and $c_1$ take the numerical values,
$-0.001763358117$ and $-0.9329227851$ respectively.}}}
\label{Fig:2}
\end{figure}

To test the behavior of the  sound velocities, we evaluate the
derivative of the energy--density, and the radial and tangential
pressures, the final expressions of which we present in the
Appendix B. The equations given in Appendix B do not inform us if
the gradients of the components of the energy-momentum tensor have
a positive or negative behavior, therefore, we plot them in  Fig.
\ref{Fig:2}  \subref{fig:2a} and from the plots it is ensured that
the gradients of the energy-momentum tensor are negative, as it is
required for any realistic physical stellar.
To  verify the causality conditions we must prove that the radial
and transverse sound speeds, $v_r{}^2$ and $v_\bot{}^2$,  have
values less than the speed of light. We evaluate these quantities,
$v_r{}^2$ and $v_\bot{}^2$,  and we present their functional forms
in the Appendix C :
We plot the equations given in Appendix  C  in Fig. \ref{Fig:2}
\subref{fig:2b} to ensure  the validity of the conditions
$1>v_r{}^2>0$ and $1>v_t{}^2>0$.

The appearance of non--vanishing radial force with different signs
in different regions of the fluid is called {\it gravitational
cracking}. This happen when the radial force is directed inwards
in the inner part of the sphere for all values of the radial
coordinate $r$ between the center, and some value beyond which the
force reverses its direction \cite{1994PhLA..188..402H}. It is
shown in Ref. \cite{Abreu:2007ew}, that a simple requirement to
avoid gravitational cracking is $0>v_\theta{}^2- v_r{}^2>-1$. In
Fig.  \ref{Fig:2} \subref{fig:2b},  we show that the solution
given in Appendix C is stable against cracking.
\begin{figure}
\centering
\subfigure[~The DEC,  $\rho-p_r$ and $\rho-p_\theta$]{\label{fig:3a}\includegraphics[scale=0.3]{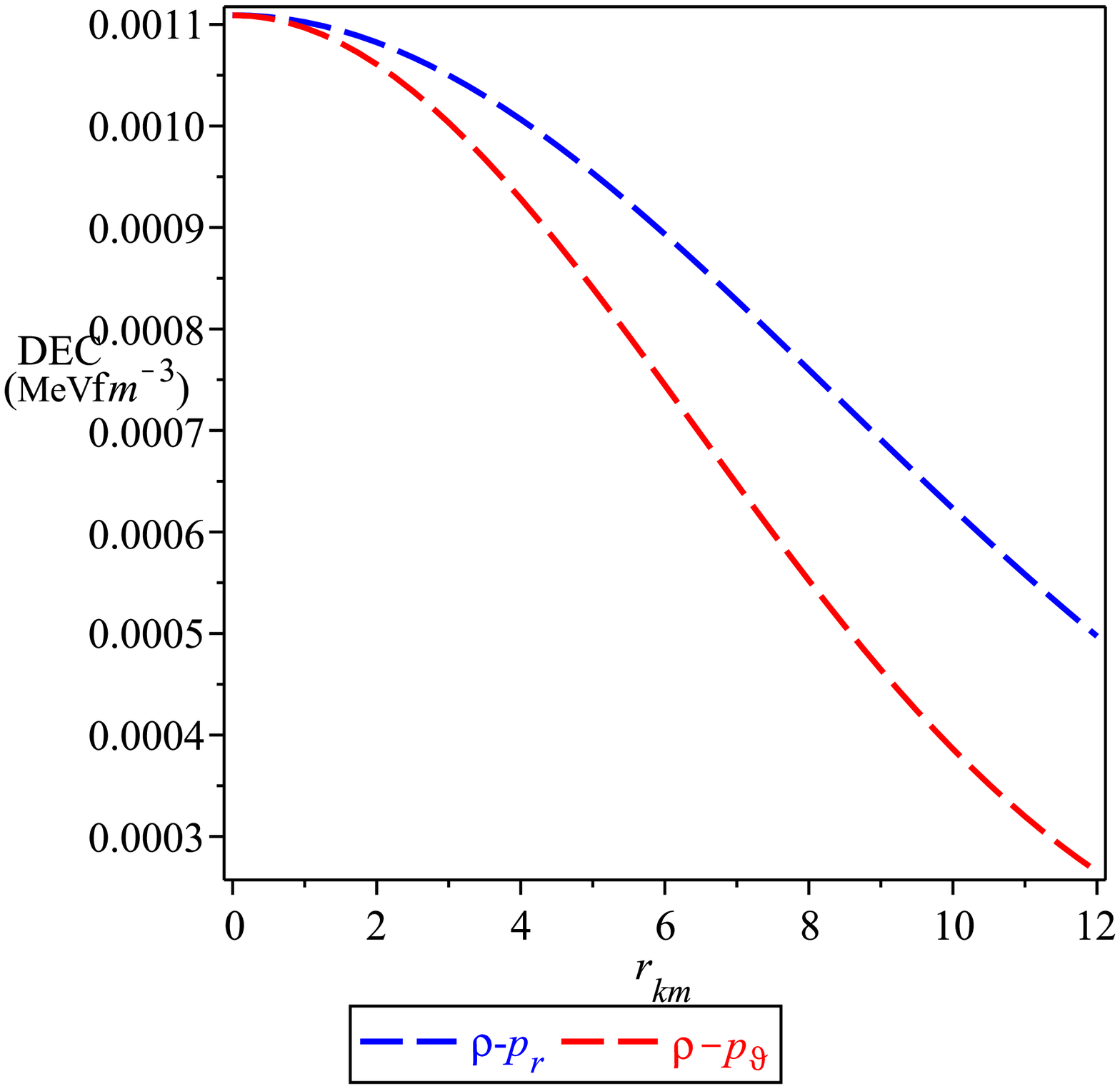}}
\subfigure[~The WEC,   $\rho$ and $\rho+p_r$]{\label{fig:3b}\includegraphics[scale=.3]{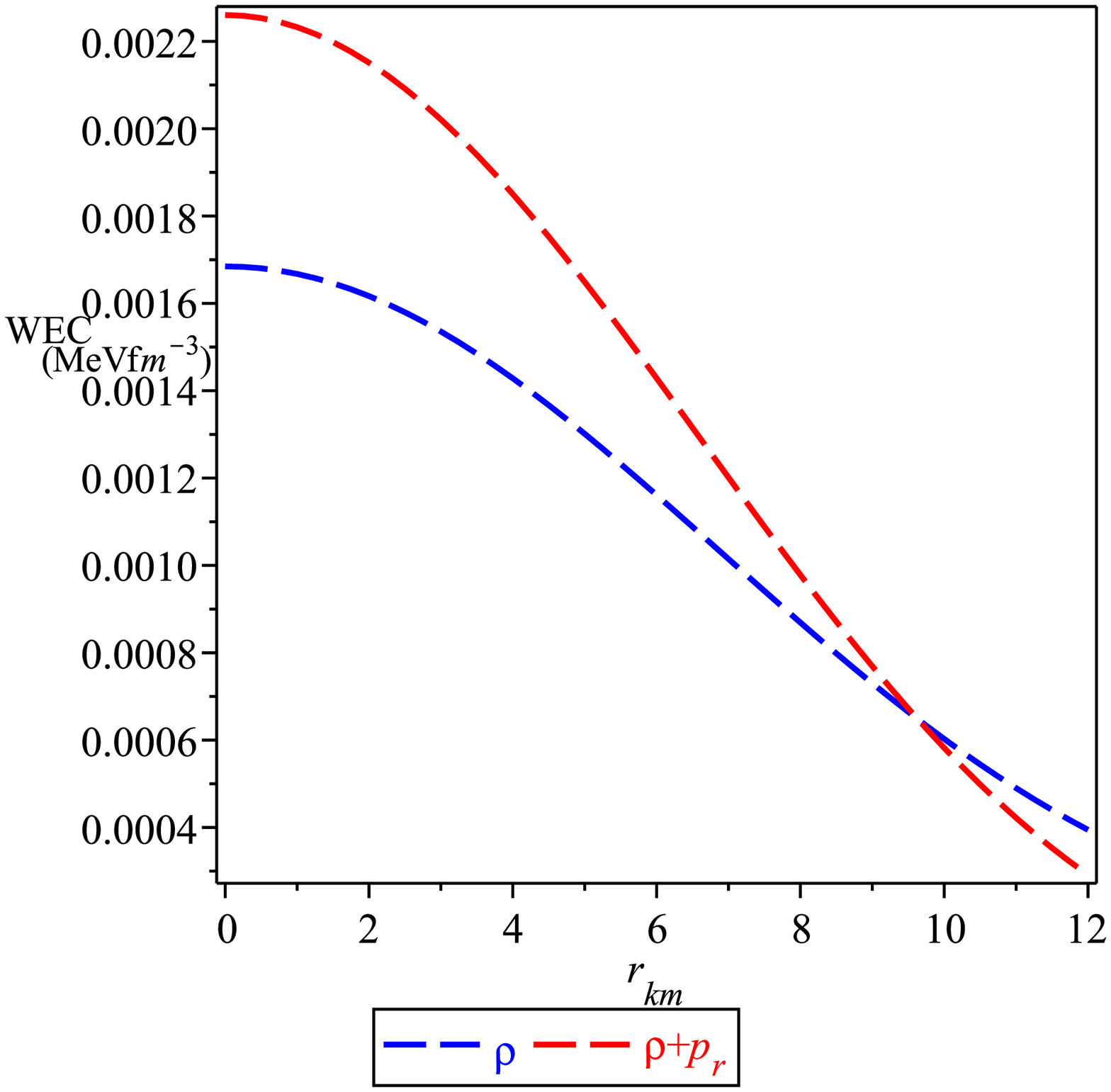}}
\subfigure[~The NEC,  $\rho$ and $\rho+p_\theta$]{\label{fig:3c}\includegraphics[scale=.3]{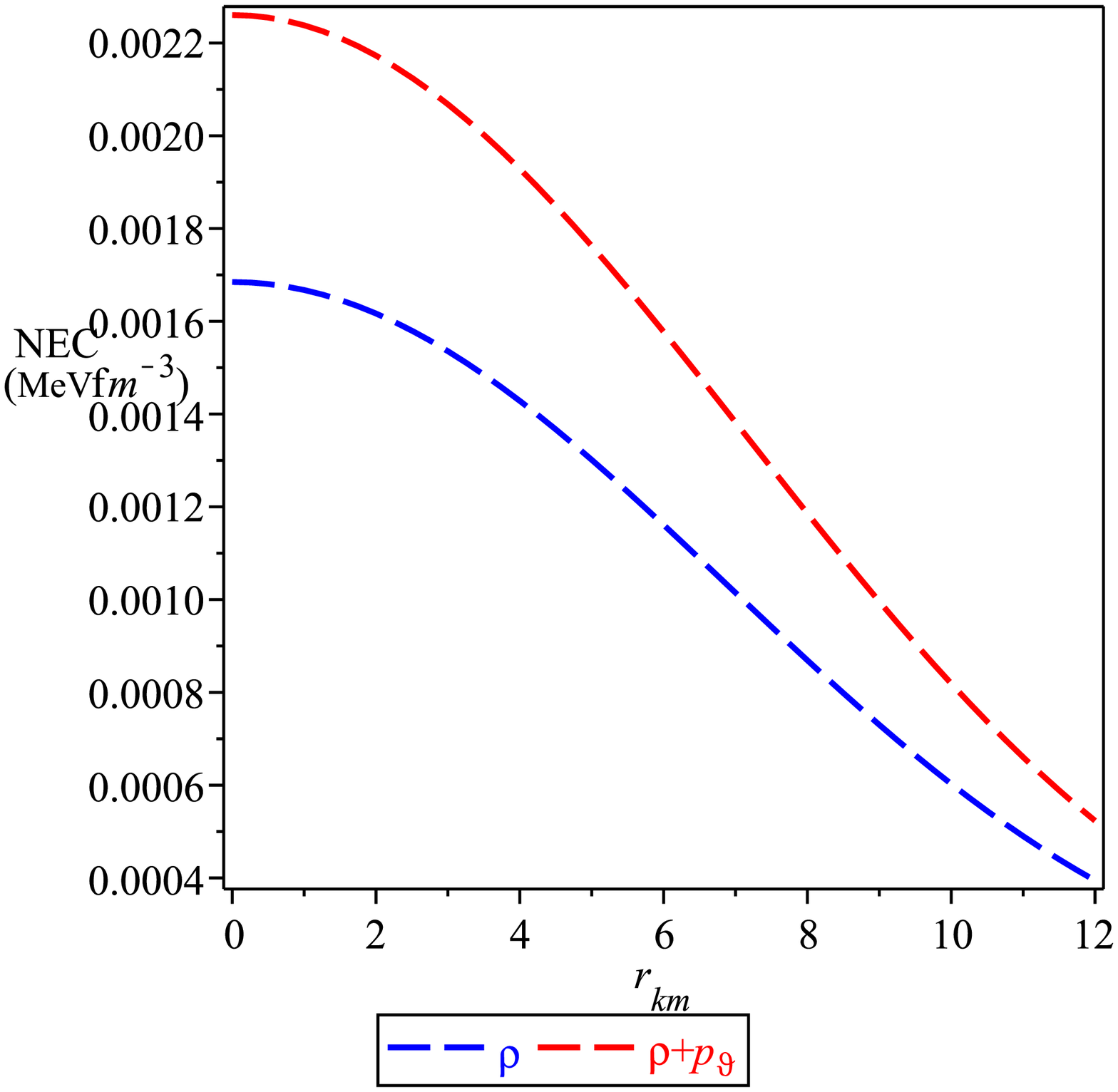}}
\subfigure[~The SEC,   $\rho$, $\rho+p_r$ and $\rho-p_r-2p_\theta$ ]{\label{fig:3d}\includegraphics[scale=.3]{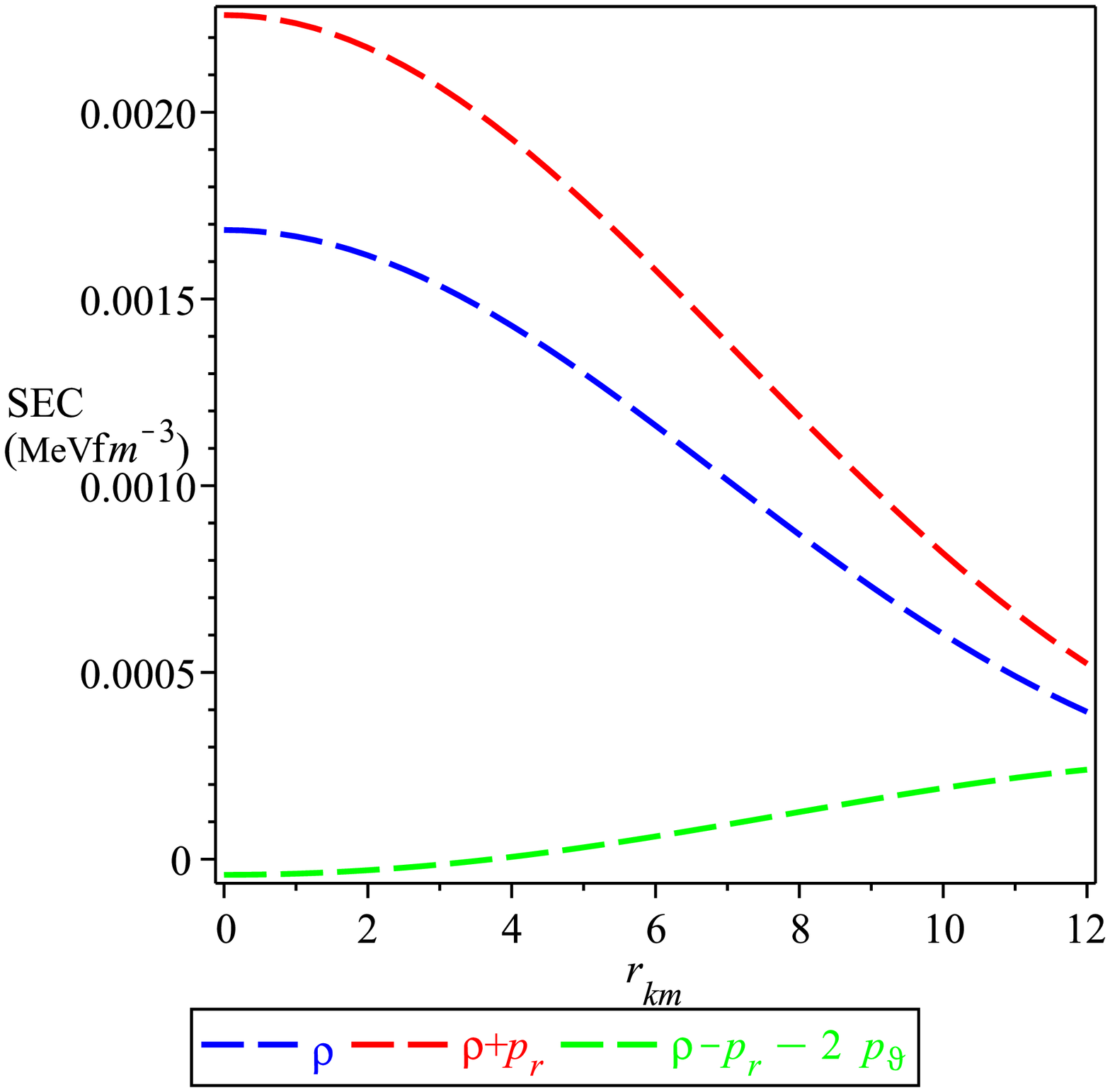}}
\caption[figtopcap]{\small{{DEC, WEC, NEC and SEC for $\beta=-0.0001$,  $c_0=-0.001763358117$ and $c_1=-0.9329227851$.}}}
\label{Fig:3}
\end{figure}

\subsection{Energy conditions}

Energy conditions are considered as important tests for
non--vacuum solutions. To satisfy the dominant energy condition
(DEC), we have to prove that  $\rho-p_r >0$ $\&$
$\rho-p_\theta>0$.  As shown in   Fig.  \ref{Fig:3}
\subref{fig:3a} the DEC is satisfied for suitable choices of the
parameters $\beta$, $c_0$, and $c_1$. Also, in   Fig.  \ref{Fig:3}
\subref{fig:3b}, \subref{fig:3c} and \subref{fig:3d} we show that
the Weak Energy Condition (WEC), the Null Energy Condition (NEC),
and the Strong Energy Condition (SEC)  are all satisfied. { It is interest to note that the problem of energy conditions in the frame of $f(G)$ has been studied  and the results presented in this study are consistent
with the results presented in \cite{Bamba:2017cjr}.}
\subsection{Mass-Radius relation}

\begin{figure}
\centering
\subfigure[~The gravitational mass and compact parameter]{\label{fig:mass}\includegraphics[scale=0.3]{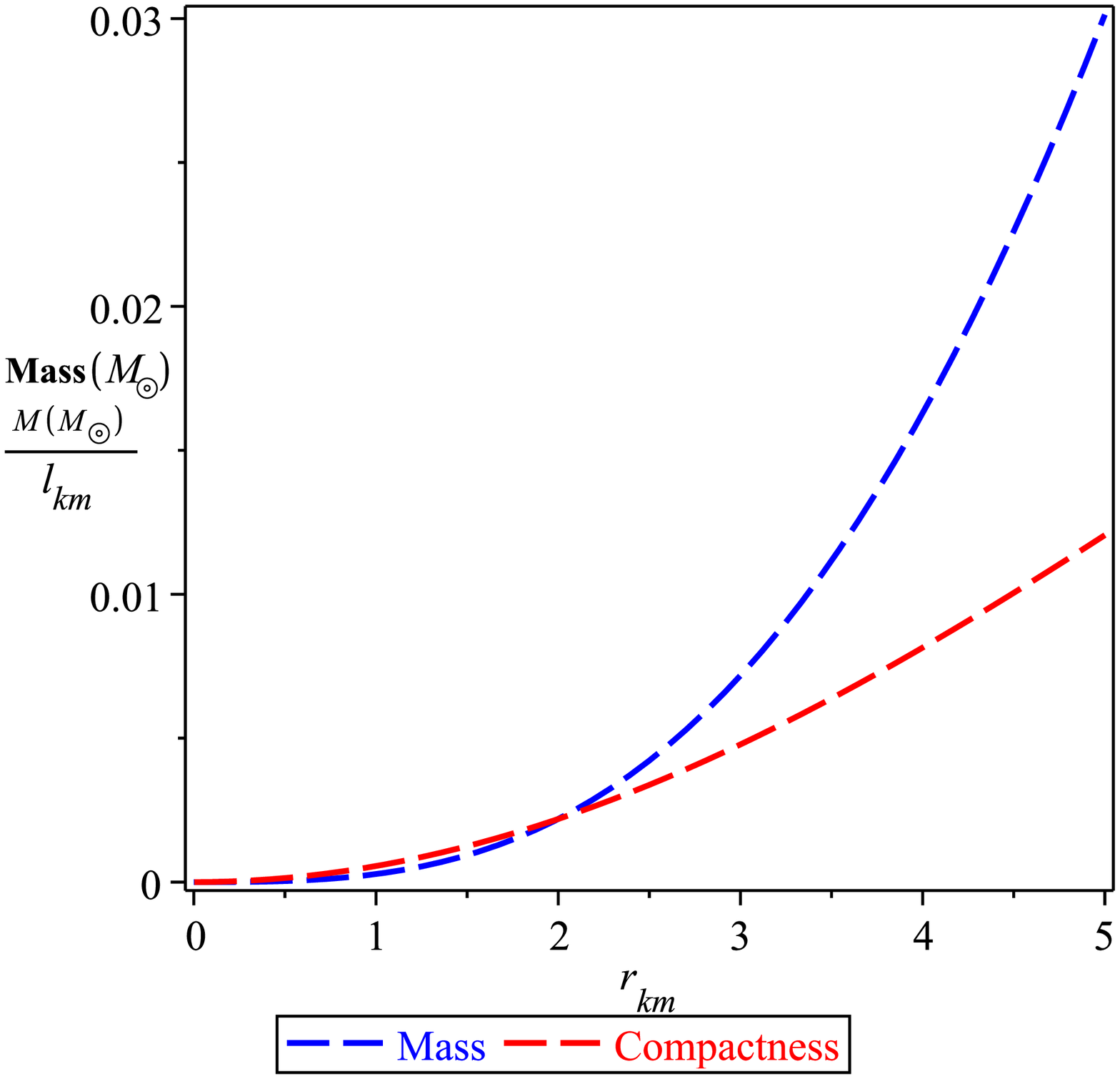}}
\subfigure[~The EoS parameter ]{\label{fig:u}\includegraphics[scale=.3]{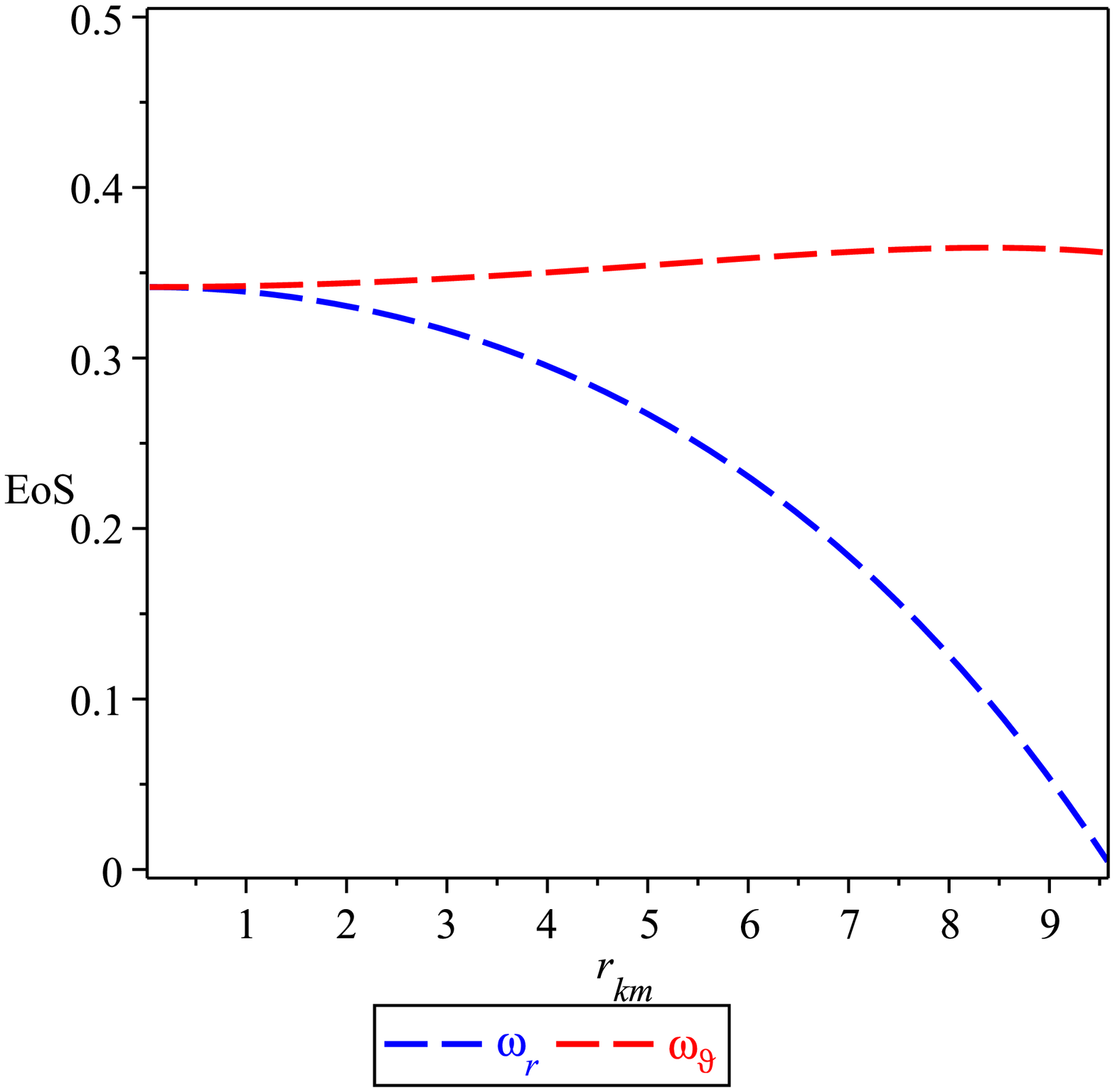}}
\subfigure[~The radial EOS as a function of the energy-density]{\label{fig:pr5}\includegraphics[scale=.3]{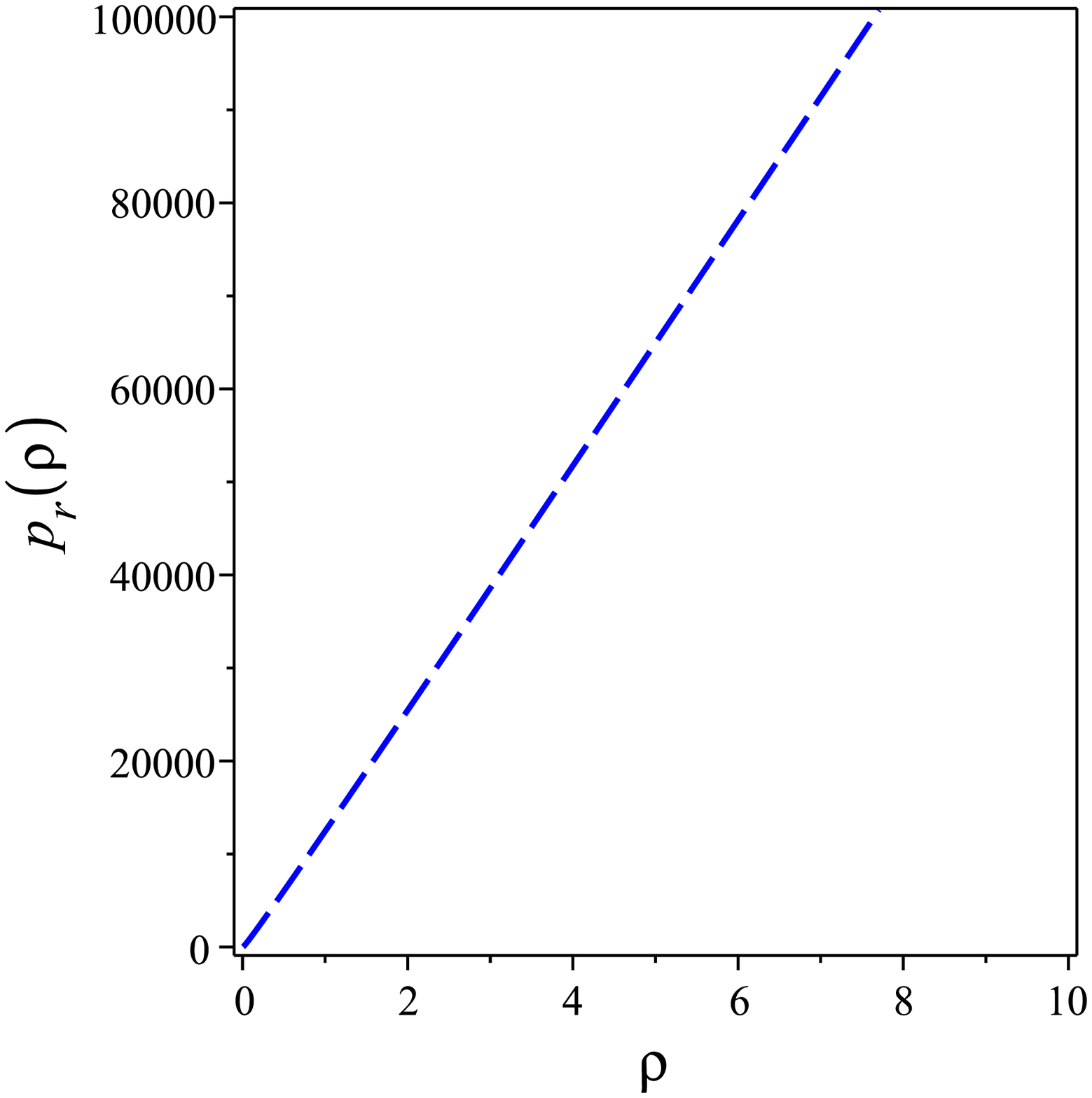}}
\subfigure[~The tangential EOS as a function of the energy-density]{\label{fig:pt5}\includegraphics[scale=.3]{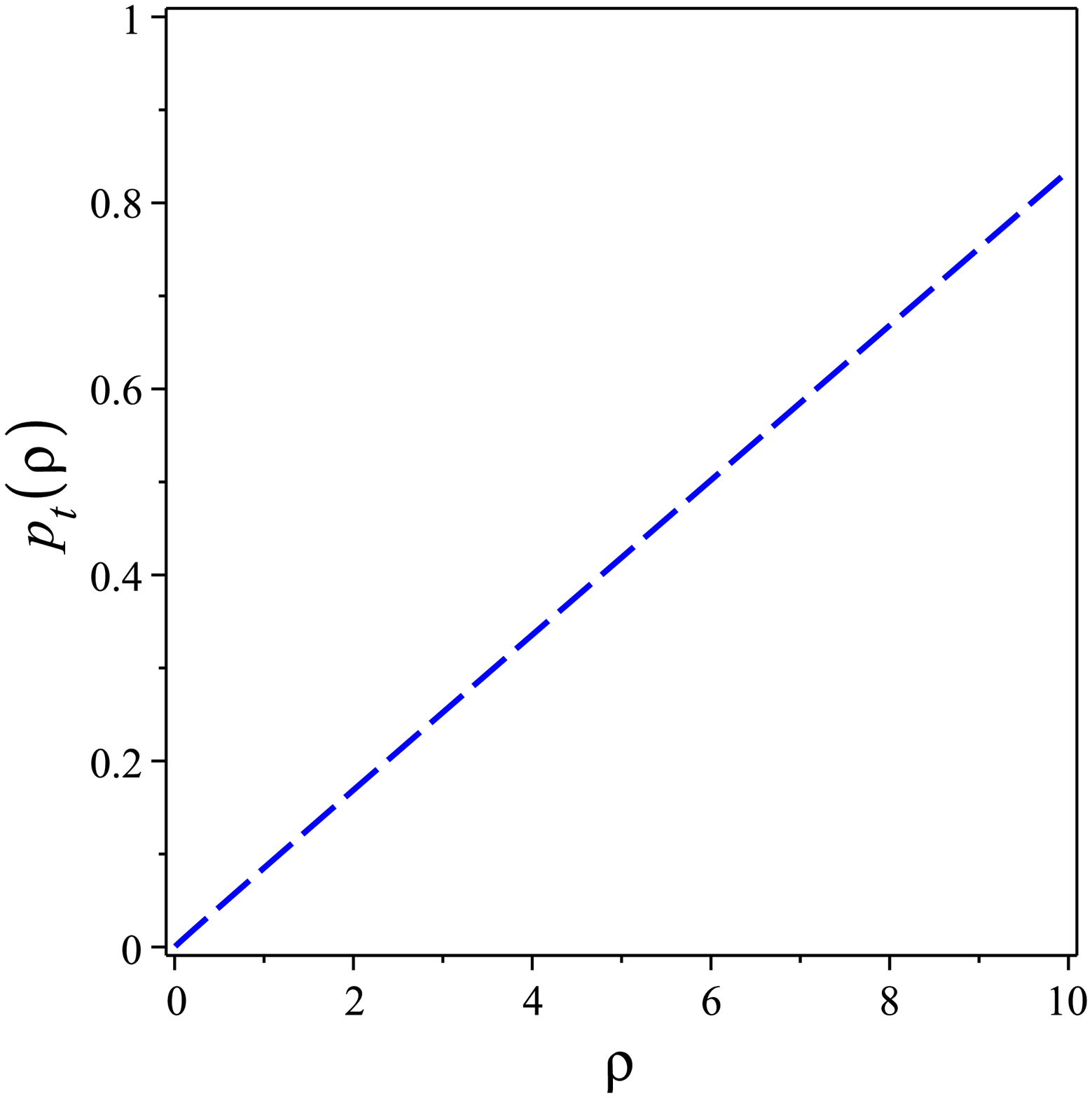}}
\subfigure[~The EOS  as a function of the energy-density]{\label{fig:p5}\includegraphics[scale=.3]{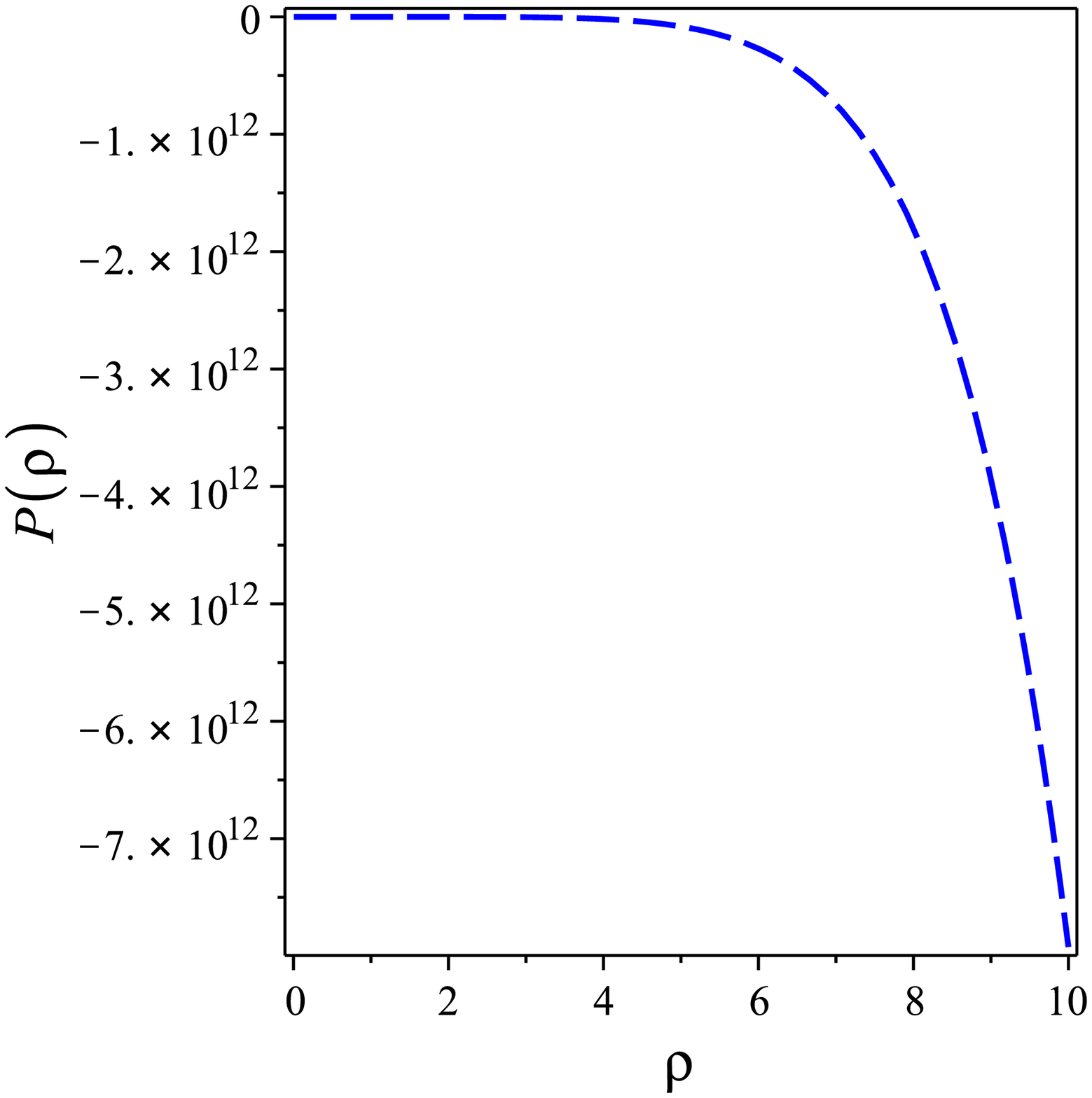}}
\caption[figtopcap]{\small{{Schematic plots of~\subref{fig:mass}
the  gravitational mass and compactification factor  of the
solution  given by Eqs. (\ref{mass}), and (\ref{comp});
~\subref{fig:u} the components of radial and tangential EoS
parameters as function of the radial coordinate; ~\subref{fig:pr5}
the radial EoS as a function of the energy-density;
~\subref{fig:pt5}  the tangential EoS  as a function of the
energy-density and; ~\subref{fig:p5}  the  EoS  as a function of
the energy-density. The values the GB parameter and the constants
$c_0$ and $c_1$ are taken as $\beta=-0.0001$,
$c_0=-0.001763358117$ and $c_1=-0.9329227851$ in these plots.}}}
\label{Fig:4}
\end{figure}

The compactification factor, $u(r)$, is the one  defined as the
ratio between the mass and radius. It has an important role
towards to revealing the physical properties of compact objects.
Starting from the solution  given in Appendix A, we define  the
gravitational mass by the following expression:
 \begin{eqnarray}\label{mass}
&& M(r)={\int_0}^r  \rho(\xi) \xi^2 d\xi=\frac {c_0\,{r}^{3} }{16\pi } \left( c_0\,{r}^{2}+2 \right)
 \left( {c_0}^{2}{r}^{4}+2+2\,c_0\,{r}^{2} \right)  \left( {
c_0}^{4}{r}^{8}+4\,{c_0}^{3}{r}^{6}+6\,{c_0}^{2}{r}^{4}
+4\,c_0\,{r}^{2}+2 \right)\nonumber\\
 &&  \times\left( \beta\,{c_0}^{8}{r}^{14}
+8\,\beta\,{c_0}^{7}{r}^{12}+28\,\beta\,{c_0}^{6}{r}^{10}+
56\,\beta\,{c_0}^{5}{r}^{8}+70\,\beta\,{c_0}^{4}{r}^{6}+56
\,\beta\,{c_0}^{3}{r}^{4}+28\,\beta\,{c_0}^{2}{r}^{2}+8\,
\beta\,c_0-1 \right)\,,
\end{eqnarray}
where we substitute the value of energy-density from the first
equation presented in Appendix A. The compactification factor
$C(r)$ is then defined as:
\begin{eqnarray}\label{comp}
&&C(r)=\frac{M(r)}{l}\,.\end{eqnarray} Substituting Eq.
(\ref{mass}) into (\ref{comp}), one can get the explicit form of
the compactification factor. The behavior of the gravitational
mass and the  compactification factor are plotted in Fig.
\ref{Fig:4} which indicates in a clear way that the gravitational
mass and the compactification factor are directly proportional to
the radial coordinate.


\subsection{Equation of State}

Finally, let us  study the  equation of state (EoS) of a compact
stellar object as presented in  \cite{Das:2019dkn} which uses a
linear EoS.  In the present case, we  show that  the EoS is not
linear due to the contribution of the GB term. To show this, we
define the radial and transverse  EoS  as:
  \begin{eqnarray} \label{sol3}
 && \omega_r=\frac{p_r(r)}{\rho(r)}\,, \qquad \qquad \omega_t=\frac{p_\theta(r)}{\rho(r)}\,,
  \end{eqnarray}
with $\omega_r$ and $\omega_t$ being the radial and transverse EoS
parameters. Using the form of energy-density, radial and
tangential pressures presented in Appendix A we get the EoSs of
our model as presented in Appendix D. As it can be seen in Fig.
\ref{Fig:4} \subref{fig:u}, the EoS is non-linear because of the
contribution of the GB term.

Now, we are going to calculate the EoS as $P(\rho)=\omega \rho$,
for such purpose we write  $r=r(\rho)$. Using the form of
energy-density given in Appendix A we get the form of the radial,
tangential and pressure components in Appendix  D. The behavior of
the EoS as a function of the energy-density is presented in  Fig.
\ref{Fig:4} \subref{fig:pr5}, \subref{fig:pt5}  and
\subref{fig:p5}. As it can be seen in Figs. \ref{Fig:4}
\subref{fig:pr5}, \subref{fig:pt5}  and \subref{fig:p5}, the
radial and tangential EoS parameters are positive, while the total
EoS, i.e., $P=\frac{p_r+2p_\theta}{3}$ is negative.

\section{Stability of the Model} \label{S5}

We  use   two different procedures to study the stability problem,
the first is  the Tolman-Oppenheimer-Volkoff (TOV) equation and
the second  is to study the adiabatic index. Now we are going to
discuss the TOV equation in the frame of our solution in presented
in Appendix A.

\subsection{The Tolman-Oppenheimer-Volkoff equation}

To study the stability of our solution  displayed in Appendix A, we are going to suppose the hydrostatic equilibrium
by using the TOV equation  \cite{PhysRev.55.364,PhysRev.55.374} as  given  in \cite{PoncedeLeon1993}:
\begin{eqnarray}\label{TOV} {\frac{2[p_\theta-{p_r}]}{r}-\frac{M_g(r)[{\rho}(r)+{p}_r(r)]e^{[h(r)-h_1(r)]/2}}{r}-\frac{d{p_r}(r)}{dr}=0}\,.
 \end{eqnarray}
Here  $M_g(r)$ represents the gravitational mass evaluated at the
radius $r$ and  has the following form:
\begin{eqnarray}\label{ma} M_g(r)=4\pi{\int_0}^r\Big({{\mathcal T}_t}^t-{{\mathcal T}_r}^r-{{\mathcal T}_\theta}^\theta-{{\mathcal T}_\phi}^\phi\Big)y^2e^{[h(y)+h_1(y)]}dy={\frac{r h' e^{[h_1(r)-h(r)]/2}}{2}}\,,
 \end{eqnarray}
Using Eq. (\ref{ma}) in Eq. (\ref{TOV}) gives:
\begin{eqnarray}\label{ma1} \frac{2({p}_\theta-{p_r})}{r}-\frac{d{p_r}}{dr}-\frac{h'[{\rho}(r)+{p_r}(r)]}{2}=F_g+F_a+F_h=0\,.
 \end{eqnarray}
Here $F_g=-\frac{h'[{\rho}+{p_r}]}{2}$,
$F_a=\frac{2({p}_\theta-{p_r})}{r}$ and
$F_h=-\frac{d{p_r}(r)}{dr}$ are the gravitational mass, the
anisotropic and the hydrostatic forces, respectively. The explicit
forms of these forces can be found in Appendix E. We plot the
different forces presented in Appendix E in Fig. \ref{Fig:5}
\subref{fig:TOV}, and it can be seen, it is ensured that  the
hydrostatic and anisotropic forces are positive and dominate over
the gravitational force, which has a  negative value, hence
opposite direction, and the system is kept in static equilibrium.
\begin{figure}
\centering
\subfigure[~The different forces acting on the model]{\label{fig:TOV}\includegraphics[scale=0.3]{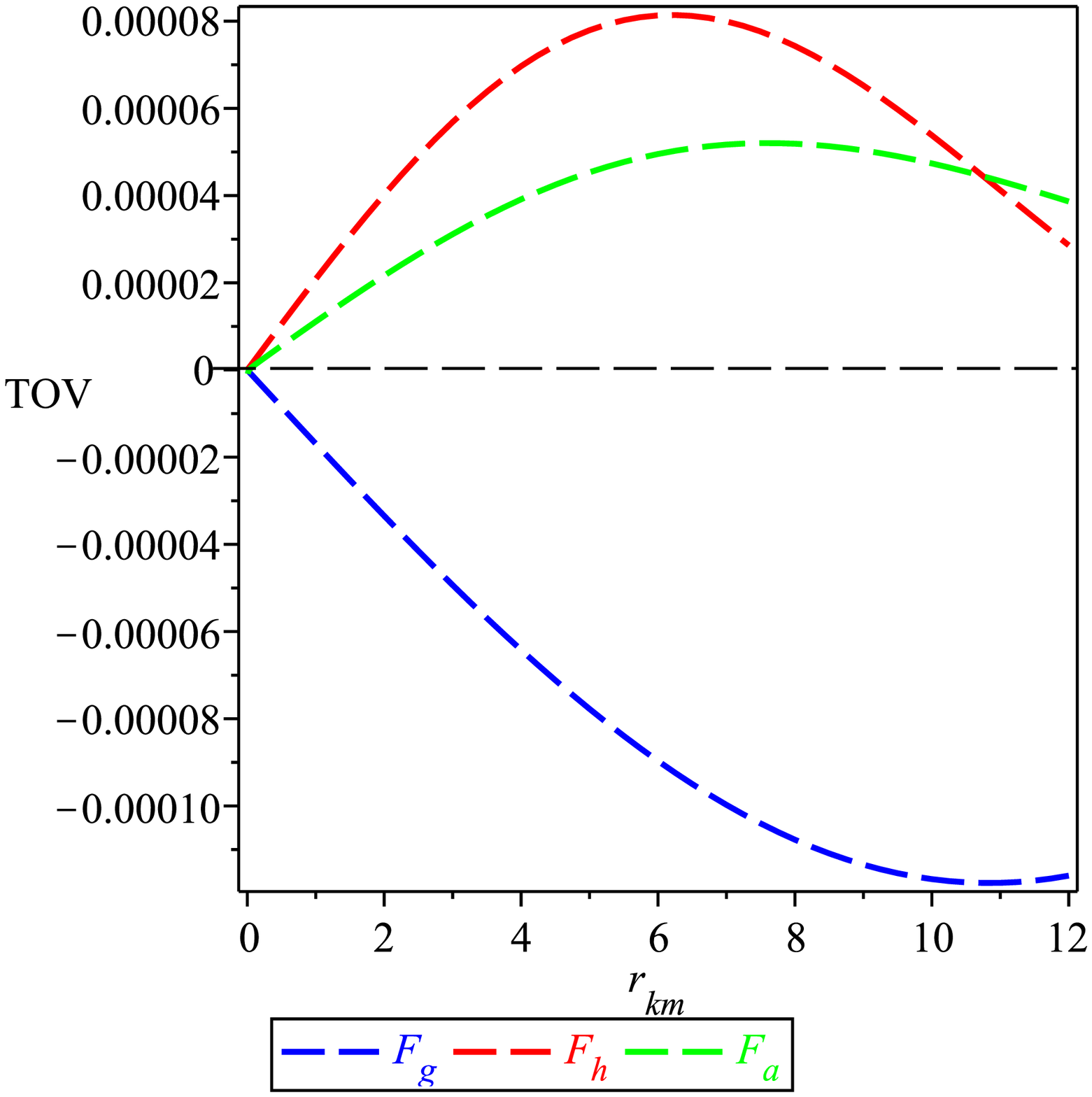}}
\subfigure[~The adiabatic index of the stellar under consideration]{\label{fig:adib1}\includegraphics[scale=.3]{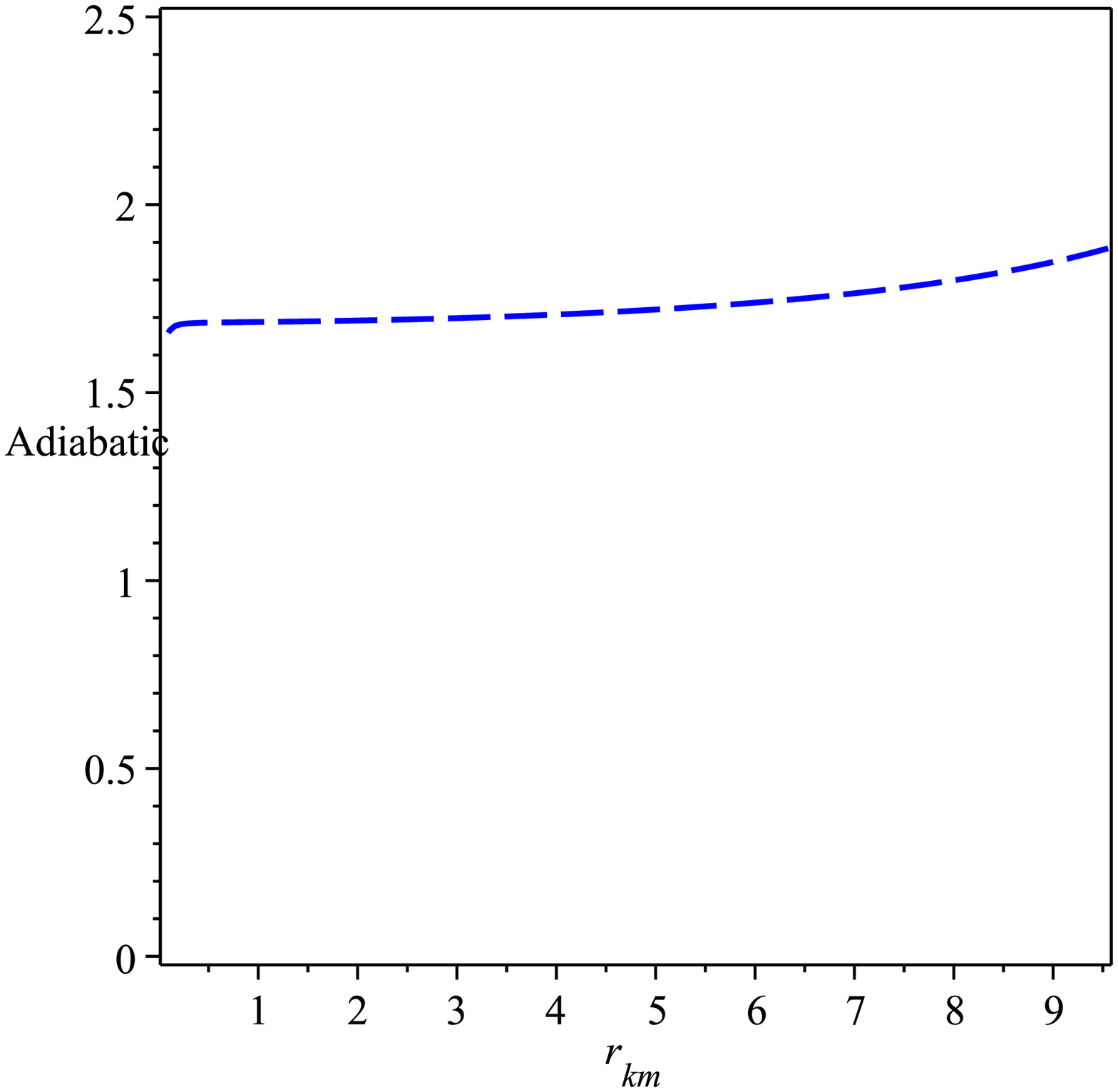}}
\subfigure[~{ The  mass as a function of the central density} ]{\label{fig:mrhon}\includegraphics[scale=.3]{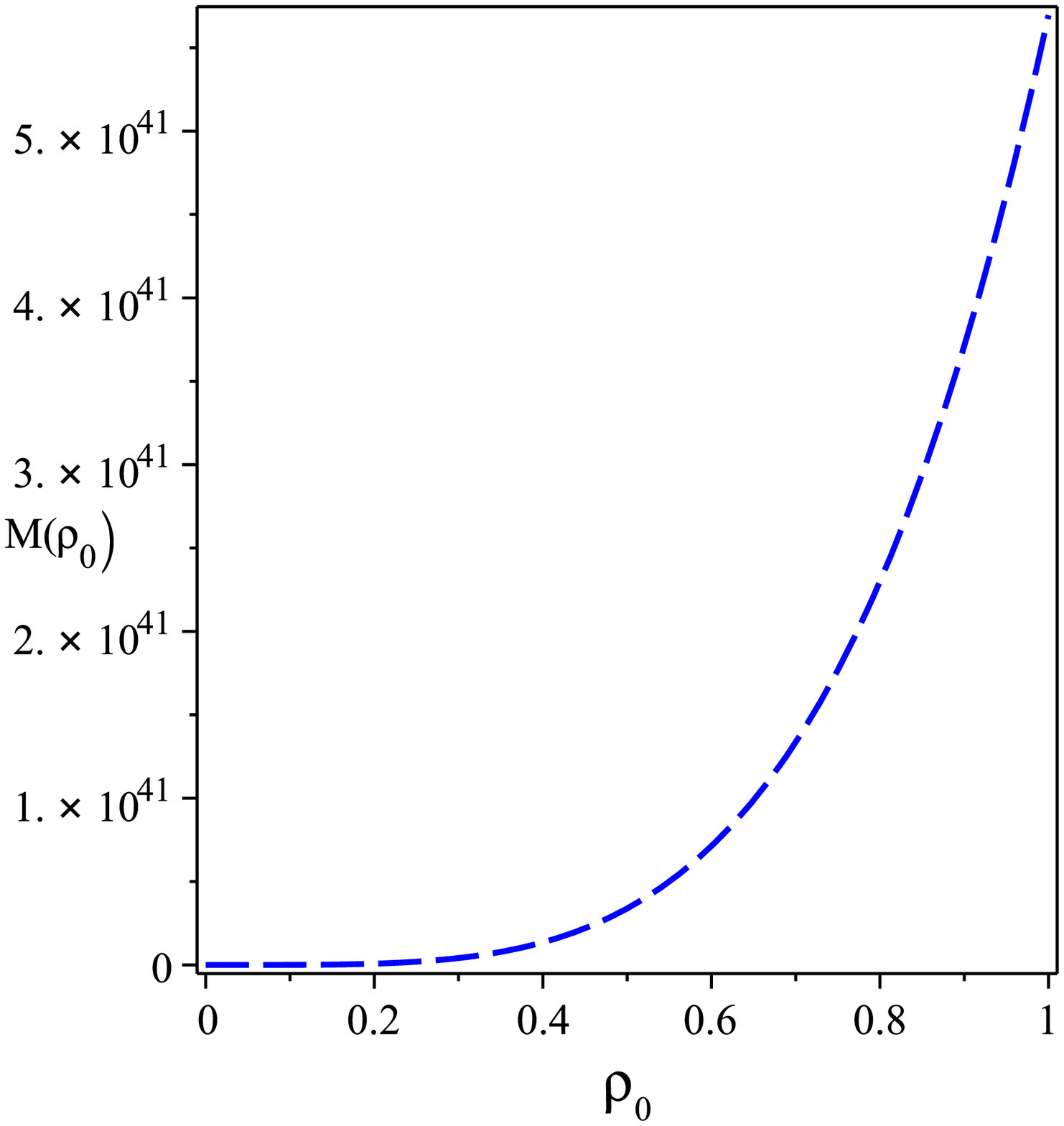}}
\subfigure[~The derivative of mass w.r.t. the central density ]{\label{fig:mrho}\includegraphics[scale=.3]{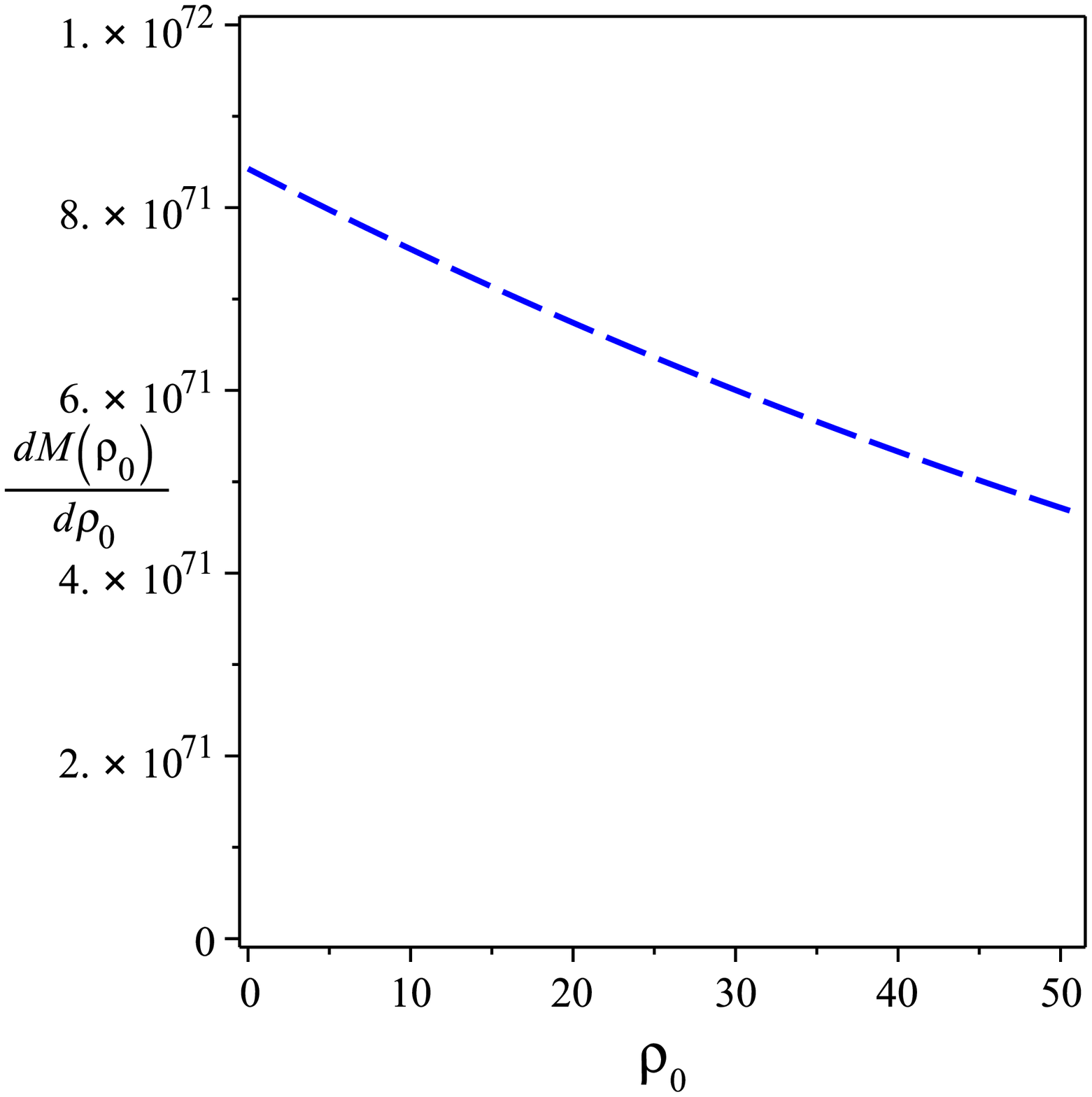}}
\caption[figtopcap]{\small{Schematic plots of~\subref{fig:TOV} which represents the different forces acting on the model under consideration; ~\subref{fig:adib1} which represents the adiabatic index of the model under consideration; \subref{fig:mrhon} which represents the variation of the mass (which is a function of the central density) via the central density; \subref{fig:mrho} which represents the stability of the model in the static state. The numerical values of the GB and the constants $c_0$ and $c_1$ used in these plots are given as  $\beta=-0.0001$,  $c_0=-0.001763358117$ and $c_1=-0.9329227851$ in this figure.}}
\label{Fig:5}
\end{figure}

\subsection{The Adiabatic Index}

The adiabatic index connects the structure of a spherically
symmetric static object to the EoS of the interior solution. One
can use the adiabatic index to discuss the stability of the
interior solution \cite{Moustakidis:2016ndw}. For any  interior
solution, it becomes stable if its adiabatic index is  greater
than $4/3$ \cite{1975A&A....38...51H} however, if
$\gamma=\frac{4}{3}$, then the isotropic sphere is in neutral
equilibrium.   Following Chan et al.
\cite{10.1093/mnras/265.3.533},  the stability condition of a
relativistic anisotropic sphere, $\gamma >\Gamma$, must be
satisfied. Here $\Gamma$ is defined as
\begin{eqnarray}\label{ai}
\Gamma=\frac{4}{3}-\left\{\frac{4(p_r-p_\theta)}{3\lvert
p'_r\lvert}\right\}_{max}\,,
 \end{eqnarray}
where the collapse  happens according to the nature of the
anisotropy. When $p_r>p_\theta$ one can  have $\gamma<4/3$ and the
system still stable and  when  $p_\theta >p_r$ one can have
unstable system even if $\gamma>4/3$.  The behavior of the
adiabatic index is shown in Fig. \ref{Fig:5} \subref{fig:adib1}
which ensures the  stability condition of our solution.

 \subsection{Stability in the Static State}

{ Finally, let us study the stability through the procedure given by
Harrison, Zeldovich and Novikov
\cite{1965gtgc.book.....H,1971reas.book.....Z,1983reas.book.....Z}
who showed that for stable compact stars the mass, (which is a function of the central density) must be positive and increasing and also the derivative of  the
mass with respect to the central density must have a positive
 value, i.e., $\frac{\partial M}{\partial \rho_0}> 0$.}
Applying this condition to our model we get the density at the
center of the star:
\begin{eqnarray}\label{rm}
\rho_0(r\to 0)=\frac{3(8\beta c_0-1)c_0}{\pi} \Longrightarrow c_0=\frac{3\pm\sqrt{9+96\pi \beta \rho_0}}{48\beta}\,.
\end{eqnarray}
Using Eq. (\ref{rm}) in  Eq. (\ref{mass}) we get the form of mass
in terms of the central density which is written in Appendix F.
The pattern of the derivative of mass with respect to the central
density given by the equation presented in Appendix $F_1$ and it
is plotted in Fig. \ref{Fig:5} \subref{fig:mrho} which ensures the
stability of our model.
\begin{table*}[t!]
\caption{\label{Table1}%
Values of model parameters \cite{2016ApJ...820...28O}}
\begin{ruledtabular}
\begin{tabular*}{\textwidth}{lcccc}
{{Pulsar}}                              & Mass ($M_{\odot}$) &      {Radius (km)} &   {$c_0$}  &    {$c_1$}   \\ \hline
&&&&\\
4U 1724-207       &$1.81^{+0.25}_{-0.37}$    & $12.2^{+1.4}_{-1.4}$ &   $\approx$-0.75$\times10^{-3}$    &   $\approx$-0.9           \\
4U 1820-30 & $1.46^{+0.21}_{-0.21}$ & $11.1^{+1.8}_{-1.8}$ &   $\approx$-0.7$\times10^{-3}$           &$\approx$-1.5  \\
SAX J1748.9-2021 & $1.81^{+0.25}_{-0.37}$ & $11.7^{+1.7}_{-1.7}$ &   $\approx$-0.2$\times10^{-2}$        &$\approx$-1.3    \\
EXO 1745-268 & $1.65^{+0.21}_{-0.31}$ & $10.5^{+1.6}_{-1.6}$ &    $\approx$-0.1$\times10^{-2}$        &$\approx$-1.4   \\
4U 1608-52 & $1.57^{+0.30}_{-0.29}$ & $9.8^{+1.8}_{-1.8}$ &   $\approx$-0.1$\times10^{-2}$         &$\approx$-1.4  \\
KS 1731-260 & $1.61^{+0.35}_{-0.37}$ & $10.0^{+2.2}_{-2.2}$ &  $\approx$-0.1$\times10^{-2}$          &$\approx$-2.4   \\
&&&&\\
\end{tabular*}
\end{ruledtabular}
\end{table*}
\begin{table*}[t!]
\caption{\label{Table2}%
Values of physical quantities}
\begin{ruledtabular}
\begin{tabular*}{\textwidth}{lcccccccccc}
{{Pulsar}}                              &{$\rho\lvert_{_{_{0}}}$} &      {$\rho\lvert_{_{_{l}}}$} &   {$\frac{dp_r}{d\rho}\lvert_{_{_{0}}}$}  &    {$\frac{dp_r}{d\rho}\lvert_{_{_{l}}}$}
  & {$\frac{dp_\theta}{d\rho}\lvert_{_{_{0}}}$} & {$\frac{dp_\theta}{d\rho}\lvert_{_{_{l}}}$}&{$(\rho-p_r-2p_\theta)\lvert_{_{_{0}}}$}&{$(\rho-p_r-2p_\theta)\lvert_{_{_{l}}}$}&{$z\lvert_{_{_{l}}}$}&\\ \hline
&&&&&&&&&&\\
4U 1724-207 &$\approx$0.71$\times10^{-3}$     &$\approx$ 0.32$\times10^{-3}$  &$\approx$ 0.68   & $\approx$0.57      &$\approx$ 0.28&0.43&$\approx$0.42$\times10^{-3}$&$\approx$0.41$\times10^{-4}$&0.014  \\
4U 1820-30  & $\approx$0.65$\times10^{-3}$     &$\approx$ 0.3$4\times10^{-3}$ &  $\approx$0.36  &  $\approx$0.26        &$\approx$0.21& $\approx$0.15&$\approx$0.58$\times10^{-3}$&$\approx$0.28$\times10^{-3}$& 0.012   \\
SAX J1748.9-2021 & $\approx$0.2$\times10^{-2}$& $\approx$0.6$\times10^{-3}$&$\approx$0.45  &$\approx$0.24 &$\approx$0.26&$\approx$0.2 &$\approx$0.16$\times10^{-2}$&$\approx$0.35$\times10^{-3}$&$\approx$0.017    \\
EXO 1745-268  &$\approx$ 0.1$\times10^{-2}$&$\approx$0.4$\times10^{-3}$&$\approx$0.4  &$\approx$0.27 &$\approx$0.23&$\approx$0.17 &$\approx$0.8$\times10^{-3}$&$\approx$0.3$\times10^{-3}$&$\approx$0.014   \\
4U 1608-52 &  $\approx$0.1$\times10^{-2}$    &$\approx$0.44$\times10^{-3}$    &$\approx$0.38      &$\approx$0.25  &$\approx$0 .22&$\approx$0.15 &$\approx$0.92$\times10^{-3}$&$\approx$0.34$\times10^{-3}$&$\approx$0 .0148 \\
KS 1731-260 &$\approx$0.1$\times10^{-2}$&$\approx$  0.4$\times10^{-3}$  & $\approx$0.15 &$\approx$0 .034 &$\approx$0.048&$\approx$0 &$\approx$0.1$\times10^{-2}$&$\approx$0.4$\times10^{-3}$&$\approx$0.0147    \\
\end{tabular*}
\end{ruledtabular}
\end{table*}
In addition to $EXO 1785 - 248$, a similar analysis can be
developed for other pulsars. In tables \ref{Table1} and
\ref{Table2}  we report the results for other observed pulsars.


\section{Discussion  and Conclusions } \label{S7}


Among the higher curvature theories, the EGB theory offers new
possibilities for $4D$ gravity. This new proposal has received
much interest because the GB invariant may share Einstein's field
equations through the redefinition of the GB constant to be $\beta
\to \frac{\beta}{D- 4}$ in $D$-dimensions and taking the limit
$D\to 4$. Inspired by this new proposal of EGB theory, in this
work, we rigorously explained a static and spherically symmetric
interior solution. Assuming a specific form of the metric
potentials, we derived in this new proposal of EGB a new interior
solution. { In this study we showed that the GB parameter, $\beta$,must take a tiny negative value otherwise
 we will have imaginary quantities for the components of pressures. As we showed the effect of the constant $\beta$ acting
on the  EoS, yielded its behavior to behave in a non-linear form,
unlike the Einstein GR.}

We checked the impact of the anisotropy regime and showed that it
is always positive which means that we have a repulsive force.
Moreover, we listed the physical conditions that any real star
must satisfy and showed that our model yields: \vspace{0.1cm}\\i-A
well-known behavior of the energy-density, radial and tangential
pressures at the center as well as at the surface of  the star as
shown in Fig. \ref{Fig:1} \subref{fig:1a}, \subref{fig:1b} and
\subref{fig:1c}.\vspace{0.1cm}\\ii-The causality condition is
satisfied by showing that the gradient of the energy momentum
components have negative value as shown in  Fig. \ref{Fig:2}
\subref{fig:2a}. Moreover,  in Fig.  \ref{Fig:2} \subref{fig:2b}
we showed that the speed of sound of radial and tangential
components are less than one, as required for any realistic
star.\vspace{0.1cm}\\iii- In Fig. \ref{Fig:3}  we showed that the
energy conditions which must be satisfied for any realistic star,
are satisfied. \vspace{0.1cm}\\iv- Using different techniques,
TOV, adiabatic index and stability of static state, we studied the
stability of our model and  we showed that it is stable as  Fig.
\ref{Fig:5} \subref{fig:TOV}, \subref{fig:adib1} and
\subref{fig:mrho} show. The results of this study can be applied
to the observational data of various starts showing consistent
results as shown in Tables \ref{Table1} and \ref{Table2}. For
these objects, it is possible to calculate the density at the
center and at the surface, the radial and tangential speed, at the
center and the surface of the star, the SEC at the center and the
surface of the star, and the redshift at the surface of the start.
It is interesting to see that all the results of the different
stars of our solution are compatible with observations. { From the above discussions we can say that our results presented in this study are in agreement with that presented in \cite{Bhatti:2017fov} from the viewpoint of the structure of compact star.}

In summary, we derived a new interior solution in the framework of
EGB in the limit $D\to 4$, assuming physically motivated metric
potentials. This solution has a non-trivial form of the Ricci
scalar as well as the GB term. The internal solution can be
matched with the external in the limit $D\to 4$. In a forthcoming
paper, we will consider a similar situation assuming a charged
interior solution.

\section*{Acknowledgements}

The Authors would like to thank the referees for their valuable comments. This work was supported by MINECO (Spain), project
PID2019-104397GB-I00.

\newpage
{\centerline{Appendix A:  Energy density, radial and tangential
pressures.}}\vspace{0.3cm} Now, let us present  the quantities
related to the interior solution discussed in the present study
and write them as:
\renewcommand{\eqref}{}
\begin{eqnarray*}
 &&8\pi\,\rho(r)=\Bigg\{ 192\beta\,c_0-140c_0\,{r}^{2}-392{c_0}^{2}{r}^{4}-630{c_0}^{3}{r}^{6}-616{c_0}^{4}{r}^{8}-364{c_0}^{5}{r}^{10}-120{c_0}^{6}{r}^{12}-
 17{c_0}^{7}{r}^{14}-24+11760\beta\,{c_0}^{3}{r}^{4}\nonumber\\
 &&+38304\,\beta\,{c_0}^{4}{r}^{6}+87472\,\beta\,{c_0}^{5}{r}^{8}+148512\,\beta\,{c_0}^{6}{r}^{10}
 +193020\,\beta\,{c_0}^{7}{r}^{12}+194480\,
\beta\,{c_0}^{8}{r}^{14}+152152\,\beta\,{c_0}^{9}{r}^{16}+
91728\,\beta\,{c_0}^{10}{r}^{18}\nonumber\\
 &&+41860\,\beta\,{c_0}^{11}{
r}^{20}+14000\,\beta\,{c_0}^{12}{r}^{22}+3240\,\beta\,{c_0
}^{13}{r}^{24}+464\,\beta\,{c_0}^{14}{r}^{26}+31\,\beta\,{c_0}^{15}{r}^{28}+2240\,\beta\,{c_0}^{2}{r}^{2} \Bigg\} c_0\,,\nonumber\\
 &&8\pi\,p_r(r)=\frac{c_0}{c_0\,{r}^{2}-c_1}\Bigg\{ -16-420\,{c_0}^{2}{r}^{4}-840\,{c_0}^{3}{r}^{6}-1050\,{c_0}^{4}{r}^{8}-840\,{c_0}^{5}{r}^{10}- 420\,{c_0}^{6}{r}^{12}-120\,{c_0}^{7}{r}^{14}-{c_0}^{7} {r}^{14}c_1\nonumber\\
 &&-120\,c_0\,{r}^{2}+256\,\beta\,c_0-70\,{c_0}^{3}{r}^{6}c_1-\beta\,{c_0}^{15}{r}^{28}c_1- 448\,\beta\,{c_0}^{2}{r}^{2}c_1-11440\,\beta\,{c_0}^{ 8}{r}^{14}c_1-8008\,\beta\,{c_0}^{9}{r}^{16}c_1-1680\, \beta\,{c_0}^{3}{r}^{4}c_1\nonumber\\
 &&-4256\beta\,{c_0}^{4}{r}^ {6}c_1-7952\beta\,{c_0}^{5}{r}^{8}c_1-11424\beta \,{c_0}^{6}{r}^{10}c_1-12868\beta\,{c_0}^{7}{r}^{12} c_1-4368\beta\,{c_0}^{10}{r}^{18}c_1-1820\beta\,{ c_0}^{11}{r}^{20}c_1-560\beta{c_0}^{12}{r}^{22}c_1\nonumber\\
 &&-120\,\beta\,{c_0}^{13}{r}^{24}c_1-16\,\beta\,{c0}^{14}{r}^{26}c_1+33\,\beta\,{c_0}^{16}{r}^{30}-28\, {c_0}^{5}{r}^{10}c_1-56\,{c_0}^{4}{r}^{8}c_1-64\, \beta\,c_0\,c_1-8\,{c_0}^{6}{r}^{12}c_1-28\,c_0\,{r}^{2}c_1-56\,{c_0}^{2}{r}^{4}c_1\nonumber\\
 &&+3008\,\beta\,{ c_0}^{2}{r}^{2}+528\,\beta\,{c_0}^{15}{r}^{28}+3960\,\beta \,{c_0}^{14}{r}^{26}+60060\,\beta\,{c_0}^{12}{r}^{22}+18480 \,\beta\,{c_0}^{13}{r}^{24}+264264\,\beta\,{c_0}^{10}{r}^{ 18}+144144\,\beta\,{c_0}^{11}{r}^{20}\nonumber\\
 &&+377248\,\beta\,{c_0} ^{7}{r}^{12}+263312\,\beta\,{c_0}^{6}{r}^{10}+377520\,\beta\,{c_0}^{9}{r}^{16}+424676\,\beta\,{c_0}^{8}{r}^{14}+16576\, \beta\,{c_0}^{3}{r}^{4}+57680\,\beta\,{c_0}^{4}{r}^{6}+ 142240\,\beta\,{c_0}^{5}{r}^{8}\nonumber\\
 &&-8\,c_1-15\,{c_0}^{8}{r }^{16}\Bigg\}\,,\nonumber\\
&&8\pi\,p_\theta(r)=\frac {2c_0 }{ \left( c_0{r}^{2}-c_1  \right) ^{2}}\Bigg\{228{c_0}^{2}{r}^{4}+700{c_0 }^{3}{r}^{6}+1204{c_0}^{4}{r}^{8}+1260{c_0}^{5}{r}^{10}+ 812{c_0}^{6}{r}^{12}+308{c_0}^{7}{r}^{14}+28{c_1} ^{2}c_0{r}^{2}+232{c_0}^{7}{r}^{14}c_1\nonumber\\
&&+32\,c_0 \,{r}^{2}+952\,{c_0}^{3}{r}^{6}c_1+224\,\beta\,{c_0}^{ 2}{c_1}^{2}{r}^{2}+1680\,\beta\,{c_0}^{3}{c_1}^{2}{r}^ {4}+6384\,\beta\,{c_0}^{4}{c_1}^{2}{r}^{6}+15904\,\beta\,{ c_0}^{5}{c_1}^{2}{r}^{8}+28560\,\beta\,{c_0}^{6}{c_1}^{2}{r}^{10}\nonumber\\
&&+38604\,\beta\,{c_0}^{7}{c_1}^{2}{r}^{12}+ 40040\,\beta\,{c_0}^{8}{c_1}^{2}{r}^{14}+32032\,\beta\,{c_0}^{9}{r}^{16}{c_1}^{2}+19656\,\beta\,{c_0}^{10}{r}^{ 18}{c_1}^{2}+9100\,\beta\,{c_0}^{11}{r}^{20}{c_1}^{2}+ 3080\,\beta\,{c_0}^{12}{r}^{22}{c_1}^{2}\nonumber\\
&&+720\,\beta\,{c_0}^{13}{r}^{24}{c_1}^{2}+104\,\beta\,{c_0}^{14}{r}^{ 26}{c_1}^{2}+7\,\beta\,{c_0}^{15}{r}^{28}{c_1}^{2}-406 \,\beta\,{c_0}^{16}{r}^{30}c_1-6096\,\beta\,{c_0}^{15 }{r}^{28}c_1-5600\,\beta\,{c_0}^{2}{r}^{2}c_1\nonumber\\
&&-2651072 \,\beta\,{c_0}^{8}{r}^{14}c_1-2642640\,\beta\,{c_0}^{ 9}{r}^{16}c_1-42112\,\beta\,{c_0}^{3}{r}^{4}c_1-187600 \,\beta\,{c_0}^{4}{r}^{6}c_1-566944\,\beta\,{c_0}^{5} {r}^{8}c_1-1246112\,\beta\,{c_0}^{6}{r}^{10}c_1\nonumber\\
&&- 2069600\,\beta\,{c_0}^{7}{r}^{12}c_1-2050048\,\beta\,{c_0}^{10}{r}^{18}c_1-1227408\,\beta\,{c_0}^{11}{r}^{20} c_1-556920\,\beta\,{c_0}^{12}{r}^{22}c_1-185360\, \beta\,{c_0}^{13}{r}^{24}c_1\nonumber\\
&&-42720\,\beta\,{c_0}^{14} {r}^{26}c_1+3432\,\beta\,{c_0}^{16}{r}^{30}+1288\,{c_0 }^{5}{r}^{10}c_1+1400\,{c_0}^{4}{r}^{8}c_1-320\,\beta \,c_0\,c_1+728\,{c_0}^{6}{r}^{12}c_1+88\,c_0 \,{r}^{2}c_1+392\,{c_0}^{2}{r}^{4}c_1\nonumber\\
&&-960\beta\,{c_0}^{2}{r}^{2}+22800\beta\,{c_0}^{15}{r}^{28}+92680 \beta\,{c_0}^{14}{r}^{26}+508872\beta\,{c_0}^{12}{r}^{22 }+256620\beta{c_0}^{13}{r}^{24}+772200\beta\,{c_0}^{ 10}{r}^{18}+736736\beta\,{c_0}^{11}{r}^{20}\nonumber\\
&&-46592\beta{c_0}^{7}{r}^{12}-129360\beta{c_0}^{6}{r}^{10}+553428 \beta{c_0}^{9}{r}^{16}+211920\beta{c_0}^{8}{r}^{14}- 9120\beta{c_0}^{3}{r}^{4}-38752\beta{c_0}^{4}{r}^{ 6}-94080\beta{c_0}^{5}{r}^{8}\nonumber\\
&&+84\,{c_1}^{2}{c_0}^{ 5}{r}^{10}+140\,{c_1}^{2}{c_0}^{4}{r}^{8}+4\,{c_1}^{2}{ c_0}^{7}{r}^{14}+28\,{c_1}^{2}{c_0}^{6}{r}^{12}+239\, \beta\,{c_0}^{17}{r}^{32}+140\,{c_1}^{2}{c_0}^{3}{r}^{ 6}+84\,{c_1}^{2}{c_0}^{2}{r}^{4}+32\,{c_0}^{8}{r}^{16}c_1+8\,c_1\nonumber\\
&&+60\,{c_0}^{8}{r}^{16}+4\,{c_0}^{9}{r}^{ 18}+4\,{c_1}^{2} \Bigg\}\,,\nonumber\\
&&8\pi\,\Delta(r)=\frac {{c_0}^{2} }{ \left(c_0\,{r}^{2}-c_1\right) ^{2}} \Bigg\{ 80{r}^{2}+1120{c_0}^{2}{r}^{6}c_1+6336\beta{c_0}^{15}{r}^{30}+352{c_0}^{6}{r}^{ 14}c_1+1064{c_0}^{5}{r}^{12}c_1+1792{c_0}^{4}{ r}^{10}c_1+1820{c_0}^{3}{r}^{8}c_1\nonumber\\
&&-400960\,\beta\,{c_0}^{5}{r}^{10}+682180\,\beta\,{c_0}^{8}{r}^{16}+1680\, \beta\,{c_0}^{2}{c_1}^{2}{r}^{4}+8512\,\beta\,{c_0}^{ 3}{c_1}^{2}{r}^{6}+23856\,\beta\,{c_0}^{4}{c_1}^{2}{r} ^{8}+45696\,\beta\,{c_0}^{5}{c_1}^{2}{r}^{10}\nonumber\\
&&+64340\,\beta \,{c_0}^{6}{c_1}^{2}{r}^{12}+68640\,\beta\,{c_0}^{7}{c_1}^{2}{r}^{14}+56056\,\beta\,{c_0}^{8}{r}^{16}{c_1}^{ 2}+34944\,\beta\,{c_0}^{9}{r}^{18}{c_1}^{2}+16380\,\beta\, {c_0}^{10}{r}^{20}{c_1}^{2}+5600\,\beta\,{c_0}^{11}{r} ^{22}{c_1}^{2}\nonumber\\
&&+1320\,\beta\,{c_0}^{12}{r}^{24}{c_1}^{2 }+192\,\beta\,{c_0}^{13}{r}^{26}{c_1}^{2}+13\,\beta\,{c_0}^{14}{r}^{28}{c_1}^{2}-778\,\beta\,{c_0}^{15}{r}^{ 30}c_1-11648\,\beta\,{c_0}^{14}{r}^{28}c_1-8128\, \beta\,c_0\,{r}^{2}c_1\nonumber\\
&&-4864600\,\beta\,{c_0}^{7}{r}^{ 14}c_1-4896320\,\beta\,{c_0}^{8}{r}^{16}c_1-67200\, \beta\,{c_0}^{2}{r}^{4}c_1-315840\,\beta\,{c_0}^{3}{r }^{6}c_1-987392\,\beta\,{c_0}^{4}{r}^{8}c_1-2220960\, \beta\,{c_0}^{5}{r}^{10}c_1\nonumber\\
&&-3750528\beta\,{c_0}^{6} {r}^{12}c_1-3827824\,\beta{c_0}^{9}{r}^{18}c_1- 2306304\beta{c_0}^{10}{r}^{20}c_1-1051960\beta{c_0}^{11}{r}^{22}c_1-351680\beta\,{c_0}^{12}{r}^{24}c_1-81360\beta{c_0}^{13}{r}^{26}c_1\nonumber\\
&&
-384\,\beta\,c_1+64\,{r}^{2}c_1-64\,\beta\,{c_1}^{2}+28\,{r}^{2}{c_1}^{2}+1209208\,\beta\,{c_0}^{10}{r}^{20}-356496\,\beta\, {c_0}^{6}{r}^{12}-94080\,\beta\,{c_0}^{3}{r}^{6}-245840\, \beta\,{c_0}^{4}{r}^{8}\nonumber\\
&&+46592\,\beta\,{c_0}^{7}{r}^{14}- 21248\,\beta\,{c_0}^{2}{r}^{4}-2176\,\beta\,c_0\,{r}^{2}+ 41640\,\beta\,{c_0}^{14}{r}^{28}+392\,c_0\,{r}^{4}c_1+ 453180\,\beta\,{c_0}^{12}{r}^{24}+1166880\,\beta\,{c_0}^{9 }{r}^{18}\nonumber\\
&&+166880\,\beta\,{c_0}^{13}{r}^{26}+873600\,\beta\,{c_0}^{11}{r}^{22}+140\,{c_1}^{2}{c_0}^{4}{r}^{10}+224\,{ c_1}^{2}{c_0}^{3}{r}^{8}+7\,{c_1}^{2}{c_0}^{6}{r}^ {14}+48\,{c_1}^{2}{c_0}^{5}{r}^{12}+445\,\beta\,{c_0}^ {16}{r}^{32}\nonumber\\
&&+210\,{c_1}^{2}{c_0}^{2}{r}^{6}+112\,{c_1}^ {2}c_0\,{r}^{4}+50\,{c_0}^{7}{r}^{16}c_1+576\,c_0 \,{r}^{4}+1820\,{c_0}^{2}{r}^{6}+3248\,{c_0}^{3}{r}^{8}+3570 \,{c_0}^{4}{r}^{10}+2464\,{c_0}^{5}{r}^{12}+1036\,{c_0} ^{6}{r}^{14}\nonumber\\
&&+240\,{c_0}^{7}{r}^{16}+23\,{c_0}^{8}{r}^{18} \Bigg\}\,.  \hspace*{13cm}  {\color{blue} ({\mathbf A})} \nonumber
  \end{eqnarray*}
  where $\Delta$ is the anisotropy defined as $\Delta=p_\theta-p_r$.\\
  {\centerline{Appendix B:  The gradients of   energy--density, radial and transverse  pressures.}}\vspace{0.3cm}
  The gradients of the components of the energy-momentum tensor given in the Appendix A  take the following form:
\renewcommand{\eqref}{}
  \begin{eqnarray*}
 && \rho'=\frac{d\rho}{dr}=\frac{{c_0}^{2}r}{4\pi }\Bigg\{23520 \,\beta\,{c_0}^{2}{r}^{2}-140-784\,c_0\,{r}^{2}-1890\, {c_0}^{2}{r}^{4}-2464\,{c_0}^{3}{r}^{6}-1820\,{c_0}^{4} {r}^{8}-720\,{c_0}^{5}{r}^{10}-119\,{c_0}^{6}{r}^{12}\nonumber\\
&&+114912\,\beta\,{c_0}^{3}{r}^{4}+ 349888\,\beta\,{c_0}^{4}{r}^{6}+742560\,\beta\,{c_0}^{5}{r }^{8}+1158120\,\beta\,{c_0}^{6}{r}^{10}+1361360\,\beta\,{c_0}^{7}{r}^{12}+1217216\,\beta\,{c_0}^{8}{r}^{14}+2240\,\beta\,c_0 \nonumber\\
&&+825552\, \beta\,{c_0}^{9}{r}^{16}+418600\,\beta\,{c_0}^{10}{r}^{18} +154000\,\beta\,{c_0}^{11}{r}^{20}+38880\,\beta\,{c_0}^{12 }{r}^{22}+6032\,\beta\,{c_0}^{13}{r}^{24}+434\,\beta\,{c_0 }^{14}{r}^{26}\Bigg\}
\,,\nonumber\\
&&p'_r=\frac {{c_0}^{2}r}{4\pi\left( c_0\,{r}^{2}-c_1 \right) ^{2}}\Bigg\{ 128c_1+3360\beta{c_0}^{2}{c_1}^{2}{r}^{2}+12768\beta{c_0}^{3}{c_1}^{2}{r}^{4}+31808\beta{c_0}^{4}{c_1}^{2}{r}^{6}+ 57120\beta{c_0}^{5}{c_1}^{2}{r}^{8}+77208\beta{c_0}^{6}{c_1}^{2}{r}^{10}\nonumber\\
&&+80080\,\beta\,{c_0}^{7}{c_1}^{2}{r}^{12}+64064\,\beta\,{c_0}^{8}{c_1}^{2}
{r}^{14}+ 39312\,\beta\,{c_0}^{9}{r}^{16}{c_1}^{2}+18200\,\beta\,{c_0}^{10}{r}^{18}{c_1}^{2}+6160\,\beta\,{c_0}^{11}{r}^{ 20}{c_1}^{2}+1440\,\beta\,{c_0}^{12}{r}^{22}{c_1}^{2}\nonumber\\
&&+ 208\,\beta\,{c_0}^{13}{r}^{24}{c_1}^{2}+14\,\beta\,{c_0}^{14}{r}^{26}{c_1}^{2}-508\,\beta\,{c_0}^{15}{r}^{28}c_1-33152\,\beta\,{c_0}^{2}{r}^{2}c_1
-3088800\,\beta\, {c_0}^{8}{r}^{14}c_1-2434432\,\beta\,{c_0}^{9}{r}^{16} c_1\nonumber\\
&& -577472\,\beta \,{c_0}^{4}{r}^{6}c_1-1340416\,\beta\,{c_0}^{5}{r}^{8} c_1-2309184\,\beta\,{c_0}^{6}{r}^{10}c_1-3037072\, \beta\,{c_0}^{7}{r}^{12}c_1-1476384\,\beta\,{c_0}^{10 }{r}^{18}c_1-174720\,\beta\,{c_0}^{3}{r}^{4}c_1\nonumber\\
&&+16-677040\,\beta\,{c_0}^{11}{r}^{20}c_1- 227360\,\beta\,{c_0}^{12}{r}^{22}c_1-52800\,\beta\,{c_0}^{13}{r}^{24}c_1-7584\,\beta\,{c_0}^{14}{r}^{26}c_1-256\,\beta\,c_0-420\,{c_0}^{2}{r}^{4}
-1680\,{c_0}^{3 }{r}^{6}\nonumber\\
&&-3150\,{c_0}^{4}{r}^{8}-3360\,{c_0}^{5}{r}^{10}-2100 \,{c_0}^{6}{r}^{12}-720\,{c_0}^{7}{r}^{14}-105\,{c_0}^{ 8}{r}^{16}+28\,{c_1}^{2}+16576\,\beta\,{c_0}^{3}{r}^{4}+ 115360\,\beta\,{c_0}^{4}{r}^{6}\nonumber\\
&&+426720\,\beta\,{c_0}^{5}{r }^{8}+1053248\,\beta\,{c_0}^{6}{r}^{10}+1886240\,\beta\,{c_0}^{7}{r}^{12}+2548056\,\beta\,{c_0}^{8}{r}^{14}+2642640\, \beta\,{c_0}^{9}{r}^{16}+2114112\,\beta\,{c_0}^{10}{r}^{18 }\nonumber\\
&&+1297296\,\beta\,{c_0}^{11}{r}^{20}+600600\,\beta\,{c_0}^ {12}{r}^{22}+203280\,\beta\,{c_0}^{13}{r}^{24}+47520\,\beta\,{c_0}^{14}{r}^{26}+6864\,\beta\,{c_0}^{15}{r}^{28}+114\,{c_0}^{7}{r}^{14}c_1
+800\,{c_0}^{6}{r}^{12}c_1\nonumber\\
&&+840\, c_0\,{r}^{2}c_1+2464\,{c_0}^{2}{r}^{4}c_1+4060\,{c_0}^{3}{r}^{6}c_1+4032\,{c_0}^{4}{r}^{8}c_1+2408\, {c_0}^{5}{r}^{10}c_1-2944\,\beta\,c_0\,c_1+462\, \beta\,{c_0}^{16}{r}^{30}+112\,{c_1}^{2}c_0\,{r}^{2}\nonumber\\
&&+ 210\,{c_1}^{2}{c_0}^{2}{r}^{4}+224\,{c_1}^{2}{c_0} ^{3}{r}^{6}+140\,{c_1}^{2}{c_0}^{4}{r}^{8}+48\,{c_1}^{2 }{c_0}^{5}{r}^{10}+7\,{c_1}^{2}{c_0}^{6}{r}^{12}+448\, \beta\,c_0\,{c_1}^{2} \Bigg\}
\,,\nonumber\\
&&p'_\theta=\frac {{c_0}^{2}r }{\pi \, \left( -c_1+c_0\,{r}^{2} \right) ^{3}} \Bigg\{42000\,\beta\,{c_0} ^{2}{c_1}^{2}{r}^{2} -24\,c_1+281400\,\beta\,{c_0}^{3}{c_1}^{2} {r}^{4}+1137080\,\beta\,{c_0}^{4}{c_1}^{2}{r}^{6}+3131184\, \beta\,{c_0}^{5}{c_1}^{2}{r}^{8}\nonumber\\
&&+6251640\,\beta\,{c_0 }^{6}{c_1}^{2}{r}^{10}+9355960\,\beta\,{c_0}^{7}{c_1}^ {2}{r}^{12}+10670660\,\beta\,{c_0}^{8}{c_1}^{2}{r}^{14}+ 9321312\,\beta\,{c_0}^{9}{r}^{16}{c_1}^{2}+6205836\,\beta \,{c_0}^{10}{r}^{18}{c_1}^{2}\nonumber\\
&&+3099460\,\beta\,{c_0}^{ 11}{r}^{20}{c_1}^{2}+1126020\,\beta\,{c_0}^{12}{r}^{22}{c_1}^{2}+281280\,\beta\,{c_0}^{13}{r}^{24}{c_1}^{2}+ 43244\,\beta\,{c_0}^{14}{r}^{26}{c_1}^{2}+3087\,\beta\,{c_0}^{15}{r}^{28}{c_1}^{2}\nonumber\\
&&-4551\,\beta\,{c_0}^{16}{r}^{ 30}c_1-16\,c_0\,{r}^{2}-62316\,\beta\,{c_0}^{15}{r}^{ 28}c_1+11920\,\beta\,{c_0}^{2}{r}^{2}c_1-8841392\, \beta\,{c_0}^{8}{r}^{14}c_1-11402820\,\beta\,{c_0}^{9 }{r}^{16}c_1\nonumber\\
&&+58128\,\beta\,{c_0}^{3}{r}^{4}c_1+94360\, \beta\,{c_0}^{4}{r}^{6}c_1-243544\,\beta\,{c_0}^{5}{r }^{8}c_1-1729392\,\beta\,{c_0}^{6}{r}^{10}c_1-4880920 \,\beta\,{c_0}^{7}{r}^{12}c_1-10858848\,\beta\,{c_0}^ {10}{r}^{18}c_1\nonumber\\
&&-7708428\,\beta\,{c_0}^{11}{r}^{20}c_1- 4045860\,\beta\,{c_0}^{12}{r}^{22}c_1-1529220\,\beta\,{c_0}^{13}{r}^{24}c_1-394560\,\beta\,{c_0}^{14}{r}^{26}c_1+350\,{c_0}^{3}{r}^{6}+1204\,{c_0}^{4}{r}^{8}\nonumber\\
&&
+1890\,{ c_0}^{5}{r}^{10}+1624\,{c_0}^{6}{r}^{12}+770\,{c_0}^{7} {r}^{14}+180\,{c_0}^{8}{r}^{16}+14\,{c_0}^{9}{r}^{18}-48\,{c_1}^{2}-19376\,\beta\,{c_0}^{4}{r}^{6}-94080\,\beta\,{c_0}^{5}{r}^{8}\nonumber\\
&&
-194040\,\beta\,{c_0}^{6}{r}^{10}-93184\, \beta\,{c_0}^{7}{r}^{12}+529800\,\beta\,{c_0}^{8}{r}^{14}+ 1660284\,\beta\,{c_0}^{9}{r}^{16}+2702700\,\beta\,{c_0}^{ 10}{r}^{18}+2946944\,\beta\,{c_0}^{11}{r}^{20}\nonumber\\
&&+2289924\,\beta\, {c_0}^{12}{r}^{22}+1283100\,\beta\,{c_0}^{13}{r}^{24}+ 509740\,\beta\,{c_0}^{14}{r}^{26}+136800\,\beta\,{c_0}^{15 }{r}^{28}+480\,\beta\,{c_0}^{2}{r}^{2}+340\,{c_0}^{7}{r}^{ 14}c_1\nonumber\\
&&+378\,{c_0}^{6}{r}^{12}c_1-272\,c_0\,{r}^{2} c_1-1050\,{c_0}^{2}{r}^{4}c_1-1932\,{c_0}^{3}{r}^{ 6}c_1-1750\,{c_0}^{4}{r}^{8}c_1-504\,{c_0}^{5}{r}^ {10}c_1+800\,\beta\,c_0\,c_1+22308\,\beta\,{c_0} ^{16}{r}^{30}\nonumber\\
&&+78\,{c_0}^{8}{r}^{16}c_1+1673\,\beta\,{c_0}^{17}{r}^{32}-406\,{c_1}^{2}c_0\,{r}^{2}-1428\,{c_1 }^{2}{c_0}^{2}{r}^{4}-2730\,{c_1}^{2}{c_0}^{3}{r}^{6}- 3080\,{c_1}^{2}{c_0}^{4}{r}^{8}-2058\,{c_1}^{2}{c_0}^{5}{r}^{10}\nonumber\\
&&-756\,{c_1}^{2}{c_0}^{6}{r}^{12}-118\,{c_1}^{2}{c_0}^{7}{r}^{14}
+2800\,\beta\,c_0\,{c_1}^{2}-14 \,{c_1}^{3}-1680\,\beta\,{c_0}^{2}{c_1}^{3}{r}^{2}- 9576\,\beta\,{c_0}^{3}{c_1}^{3}{r}^{4}-31808\,\beta\,{c_0}^{4}{c_1}^{3}{r}^{6}\nonumber\\
&&-71400\,\beta\,{c_0}^{5}{c_1}^{3}{r}^{8}
-115812\,\beta\,{c_0}^{6}{c_1}^{3}{r}^{10}- 140140\,\beta\,{c_0}^{7}{c_1}^{3}{r}^{12}-128128\,\beta\,{ c_0}^{8}{c_1}^{3}{r}^{14}-88452\,\beta\,{c_0}^{9}{r}^{ 16}{c_1}^{3}-45500\,\beta\,{c_0}^{10}{r}^{18}{c_1}^{3}\nonumber\\
&& -16940\,\beta\,{c_0}^{11}{r}^{20}{c_1}^{3}-4320\,\beta\,{c_0}^{12}{r}^{22}{c_1}^{3}-676\,\beta\,{c_0}^{13}{r}^{ 24}{c_1}^{3}-49\,\beta\,{c_0}^{14}{r}^{26}{c_1}^{3}-84 \,{c_1}^{3}c_0\,{r}^{2}\nonumber\\
&&-210\,{c_1}^{3}{c_0}^{2}{r} ^{4}-280\,{c_1}^{3}{c_0}^{3}{r}^{6}-210\,{c_1}^{3}{c_0}^{4}{r}^{8}-84\,{c_1}^{3}{c_0}^{5}{r}^{10}-14\,{c_1}^{3}{c_0}^{6}{r}^{12}-112\,\beta\,c_0\,{c_1}^{ 3} \Bigg\}\,, \hspace*{2cm} {\color{blue} ({\mathbf B})}
  \end{eqnarray*}
  where ${\rho}'=\frac{d{\rho}}{dr}$,\,  ${p_r}'=\frac{d{p_r}}{dr}$ \, and ${p}'_\theta=\frac{d{p}_\theta}{dr}$.\vspace*{1cm}\\
 \newpage
  {\centerline{ Appendix C:   Derivation of the radial and tangential speeds of sound}}\vspace{0.3cm}
  \renewcommand{\eqref}{}
 Using the equations of the gradient of energy-momentum components presented in Appendix B we get:
\begin{eqnarray*}
 && v_r{}^2=\frac{p'_r}{\rho'}= \Bigg\{16-256\,\beta\,c_0-420\,{c_0}^{2}{r}^{4}-1680\,{c_0}^{3}{r}^{6}-3150\,{c_0}^{4}{r}^{8}-3360\,{c_0}^{5}{r }^{10}-2100\,{c_0}^{6}{r}^{12}-720\,{c_0}^{7}{r}^{14}-105\,{ c_0}^{8}{r}^{16}\nonumber\\
&&+28\,{c_1}^{2}+426720\,\beta\,{c_0}^{5 }{r}^{8}+115360\,\beta\,{c_0}^{4}{r}^{6}+1053248\,\beta\,{c_0}^{6}{r}^{10}+16576\,\beta\,{c_0}^{3}{r}^{4}+1886240\,\beta \,{c_0}^{7}{r}^{12}+2548056\,\beta\,{c_0}^{8}{r}^{14}\nonumber\\
&&+ 2642640\,\beta\,{c_0}^{9}{r}^{16}+2114112\,\beta\,{c_0}^{ 10}{r}^{18}+1297296\,\beta\,{c_0}^{11}{r}^{20}+600600\,\beta\,{ c_0}^{12}{r}^{22}+203280\,\beta\,{c_0}^{13}{r}^{24}+47520\, \beta\,{c_0}^{14}{r}^{26}\nonumber\\
&&+6864\,\beta\,{c_0}^{15}{r}^{28}+ 114\,{c_0}^{7}{r}^{14}c_1+800\,{c_0}^{6}{r}^{12}c_1+840\,c_0\,{r}^{2}c_1+2464\,{c_0}^{2}{r}^{4}c_1+ 4060\,{c_0}^{3}{r}^{6}c_1+4032\,{c_0}^{4}{r}^{8}c_1+2408\,{c_0}^{5}{r}^{10}c_1\nonumber\\
&&-2944\,\beta\,c_0\,c_1+462\,\beta\,{c_0}^{16}{r}^{30}+112\,{c_1}^{2}c_0\,{r }^{2}+210\,{c_1}^{2}{c_0}^{2}{r}^{4}+224\,{c_1}^{2}{c_0}^{3}{r}^{6}+140\,{c_1}^{2}{c_0}^{4}{r}^{8}+48\,{c_1}^{2}{c_0}^{5}{r}^{10}+7\,{c_1}^{2}{c_0}^{6}{r}^ {12}\nonumber\\
&&+448\,\beta\,c_0\,{c_1}^{2}+3360\,\beta\,{c_0}^{2 }{c_1}^{2}{r}^{2}+12768\,\beta\,{c_0}^{3}{c_1}^{2}{r}^ {4}+31808\,\beta\,{c_0}^{4}{c_1}^{2}{r}^{6}+57120\,\beta\, {c_0}^{5}{c_1}^{2}{r}^{8}+77208\,\beta\,{c_0}^{6}{c_1}^{2}{r}^{10}\nonumber\\
&&+80080\,\beta\,{c_0}^{7}{c_1}^{2}{r}^{ 12}+64064\,\beta\,{c_0}^{8}{c_1}^{2}{r}^{14}+39312\,\beta \,{c_0}^{9}{r}^{16}{c_1}^{2}+18200\,\beta\,{c_0}^{10}{ r}^{18}{c_1}^{2}+6160\,\beta\,{c_0}^{11}{r}^{20}{c_1}^ {2}+1440\,\beta\,{c_0}^{12}{r}^{22}{c_1}^{2}\nonumber\\
&&+208\,\beta\,{ c_0}^{13}{r}^{24}{c_1}^{2}+14\,\beta\,{c_0}^{14}{r}^{ 26}{c_1}^{2}-508\,\beta\,{c_0}^{15}{r}^{28}c_1-33152\, \beta\,{c_0}^{2}{r}^{2}c_1-3088800\,\beta\,{c_0}^{8}{ r}^{14}c_1-2434432\,\beta\,{c_0}^{9}{r}^{16}c_1\nonumber\\
&&-174720 \,\beta\,{c_0}^{3}{r}^{4}c_1-577472\,\beta\,{c_0}^{4} {r}^{6}c_1-1340416\,\beta\,{c_0}^{5}{r}^{8}c_1-2309184 \,\beta\,{c_0}^{6}{r}^{10}c_1-3037072\,\beta\,{c_0}^{ 7}{r}^{12}c_1-1476384\,\beta\,{c_0}^{10}{r}^{18}c_1\nonumber\\
&&- 677040\,\beta\,{c_0}^{11}{r}^{20}c_1-227360\,\beta\,{c_0}^{12}{r}^{22}c_1-52800\,\beta\,{c_0}^{13}{r}^{24}c_1-7584\,\beta\,{c_0}^{14}{r}^{26}c_1+128\,c_1\Bigg\} \Bigg[\left(c_0\,{r}^{2}-c_1\right) ^{2} \Big\{23520\,\beta\,{c_0}^{2}{r}^{2}\nonumber\\
&&140-784\,c_0\,{r}^{2}-1890\,{c_0}^{2}{r}^{4}-2464\,{c_0}^{3}{r}^{ 6}-1820\,{c_0}^{4}{r}^{8}-720\,{c_0}^{5}{r}^{10}-119\,{c_0}^{6}{r}^{12}+114912\,\beta\, {c_0}^{3}{r}^{4}+349888\,\beta\,{c_0}^{4}{r}^{6}\nonumber\\
&&+742560\, \beta\,{c_0}^{5}{r}^{8}+1158120\,\beta\,{c_0}^{6}{r}^{10}+ 1361360\,\beta\,{c_0}^{7}{r}^{12}+1217216\,\beta\,{c_0}^{8 }{r}^{14}+825552\,\beta\,{c_0}^{9}{r}^{16}+418600\,\beta\,{c_0}^{10}{r}^{18}\nonumber\\
&&+154000\,\beta\,{c_0}^{11}{r}^{20}+38880\, \beta\,{c_0}^{12}{r}^{22}+6032\,\beta\,{c_0}^{13}{r}^{24}+ 434\,\beta\,{c_0}^{14}{r}^{26}+2240\,\beta\,c_0 \Big\}\Bigg]^{-1}
\,, \nonumber\\
 &&
v_t{}^2=\frac{{p}'_\theta}{\rho'}=-4\,\Bigg\{770\,{c_0}^{7}{r}^{14}-280\,{c_1}^{3}{c_0}^{3}{r}^{6}+78\,{c_0}^{ 8}{r}^{16}c_1+180\,{c_0}^{8}{r}^{ 16}+350\,{c_0}^{3}{r}^{6}+ 1204\,{c_0}^{4}{r}^{8}+1890\,{c_0}^{5}{r}^{10}+1624\,{c_0}^{6}{r}^{12}\nonumber\\
 &&-16\,c_0\,{r}^{2}-48\,{c_1}^{2}-94080\,\beta\,{c_0}^{5}{r}^{8}-19376\, \beta\,{c_0}^{4}{r}^{6}-194040\,\beta\,{c_0}^{6}{r}^{10}- 93184\,\beta\,{c_0}^{7}{r}^{12}+529800\,\beta\,{c_0}^{8}{r }^{14}+1660284\,\beta\,{c_0}^{9}{r}^{16}\nonumber\\
 &&+2702700\,\beta\,{c_0}^{10}{r}^{18}+2946944\,\beta\,{c_0}^{11}{r}^{20}+2289924\, \beta\,{c_0}^{12}{r}^{22}+1283100\,\beta\,{c_0}^{13}{r}^{ 24}+509740\,\beta\,{c_0}^{14}{r}^{26}+136800\,\beta\,{c_0} ^{15}{r}^{28}\nonumber\\
 &&+340\,{c_0}^{7}{r}^{14}c_1+378\,{c_0}^{6}{ r}^{12}c_1-272\,c_0\,{r}^{2}c_1-1050\,{c_0}^{2}{r} ^{4}c_1-1932\,{c_0}^{3}{r}^{6}c_1-1750\,{c_0}^{4}{ r}^{8}c_1-504\,{c_0}^{5}{r}^{10}c_1+800\,\beta\,c_0\,c_1\nonumber\\
 &&+22308\,\beta\,{c_0}^{16}{r}^{30}-406\,{c_1}^{ 2}c_0\,{r}^{2}-1428\,{c_1}^{2}{c_0}^{2}{r}^{4}-2730\,{c_1}^{2}{c_0}^{3}{r}^{6}-3080\,{c_1}^{2}{c_0}^{4}{r }^{8}-2058\,{c_1}^{2}{c_0}^{5}{r}^{10}-756\,{c_1}^{2}{c_0}^{6}{r}^{12}\nonumber\\
 &&+2800\,\beta\,c_0\,{c_1}^{2}+480\, \beta\,{c_0}^{2}{r}^{2}+42000\,\beta\,{c_0}^{2}{c_1}^ {2}{r}^{2}+281400\,\beta\,{c_0}^{3}{c_1}^{2}{r}^{4}+1137080 \,\beta\,{c_0}^{4}{c_1}^{2}{r}^{6}+3131184\,\beta\,{c_0}^{5}{c_1}^{2}{r}^{8}\nonumber\\
 &&+6251640\,\beta\,{c_0}^{6}{c_1 }^{2}{r}^{10}+9355960\,\beta\,{c_0}^{7}{c_1}^{2}{r}^{12}+ 10670660\,\beta\,{c_0}^{8}{c_1}^{2}{r}^{14}+9321312\,\beta \,{c_0}^{9}{r}^{16}{c_1}^{2}+6205836\,\beta\,{c_0}^{10 }{r}^{18}{c_1}^{2}\nonumber\\
 &&+3099460\,\beta\,{c_0}^{11}{r}^{20}{c_1}^{2}+1126020\,\beta\,{c_0}^{12}{r}^{22}{c_1}^{2}+281280 \,\beta\,{c_0}^{13}{r}^{24}{c_1}^{2}+43244\,\beta\,{c_0}^{14}{r}^{26}{c_1}^{2}-62316\,\beta\,{c_0}^{15}{r}^{28} c_1\nonumber\\
 &&+11920\,\beta\,{c_0}^{2}{r}^{2}c_1-8841392\,\beta \,{c_0}^{8}{r}^{14}c_1-11402820\,\beta\,{c_0}^{9}{r}^{ 16}c_1+58128\,\beta\,{c_0}^{3}{r}^{4}c_1+94360\,\beta \,{c_0}^{4}{r}^{6}c_1-243544\,\beta\,{c_0}^{5}{r}^{8}c_1\nonumber\\
 &&-1729392\,\beta\,{c_0}^{6}{r}^{10}c_1-4880920\, \beta\,{c_0}^{7}{r}^{12}c_1-10858848\,\beta\,{c_0}^{ 10}{r}^{18}c_1-7708428\,\beta\,{c_0}^{11}{r}^{20}c_1- 4045860\,\beta\,{c_0}^{12}{r}^{22}c_1\nonumber\\
 &&-1529220\,\beta\,{c_0}^{13}{r}^{24}c_1-394560\,\beta\,{c_0}^{14}{r}^{26}c_1-24\,c_1-112\,\beta\,c_0\,{c_1}^{3}-210\,{c_1}^{3}{c_0}^{4}{r}^{8}-84\,{c_1}^{3}{c_0}^{5}{r}^{10} -210\,{c_1}^{3}{c_0}^{2}{r}^{4}-14\,{c_1}^{3}{c_0} ^{6}{r}^{12}\nonumber\\
 &&-118\,{c_1}^{2}{c_0}^{7}{r}^{14}+1673\,\beta\,{ c_0}^{17}{r}^{32}+3087\,\beta\,{c_0}^{15}{r}^{28}{c_1} ^{2}-4551\,\beta\,{c_0}^{16}{r}^{30}c_1-1680\,\beta\,{c_0}^{2}{c_1}^{3}{r}^{2}-9576\,\beta\,{c_0}^{3}{c_1}^{3}{r}^{4}-31808\,\beta\,{c_0}^{4}{c_1}^{3}{r}^{6}\nonumber\\
 &&-71400 \,\beta\,{c_0}^{5}{c_1}^{3}{r}^{8}-115812\,\beta\,{c_0}^{6}{c_1}^{3}{r}^{10}-140140\,\beta\,{c_0}^{7}{c_1}^ {3}{r}^{12}-128128\,\beta\,{c_0}^{8}{c_1}^{3}{r}^{14}-88452 \,\beta\,{c_0}^{9}{r}^{16}{c_1}^{3}-45500\,\beta\,{c_0}^{10}{r}^{18}{c_1}^{3}\nonumber\\
 &&-16940\,\beta\,{c_0}^{11}{r}^{20}{c_1}^{3}-4320\,\beta\,{c_0}^{12}{r}^{22}{c_1}^{3}-676\, \beta\,{c_0}^{13}{r}^{24}{c_1}^{3}-49\,\beta\,{c_0}^{ 14}{r}^{26}{c_1}^{3}-84\,{c_1}^{3}c_0\,{r}^{2}+14\,{c_0}^{9}{r}^{18}-14\,{c_1}^{3}\Bigg\}\nonumber\\
 &&\times\Bigg[ \left( -c_1+c_0\,{ r}^{2} \right) ^{3} \Big[ -140-784\,c_0\,{r}^{2}-1890\,{c_0 }^{2}{r}^{4}-2464\,{c_0}^{3}{r}^{6}-1820\,{c_0}^{4}{r}^{8}- 720\,{c_0}^{5}{r}^{10}-119\,{c_0}^{6}{r}^{12}+23520\,\beta \,{c_0}^{2}{r}^{2}\nonumber\\
 &&+114912\,\beta\,{c_0}^{3}{r}^{4}+349888\, \beta\,{c_0}^{4}{r}^{6}+742560\,\beta\,{c_0}^{5}{r}^{8}+ 1158120\,\beta\,{c_0}^{6}{r}^{10}+1361360\,\beta\,{c_0}^{7 }{r}^{12}+1217216\,\beta\,{c_0}^{8}{r}^{14}\nonumber\\
 &&+825552\,\beta\,{c_0}^{9}{r}^{16}+418600\,\beta\,{c_0}^{10}{r}^{18}+154000\, \beta\,{c_0}^{11}{r}^{20}+38880\,\beta\,{c_0}^{12}{r}^{22} +6032\,\beta\,{c_0}^{13}{r}^{24}+434\,\beta\,{c_0}^{14}{r} ^{26}+2240\,\beta\,c_0 \Big]\Bigg]^{-1}\,.\nonumber\\
 &&\hspace*{15cm}{\color{blue} ({\mathbf C})}
\end{eqnarray*}
\newpage
 {\centerline{Appendix D:   Derivation of the EoS's}}\vspace{0.3cm}
  \renewcommand{\eqref}{}
 Using the form of energy-density, radial and tangential pressures presented in Appendix A, we get the  EoS as:
\begin{eqnarray*}
&& \omega_r=  \Bigg\{-16-120\,c_0\,{r}^{2}+256\,\beta\,c_0-420\,{c_0}^{2}{r}^{4}-840\,{c_0}^{3}{r}^{6}-1050\,{c_0}^{4}{r}^{8}- 840\,{c_0}^{5}{r}^{10}-420\,{c_0}^{6}{r}^{12}-120\,{c_0 }^{7}{r}^{14}-15\,{c_0}^{8}{r}^{16}\nonumber\\
&&+142240\,\beta\,{c_0}^{5 }{r}^{8}+57680\,\beta\,{c_0}^{4}{r}^{6}+263312\,\beta\,{c_0}^{6}{r}^{10}+16576\,\beta\,{c_0}^{3}{r}^{4}+377248\,\beta\,{c_0}^{7}{r}^{12}+424676\,\beta\,
{c_0}^{8}{r}^{14}+377520\, \beta\,{c_0}^{9}{r}^{16}\nonumber\\
&&+264264\,\beta\,{c_0}^{10}{r}^{18} +144144\,\beta\,{c_0}^{11}{r}^{20}+60060\,\beta\,{c_0}^{12 }{r}^{22}+18480\,\beta\,{c_0}^{13}{r}^{24}+3960\,\beta\,{c_0}^{14}{r}^{26}+528\,\beta\,{c_0}^{15}{r}^{28}-{c_0}^{7}{ r}^{14}c_1\nonumber\\
&&-8\,{c_0}^{6}{r}^{12}c_1-28\,c_0\,{r}^{2 }c_1-56\,{c_0}^{2}{r}^{4}c_1-70\,{c_0}^{3}{r}^{6}c_1-56\,{c_0}^{4}{r}^{8}c_1-28\,{c_0}^{5}{r}^{10}c_1-64\,\beta\,c_0\,c_1+33\,\beta\,{c_0}^{16}{r}^ {30}+3008\,\beta\,{c_0}^{2}{r}^{2}\nonumber\\
&&-\beta\,{c_0}^{15}{r}^{ 28}c_1-448\,\beta\,{c_0}^{2}{r}^{2}c_1-11440\,\beta\, {c_0}^{8}{r}^{14}c_1-8008\,\beta\,{c_0}^{9}{r}^{16}c_1-1680\,\beta\,{c_0}^{3}{r}^{4}c_1-4256\,\beta\,{c_0}^{4}{r}^{6}c_1-7952\,\beta\,{c_0}^{5}{r}^{8}c_1\nonumber\\
&&
-11424\,\beta\,{c_0}^{6}{r}^{10}c_1-12868\,\beta\,{c_0}^{7}{r}^{12}c_1-4368\,\beta\,{c_0}^{10}{r}^{18}c_1 -1820\,\beta\,{c_0}^{11}{r}^{20}c_1-560\,\beta\,{c_0} ^{12}{r}^{22}c_1-120\,\beta\,{c_0}^{13}{r}^{24}c_1\nonumber\\
&&-16 \,\beta\,{c_0}^{14}{r}^{26}c_1-8\,c_1\Bigg\}\Bigg[\left( -c_1+c_0\,{r}^{2} \right)  \Big\{ 192\,\beta\,c_0-140\,c_0\,{r}^{2}-392\,{c_0}^{2}{r}^{4}-630\,{c_0}^{3}{r}^{6}-616 \,{c_0}^{4}{r}^{8}-364\,{c_0}^{5}{r}^{10}-120\,{c_0}^{6 }{r}^{12}\nonumber\\
&&-17\,{c_0}^{7}{r}^{14}-24+11760\,\beta\,{c_0}^{3}{ r}^{4}+38304\,\beta\,{c_0}^{4}{r}^{6}+87472\,\beta\,{c_0}^ {5}{r}^{8}+148512\,\beta\,{c_0}^{6}{r}^{10}+193020\,\beta\,{c_0}^{7}{r}^{12}+194480\,\beta\,{c_0}^{8}{r}^{14}\nonumber\\
&&+152152\, \beta\,{c_0}^{9}{r}^{16}+91728\,\beta\,{c_0}^{10}{r}^{18}+ 41860\,\beta\,{c_0}^{11}{r}^{20}+14000\,\beta\,{c_0}^{12}{ r}^{22}+3240\,\beta\,{c_0}^{13}{r}^{24}+464\,\beta\,{c_0}^ {14}{r}^{26}+31\,\beta\,{c_0}^{15}{r}^{28}\nonumber\\
&&+2240\,\beta\,{c_0}^{2}{r}^{2} \Big\}\Bigg]^{-1}\,,\nonumber\\
&&\omega_\theta=2\,\Bigg\{32\,{c_0}^{8}{r}^{16}c_1+32\,c_0\,{r}^{2}+ 228\,{c_0}^{2}{r}^{4}+700\,{c_0}^{3}{r}^{6}+1204\,{c_0} ^{4}{r}^{8}+1260\,{c_0}^{5}{r}^{10}+812\,{c_0}^{6}{r}^{12}+ 308\,{c_0}^{7}{r}^{14}+60\,{c_0}^{8}{r}^{16}\nonumber\\
&&-94080\,\beta\,{c_0}^{5}{r}^{8}-38752\,\beta\,{c_0}^{4}{ r}^{6}-129360\,\beta\,{c_0}^{6}{r}^{10}-9120\,\beta\,{c_0} ^{3}{r}^{4}-46592\,\beta\,{c_0}^{7}{r}^{12}+211920\,\beta\,{c_0}^{8}{r}^{14}+553428\,\beta\,{c_0}^{9}{r}^{16}\nonumber\\
&&+4\,{c_1}^{ 2}+772200\, \beta\,{c_0}^{10}{r}^{18}+736736\,\beta\,{c_0}^{11}{r}^{20 }+508872\,\beta\,{c_0}^{12}{r}^{22}+256620\,\beta\,{c_0}^{ 13}{r}^{24}+92680\,\beta\,{c_0}^{14}{r}^{26}+22800\,\beta\,{c_0}^{15}{r}^{28}\nonumber\\
&&+232\,{c_0}^{7}{r}^{14}c_1+728\,{c_0}^{6}{r}^{12}c_1+88\,c_0\,{r}^{2}c_1+392\,{c_0} ^{2}{r}^{4}c_1+952\,{c_0}^{3}{r}^{6}c_1+1400\,{c_0 }^{4}{r}^{8}c_1+1288\,{c_0}^{5}{r}^{10}c_1-320\,\beta \,c_0\,c_1\nonumber\\
&&+3432\,\beta\,{c_0}^{16}{r}^{30}+28\,{c_1}^{2}c_0\,{r}^{2}+84\,{c_1}^{2}{c_0}^{2}{r}^{4}+140 \,{c_1}^{2}{c_0}^{3}{r}^{6}+140\,{c_1}^{2}{c_0}^{4 }{r}^{8}+84\,{c_1}^{2}{c_0}^{5}{r}^{10}+28\,{c_1}^{2}{c_0}^{6}{r}^{12}-960\,\beta\,{c_0}^{2}{r}^{2}\nonumber\\
&&+224\,\beta\,{ c_0}^{2}{c_1}^{2}{r}^{2}+1680\,\beta\,{c_0}^{3}{c_1}^{2}{r}^{4}+6384\,\beta\,{c_0}^{4}{c_1}^{2}{r}^{6}+ 15904\,\beta\,{c_0}^{5}{c_1}^{2}{r}^{8}+28560\,\beta\,{c_0}^{6}{c_1}^{2}{r}^{10}+38604\,\beta\,{c_0}^{7}{c_1}^{2}{r}^{12}\nonumber\\
&&+40040\,\beta\,{c_0}^{8}{c_1}^{2}{r}^{14}+ 32032\,\beta\,{c_0}^{9}{r}^{16}{c_1}^{2}+19656\,\beta\,{c_0}^{10}{r}^{18}{c_1}^{2}+9100\,\beta\,{c_0}^{11}{r}^{ 20}{c_1}^{2}+3080\,\beta\,{c_0}^{12}{r}^{22}{c_1}^{2}+ 720\,\beta\,{c_0}^{13}{r}^{24}{c_1}^{2}\nonumber\\
&&+104\,\beta\,{c_0}^{14}{r}^{26}{c_1}^{2}-6096\,\beta\,{c_0}^{15}{r}^{28}c_1-5600\,\beta\,{c_0}^{2}{r}^{2}c_1-2651072\,\beta\,{ c_0}^{8}{r}^{14}c_1-2642640\,\beta\,{c_0}^{9}{r}^{16}c_1-42112\,\beta\,{c_0}^{3}{r}^{4}c_1\nonumber\\
&&-187600\,\beta\,{ c_0}^{4}{r}^{6}c_1-566944\,\beta\,{c_0}^{5}{r}^{8}c_1-1246112\,\beta\,{c_0}^{6}{r}^{10}c_1-2069600\, \beta\,{c_0}^{7}{r}^{12}c_1-2050048\,\beta\,{c_0}^{10 }{r}^{18}c_1-1227408\,\beta\,{c_0}^{11}{r}^{20}c_1\nonumber\\
&&- 556920\,\beta\,{c_0}^{12}{r}^{22}c_1-185360\,\beta\,{c_0}^{13}{r}^{24}c_1-42720\,\beta\,{c_0}^{14}{r}^{26}c_1+8\,c_1+4\,{c_1}^{2}{c_0}^{7}{r}^{14}+239\,\beta\,{ c_0}^{17}{r}^{32}+7\,\beta\,{c_0}^{15}{r}^{28}{c_1}^{2 }\nonumber\\
&&-406\,\beta\,{c_0}^{16}{r}^{30}c_1+4\,{c_0}^{9}{r}^{ 18}\Bigg\}\Bigg[\left( -c_1+c_0\,{r}^{2} \right) ^{2} \Bigg( 192\, \beta\,c_0-140\,c_0\,{r}^{2}-392\,{c_0}^{2}{r}^{4}-630 \,{c_0}^{3}{r}^{6}-616\,{c_0}^{4}{r}^{8}-364\,{c_0}^{5} {r}^{10}\nonumber\\
&&-120\,{c_0}^{6}{r}^{12}-17\,{c_0}^{7}{r}^{14}-24+ 11760\,\beta\,{c_0}^{3}{r}^{4}+38304\,\beta\,{c_0}^{4}{r}^ {6}+87472\,\beta\,{c_0}^{5}{r}^{8}+148512\,\beta\,{c_0}^{6 }{r}^{10}+193020\,\beta\,{c_0}^{7}{r}^{12}\nonumber\\
&&+194480\,\beta\,{c_0}^{8}{r}^{14}+152152\,\beta\,{c_0}^{9}{r}^{16}+91728\, \beta\,{c_0}^{10}{r}^{18}+41860\,\beta\,{c_0}^{11}{r}^{20} +14000\,\beta\,{c_0}^{12}{r}^{22}+3240\,\beta\,{c_0}^{13}{ r}^{24}+464\,\beta\,{c_0}^{14}{r}^{26}\nonumber\\
&&+31\,\beta\,{c_0}^{ 15}{r}^{28}+2240\,\beta\,{c_0}^{2}{r}^{2} \Bigg)\Bigg]\,.\hspace*{11cm} {\color{blue} ({\mathbf D})}
\end{eqnarray*}
Using the approximate form of the energy-density listed in
Appendix A we get:
\begin{eqnarray*} \label{solr}
&& \rho(r)\approx\left( 192\,\beta\,c_0-24 \right)c_0+ \left( -140\,c_0+2240\,\beta\,{c_0}^{2} \right) c_0\,{r}^{2}+ \left( -
392\,{c_0}^{2}+11760\,\beta\,{c_0}^{3} \right) c_0\,{r}^{4}+\mathcal{O}(r^{-6})\nonumber\\
 && \nonumber\\
 &&r=\frac {\sqrt {7}\sqrt {
 \left(5\,c_0-80\,\beta\,{c_0}^{2}+\sqrt {-5120\,{\beta}
^{2}{c_0}^{4}+1024\,\beta\,{c_0}^{3}-23\,{c_0}^{2}+480
\,\beta\,{c_0}^{2}\pi\,\rho-16\,c_0\,\pi\,\rho} \right) }}{14
c_0\, \sqrt{30\,\beta\,c_0-1} }\,, \hspace*{2cm} {\color{blue} ({\mathbf D}_1)}
\end{eqnarray*}
where the first equation of Eq. ${\color{blue} ({\mathbf D_1})}$
has four roots and we write only the real one given by the second
equation of Eq. ${\color{blue} ({\mathbf D_1})}$. The second
equation of ${\color{blue} ({\mathbf D_1})}$ ensures that the
constant $c_0$ must not equal zero and $\beta\neq
\frac{1}{30\,\,c_0}$. Using Eq. ${({\mathbf D_1})}$ in the form of
$p_r$ and $p_\theta$ displayed in Appendix A we get:
 \begin{eqnarray*} \label{solp}
&& p_r(\rho)=\frac {1}{{2c_1}^{3}\pi }\Bigg\{332416 \,\pi \,\rho\,\beta\,c_0-12983040\,\pi \,\rho\,{\beta}^{2}{c_0}^{2}-87808\,{c_1}^{2} \pi \,\rho-25088\,c_1\,\pi \,\rho-10976\,{c_1}^{3}\pi \,\rho -1551032\,
\Upsilon\beta\,c_0\,c_1\nonumber\\
&&-3136\,\pi \,\rho+252073920\, \Upsilon{\beta}^{2}{c_0}^{2}{c_1}^{2}- 2857680\,\beta{c_1}^{3}{\beta}^{2}{c_0}^{ 2}+56560000\,\Upsilon{\beta}^{2}{c_0}^{2}c_1- 6134688\,
\Upsilon\beta\,c_0\,{c_1}^{2}- 314874\,\Upsilon{c_1}^{3}\beta\,c_0\nonumber\\
&&- 4465036800\,\Upsilon{\beta}^{3}{c_0}^{3}{c_1}^{ 2}+284004000\,\Upsilon{\beta}^{3}{c_0}^{3}{c_1}^ {3}-896582400\,\Upsilon{\beta}^{3}{c_0}^{3}c_1+ 28449792000\,
\Upsilon{\beta}^{4}{c_0}^{4}{c_1}^{ 2}-2963520000\,\Upsilon{\beta}^{4}{c_0}^{4}{c_1 }^{3}\nonumber\\
&&+5193216000\,\Upsilon{\beta}^{4}{c_0}^{4}c_1-1354752000\,{\beta}^{4}{c_0}^{4}\pi \,\rho+220147200\,{ \beta}^{3}{c_0}^{3}\pi \,\rho-85349376000\,{\beta}^{4}{c_0 }^{4}\pi \,\rho\,{c_1}^{2}+8890560000\,{\beta}^{4}{c_0}^{4} \pi \,\rho\,{c_1}^{3}\nonumber\\
&&-15579648000\,{\beta}^{4}{c_0}^{4}\pi \,\rho\,c_1+
10905753600\,{\beta}^{3}{c_0}^{3}\pi \,\rho\,{c_1}^{2}-592704000\,{\beta}^{3}{c_0}^{3}\pi \,\rho\,{c_1}^{3}+2235340800\,{\beta}^{3}{c_0}^{3}\pi \,\rho\,c_1+1960 \,\Upsilon\nonumber\\
&&-3612672000\,{\beta}^{5}{c_0}^{6}-
227598336000\,{\beta}^{5}{c_0}^{6}{c_1}^{2}+23708160000\,{ \beta}^{5}{c_0}^{6}{c_1}^{3}-41545728000\,{\beta}^{5}{c_0}^{6}c_1-165580800\,{\beta}^{4}{c_0}^{5}\nonumber\\
&&-2694451200 \,{\beta}^{4}{c_0}^{5}c_1-18334310400\,{\beta}^{4}{c_0}^{5}{c_1}^{2}+3358656000\,{\beta}^{4}{c_0}^{5}{c_1}^ {3}
+451584000\,\Upsilon{\beta}^{4}{c_0}^{4}-86553600 \,\Upsilon{\beta}^{3}{c_0}^{3}\nonumber\\
&&+6028960\,\Upsilon{\beta}^{2}{c_0}^{2}-180320\,\Upsilon\beta\,c_0
+15652\,\Upsilon\,c_1+ 54656\,\Upsilon\,{c_1}^{2}+6811\,\Upsilon\,{c_1}^{3}+658560\,{c_1}^{3}\beta\,c_0\,\pi \,\rho-119669760\,{\beta}^{2}{c_0}^{2}c_1\,\pi \,\rho\nonumber\\
&&+ 2834944\,\beta\,c_0\,c_1\,\pi \,\rho-521579520\,{\beta}^{2 }{c_0}^{2}{c_1}^{2}\pi \,\rho+11063808\,\beta\,c_0\,{c_1}^{2}\pi \,\rho-129477600\,{c_1}^{3}{\beta}^{3}{c_0} ^{4}-79125312\,{\beta}^{2}{c_0}^{3}{c_1}^{2}\nonumber\\
&&+570384640\,{ \beta}^{3}{c_0}^{4}c_1+119870\,{c_1}^{3}\beta\,{c_0}^{2}-20165376\,{\beta}^{2}{c_0}^{3}c_1+2744071680\,{ \beta}^{3}{c_0}^{4}{c_1}^{2}
-4903920\,{c_1}^{3}{\beta }^{2}{c_0}^{3}+99624\,\beta\,{c_0}^{2}c_1\nonumber\\
&&+252832\,{c_1}^{2}\beta\,{c_0}^{2}+392\,c_0+14112\,\beta\,{c_0}^{2}-2356704\,{\beta}^{2}{c_0}^{3}+56949760\,{\beta}^{3}{c_0}^{4}
+2996\,c_0\,c_1+9860\,{c_1}^{2}c_0+ 1129\,{c_1}^{3}c_0\Bigg\}\,,\nonumber\\
&& p_\theta(\rho)=-\frac {1}{392{c_1}^{3}\pi \, \left(30 \,\beta\,c_0-1 \right) ^{2}}\Bigg\{ 172800\,\pi \,\rho\,{\beta}^{2}{c_0}^ {2}+23080\,\beta\,c_0\,c_1\,\Upsilon+31360\, \beta\,c_0\,{c_1}^{2}\Upsilon
-238400\,{ \beta}^{2}{c_0}^{2}c_1\,\Upsilon\nonumber\\
&&
-245280\,{ \beta}^{2}{c_0}^{2}{c_1}^{2}\Upsilon-2576\,{c_1}^{3}\beta\,c_0\,\Upsilon+10080\,{c_1}^{3 }{\beta}^{2}{c_0}^{2}\Upsilon+5360\,\beta\,c_0 \,\Upsilon-57600\,{\beta}^{2}{c_0}^{2}
\Upsilon-110\,\Upsilon-352\,c_1\,\Upsilon-  224\,{c_1}^{2}\Upsilon\nonumber\\
&&-  7\,{c_1}^{3}\Upsilon-  11040\,\pi \,\rho\,\beta\,c_0+176\,\pi \,\rho+896\,{c_1}^{2}\pi \,\rho+832\,c_1\,\pi \,\rho+168\,{c_1}^{3} \pi \,\rho
-100800\,{c_1}^{3}{\beta}^{2}{c_0}^{2}\pi \,\rho-
   1680\,{c_1}^{3}\beta\,c_0\,\pi \,\rho\nonumber\\
&&+1219200\,{\beta}^{2} {c_0}^{2}c_1\,\pi \,\rho-65600\,\beta\,c_0\,c_1\, \pi \,\rho+2499840\,{\beta}^{2}{c_0}^{2}{c_1}^{2}\pi \,\rho -
   110208\,\beta\,c_0\,{c_1}^{2}\pi \,\rho+1612800\,{c_1 }^{3}{\beta}^{3}{c_0}^{4}\nonumber\\
&&+3822112\,{\beta}^{2}{c_0}^{3}{c_1}^{2}-10188800\,{\beta}^{3}{c_0}^{4}c_1+
   2128\,{c_1}^{3}\beta\,{c_0}^{2}+2193600\,{\beta}^{2}{c_0}^{3}c_1-12149760\,{\beta}^{3}{c_0}^{4}{c_1}^{2}-358400\,{c_1}^{3}{\beta}^{2}{c_0}^{3}\nonumber\\
&&-
   73208\,\beta\,{c_0}^{2}c_1-99008\,{c_1}^{2}\beta\,{c_0}^{2}-22\,c_0-1744 \,\beta\,{c_0}^{2}+66560\,{\beta}^{2}{c_0}^{3}+460800\,{ \beta}^{3}{c_0}^{4}+
   736\,c_0\,c_1+784\,{c_1}^{2}c_0+77\,{c_1}^{3}c_0\Bigg\}\,, \nonumber\\
&&p(\rho)=\Bigg\{c_0\, \Bigg({\frac {752}{7}}\,{\frac {\beta\, \Upsilon_1 }{\Upsilon_2}} -16-\frac {\alpha\,\Upsilon_1 ^{14}c_1}{182059119829942534144{c_0}^{13} \Upsilon_2 ^{14}}-\frac { 715\beta\, \Upsilon_1 ^{7}c_1}{843308032{c_0}^{6} \Upsilon_2 ^{7}}-\frac {143\beta\, \Upsilon_1 ^{8}c_1}{6746464256{c_0}^{7} \Upsilon_2 ^{8}}-\frac {15\beta\, \Upsilon_1 ^{2}c_1}{7c_0\, \Upsilon_2 ^{2}} \nonumber\\
&&-\frac {71\beta\, \Upsilon_1 ^{4}c_1}{5488{c_0}^{3} \Upsilon_2 ^{4}}-\frac {51\beta\, \Upsilon_1 ^{5}c_1}{76832{c_0}^{4} \Upsilon_2 ^{5}}
-\frac {3217\beta\, \Upsilon_1 ^{6}c_1}{ 120472576{c_0}^{5} \Upsilon_2 ^{6}}- \frac {39\beta\, \Upsilon_1 ^{9}c_1}{94450499584{c_0}^{8}\Upsilon^{9}}-\frac {65\beta\, \Upsilon_1 ^{10}c_1}{10578455953408{c_0}^{9} \Upsilon_2 ^{10}}-\frac {5\beta\, \Upsilon_1 ^{11}c_1}{74049191673856{c_0}^{10} \Upsilon_2 ^{11}}\nonumber\\
&&-64\,\beta\,c_0 \,c_1-{\frac {19}{98}}\,{\frac {\beta\, \Upsilon_1 ^{3}c_1}{{c_0}^{2} \Upsilon^{3}}}-{\frac { 15}{29027283136151552}}\,{\frac {\beta\, \Upsilon_1 ^{12}c_1}{{c_0}^{11}\Upsilon^{12}}}-{\frac {1}{406381963906121728}}\,{\frac { \beta\, \Upsilon_1 ^{13}c_1}{{c_0}^{12 } \Upsilon_2 ^{13}}}-{\frac {30}{7}}\,{ \frac {\Upsilon_1 }{c_0\, \Upsilon_2 }}-{\frac {15}{28}}\,{\frac { \Upsilon_1 ^{2}}{{c_0}^{2} \Upsilon_2 ^ {2}}}\nonumber\\
&&-8\,c_1-\frac {15 \Upsilon_1^{3}}{392{c_0}^{3} \Upsilon_2 ^ {3}}-\frac {75\Upsilon_1^{4}}{43904{c_0}^{4} \Upsilon_2 ^ {4}}-\frac {15\Upsilon_1^{5}}{307328{c_0}^{5} \Upsilon_2 ^ {5}}-\frac {15 \Upsilon_1 ^{6}}{17210368{c_0}^{6} \Upsilon_2 ^ {6}}-\frac {15\Upsilon_1^{7}}{1686616064{c_0}^{7} \Upsilon_2 ^{7}}-\frac {15\Upsilon_1^{8}}{377801998336{c_0}^{8} \Upsilon_2^{8}}-\frac {16\beta\, \Upsilon_1 c_1}{\Upsilon_2}\nonumber\\
&&+{ \frac {148}{7}}\,{\frac {\beta\, ^{2}}{c_0\, \Upsilon_2 ^{2}} }+{\frac {515}{196}}\,{\frac {\beta\,\Upsilon_1^{3}}{{c_0}^{2} \Upsilon_2 ^ {3}}}+{\frac {635}{2744}}\,{\frac {\beta\, \Upsilon_1 ^{4}}{{c_0}^{3} \Upsilon_2 ^{4}}}+{\frac {2351}{153664}}\,{\frac {\beta\,\Upsilon_1 ^{5}}{{c_0}^{4} \Upsilon^{5}}}+{\frac {11789}{15059072}}\,{\frac { \beta\, \Upsilon_1 ^{6}}{{c_0}^{5} \Upsilon^{6}}}+{\frac {15167}{481890304}}\,{ \frac {\beta\, \Upsilon_1 ^{7}}{{c_0}^{6} \Upsilon_2 ^{7}}}+{\frac {23595}{ 23612624896}}\,{\frac {\beta\, \Upsilon_1^{8}}{ {c_0}^{7} \Upsilon_2 ^{8}}}\nonumber\\
&&+{\frac { 4719}{188900999168}}\,{\frac {\beta\, \Upsilon_1^{9}}{{c_0}^{8} \Upsilon_2 ^ {9}}}+{\frac {1287}{2644613988352}}\,{\frac {\beta\, \Upsilon_1 ^{10}}{{c_0}^{9} \Upsilon^{10}}}+{\frac {2145}{296196766695424}}\,{ \frac {\beta\, \Upsilon_1 ^{11}}{{c_0}^{ 10} \Upsilon_2 ^{11}}}+{\frac {165}{ 2073377366867968}}\,{\frac {\beta\, \Upsilon_2^{12}}{{c_0}^{11} \Upsilon_2 ^{12}}}\nonumber\\
&&+{\frac {495}{812763927812243456}}\,{\frac {\beta\, \Upsilon_1 ^{13}}{{c_0}^{12}\Upsilon^{13}}}+{\frac {33}{11378694989371408384}}\,{ \frac {\beta\, \Upsilon_1 ^{14}}{{c_0}^{ 13} \Upsilon_2 ^{14}}}-{\frac {1}{ 13492928512}}\,{\frac { \Upsilon_1^{7}c_1}{ {c_0}^{7} \Upsilon_2 ^{7}}}-{\frac {1 }{60236288}}\,{\frac { \Upsilon_1 ^{6}c_1}{{ c_0}^{6} \Upsilon_2 ^{6}}}-{\frac { \Upsilon_1 c_1}{c_0\,\Upsilon }}\nonumber\\
&&-\frac { \Upsilon_1 ^{2}c_1}{14{c_0}^{2} \Upsilon^{2}}-{\frac {5}{1568}}\,{\frac { \Upsilon_1 ^{3}c_1}{{c_0}^{3} \Upsilon^{3}}}-{\frac {1}{10976}}\,{\frac { \Upsilon_1^{4}c_1}{{c_0}^{4} \Upsilon_2 ^{4}}}-{\frac {1}{614656}}\,{\frac { \Upsilon_1 ^{5}c_1}{{c_0}^{5} \Upsilon^{5}}}+{\frac {33}{ 5097655355238390956032}}\,{\frac {\beta\, \Upsilon_1 ^{15}}{{c_0}^{14} \Upsilon_2 ^{15}}}+256\,\beta\,c_0 \Bigg)\Bigg\}\nonumber\\
&&\Bigg[24 {\pi} \left( \frac {5c_0-80\beta{c_0}^{2}+\Upsilon }{28c_0 \Upsilon}-c_1 \right)\Bigg]^{-1}
+\Bigg\{c_0 \Bigg( -{ \frac {381}{11378694989371408384}}{\frac {\beta \Upsilon_1^{14}c_1}{{c_0}^{13}\Upsilon^{14}}}-{\frac {41423}{210827008}}{ \frac {\beta\Upsilon_1 ^{7}c_1}{{c_0}^{6} \Upsilon_2 ^{7}}}-{\frac {23595}{ 3373232128}}{\frac {\beta\, \Upsilon_1 ^{8}c_1}{{c_0}^{7} \Upsilon_2 ^{8}}}\nonumber\\
&&- {\frac {376}{7}}\,{\frac {\beta\, \Upsilon_1^{2}c_1}{c_0\, \Upsilon_2 ^{2}}}-{\frac {1675}{196}}\,{\frac {\beta\, \Upsilon_1^{3}c_1}{{c_0}^{2} \Upsilon_2 ^{3}}}-{\frac {2531}{2744}}\,{\frac { \beta\, \Upsilon_1 ^{4}c_1}{{c_0}^{3} \Upsilon_2 ^{4}}}-{\frac {5563}{76832}}\, {\frac {\beta\, \Upsilon_1 ^{5}c_1}{{c_0}^{4} \Upsilon_2 ^{5}}}-{\frac {64675}{ 15059072}}\,{\frac {\beta\, \Upsilon_1 ^{6}c_1}{{c_0}^{5} \Upsilon_2 ^{6}}}- {\frac {143}{737894528}}\,{\frac {\beta\, \Upsilon_1 ^{9}c_1}{{c_0}^{8} \Upsilon^{9}}}\nonumber\\
&&-{\frac {10959}{2644613988352}}\,{\frac {\beta \, \Upsilon_1 ^{10}c_1}{{c_0}^{9} \Upsilon_2 ^{10}}}-{\frac {9945}{ 148098383347712}}\,{\frac {\beta\, \Upsilon_1^{11}c_1}{{c_0}^{10} \Upsilon_2 ^{11}}}-{\frac {1655}{2073377366867968}}\,{\frac {\beta\, \Upsilon_1 ^{12}c_1}{{c_0}^{11} \Upsilon_2 ^{12}}}-{\frac {1335}{ 203190981953060864}}\,{\frac {\beta\, \Upsilon_1^{13}c_1}{{c_0}^{12} \Upsilon_2 ^{13}}}\nonumber\\
&&+{\frac {15}{7}}\,{\frac {\beta\,{c_1}^{2} \Upsilon_1 ^{2}}{c_0\, \Upsilon^{2}}}+{\frac {57}{196}}\,{\frac {\beta\,{c_1 }^{2} \Upsilon_1 ^{3}}{{c_0}^{2} \Upsilon_2 ^{3}}}+{\frac {71}{2744}}\,{\frac {\beta \,{c_1}^{2} \Upsilon_1 ^{4}}{{c_0}^{3} \Upsilon_2 ^{4}}}+{\frac {255}{153664}}\, {\frac {\beta\,{c_1}^{2} \Upsilon_1 ^{5}}{ {c_0}^{4} \Upsilon_2 ^{5}}}+{\frac { 9651}{120472576}}\,{\frac {\beta\,{c_1}^{2}\Upsilon_1^{6}}{{c_0}^{5} \Upsilon_2 ^{6}}}+{\frac {715}{240945152}}\,{\frac {\beta\,{c_1}^{ 2} \Upsilon_1 ^{7}}{{c_0}^{6} \Upsilon^{7}}}\nonumber
\end{eqnarray*}
\begin{eqnarray*}
&&-320\,\beta\,c_0\,c_1+{\frac {143}{1686616064}}\,{\frac { \beta\, \Upsilon_1 ^{8}{c_1}^{2}}{{c_0 }^{7} \Upsilon_2 ^{8}}}+{\frac {351}{ 188900999168}}\,{\frac {\beta\,\Upsilon_1 ^{9}{c_1}^{2}}{{c_0}^{8} \Upsilon_2 ^{9}}}+{\frac {325}{10578455953408}}\,{\frac {\beta\, \Upsilon_1 ^{10}{c_1}^{2}}{{c_0}^{9} \Upsilon_2 ^{10}}}+{\frac {55}{ 148098383347712}}\,{\frac {\beta\,\Upsilon_1^{11}{c_1}^{2}}{{c_0}^{10} \Upsilon_2 ^{11}}}\nonumber\\
&&+8\,c_1+{\frac {45}{14513641568075776}}\,{\frac { \beta\, \Upsilon_1 ^{12}{c_1}^{2}}{{c_0}^{11} \Upsilon_2 ^{12}}}+{\frac {13}{ 812763927812243456}}\,{\frac {\beta\, \Upsilon_1^{13}{c_1}^{2}}{{c_0}^{12} \Upsilon_2 ^{13}}}+{\frac {1}{26008445689991790592}}\,{\frac { \beta\, \Upsilon_1 ^{14}{c_1}^{2}}{{c_0}^{13} \Upsilon_2 ^{14}}}\nonumber\\
&&-{\frac {29}{ 364118239659885068288}}\,{\frac {\beta\, \Upsilon_1 ^{15}c_1}{{c_0}^{14} \Upsilon ^{15}}}-{\frac {240 }{7}}\,{\frac {\beta\,\Upsilon_1}{\Upsilon}}+{\frac {8}{7}}\,{\frac {-80\,\beta\,{c_0}^{2} +5\,c_0+\Upsilon }{c_0\, \Upsilon_2 }}+{\frac {57}{196}} \,{\frac { \Upsilon_1 ^{2}}{{c_0}^{2} \Upsilon_2 ^{2}}}+{\frac {25}{784}}\,{ \frac { \Upsilon_1 ^{3}}{{c_0}^{3}\Upsilon^{3}}}+{\frac {43}{21952}}\,{\frac { \Upsilon_1 ^{4}}{{c_0}^{4}\Upsilon^{4}}}\nonumber\\
&&+{\frac {45}{614656}}\,{\frac { \Upsilon_1 ^{5}}{{c_0}^{5} \Upsilon^{5}}}+{\frac {29}{17210368}}\,{\frac { \Upsilon_1 ^{6}}{{c_0}^{6} \Upsilon^{6}}}+{\frac {11}{481890304}}\,{\frac { \Upsilon_1 ^{7}}{{c_0}^{7} \Upsilon^{7}}}+{\frac {15}{94450499584}}\,{\frac { \Upsilon_1 ^{8}}{{c_0}^{8} \Upsilon^{8}}}+{\frac {1}{2644613988352}}\,{\frac { \Upsilon_1 ^{9}}{{c_0}^{9} \Upsilon ^{9}}}-{\frac {200\beta\, \Upsilon_1 c_1}{\Upsilon_2}}\nonumber\\
&&-{ \frac {570}{49}}\,{\frac {\beta\, \Upsilon_1 ^{2}}{c_0\, \Upsilon_2 ^{2}} }-{\frac {173}{98}}\,{\frac {\beta\, \Upsilon_1^{3}}{{c_0}^{2} \Upsilon_2 ^ {3}}}-{\frac {15}{98}}\,{\frac {\beta\, \Upsilon_1 ^{4}}{{c_0}^{3} \Upsilon_2 ^{4}}}-{\frac {165}{21952}}\,{\frac {\beta\, \Upsilon_1^{5}}{{c_0}^{4} \Upsilon^{5}}}-{\frac {13}{134456}}\,{\frac {\beta\, \Upsilon_1 ^{6}}{{c_0}^{5}\Upsilon_2^{6}}}+{\frac {13245}{843308032}}\,{\frac { \beta\, \Upsilon_1 ^{7}}{{c_0}^{6} \Upsilon^{7}}}+{\frac {138357}{94450499584}}\,{ \frac {\beta\, \Upsilon_1 ^{8}}{{c_0}^{7} \Upsilon_2 ^{8}}}\nonumber\\
&&+{\frac {96525}{ 1322306994176}}\,{\frac {\beta\,\Upsilon_1 ^{9}}{{c_0}^{8} \Upsilon_2 ^ {9}}}+{\frac {3289}{1322306994176}}\,{\frac {\beta\, \Upsilon_1^{10}}{{c_0}^{9} \Upsilon_2^{10}}}+{\frac {9087}{148098383347712}}\,{ \frac {\beta\, \Upsilon_1 ^{11}}{{c_0}^{ 10} \Upsilon_2 ^{11}}}+{\frac {9165}{ 8293509467471872}}\,{\frac {\beta\, \Upsilon_1^{12}}{{c_0}^{11} \Upsilon_2 ^{12}}}\nonumber\\
&&+{\frac {1655}{116109132544606208}}\,{\frac {\beta\, \Upsilon_1 ^{13}}{{c_0}^{12} \Upsilon_2^{13}}}+{\frac {1425}{11378694989371408384}} \,{\frac {\beta\, \Upsilon_1 ^{14}}{{c_0}^{13} \Upsilon_2 ^{14}}}+{\frac {29}{ 1686616064}}\,{\frac { \Upsilon_1^{7}c_1}{{ c_0}^{7} \Upsilon_2 ^{7}}}+{\frac {13 }{8605184}}\,{\frac { \Upsilon_1^{6}c_1}{{ c_0}^{6} \Upsilon_2 ^{6}}}\nonumber\\
&&+{\frac {22 }{7}}\,{\frac { \Upsilon_1 c_1}{c_0\, \Upsilon_2 }}+{\frac { \Upsilon_1^{2}c_1}{2{c_0}^{2} \Upsilon_2 ^{2}}}+{\frac {17}{392}}\,{\frac { \Upsilon_1 ^{3}c_1}{{c_0}^{3}\Upsilon_2^{3}}}+{\frac {25}{10976}}\,{\frac { \Upsilon_1 ^{4}c_1}{{c_0}^{4} \Upsilon_2^{4}}}+{\frac {23}{307328}}\,{\frac { \Upsilon_1 ^{5}c_1}{{c_0}^{5}\Upsilon_2^{5}}}+{\frac {429}{ 637206919404798869504}}\,{\frac {\beta\, \Upsilon_1 ^{15}}{{c_0}^{14} \Upsilon_2 ^{15}}}\nonumber\\
&&+{\frac {1}{11806312448}}\,{\frac { \Upsilon_1^{8}c_1}{{c_0}^{8} \Upsilon_2 ^{8}}}+{\frac {239}{ 142734349946674946768896}}\,{\frac {\beta\, \Upsilon_1 ^{16}}{{c_0}^{15} \Upsilon_2 ^{16}}}+{\frac {{c_1}^{2} \Upsilon_1 }{c_0\, \Upsilon_2 }}+{\frac {3}{28}}\,{\frac {{c_1}^{2} \Upsilon_1^{2}}{{c_0}^{2} \Upsilon_2 ^{2}}}+{\frac {5}{784}}\,{\frac {{c_1}^{2}\Upsilon_1 ^{3}}{{c_0}^{3} \Upsilon_2^{3}}}+{\frac {5}{21952}}\,{\frac {{c_1} ^{2} \Upsilon_1 ^{4}}{{c_0}^{4} \Upsilon_2 ^{4}}}\nonumber\\
&&+{\frac {3}{614656}}\,{\frac {{c_1}^{2} \Upsilon_1 ^{5}}{{c_0}^{5} \Upsilon_2 ^{5}}}+{\frac {1}{17210368}}\, {\frac {{c_1}^{2} \Upsilon_1 ^{6}}{{c_0 }^{6} \Upsilon_2 ^{6}}}+{\frac {1}{ 3373232128}}\,{\frac {{c_1}^{2} \Upsilon_1 ^{7}}{{c_0}^{7} \Upsilon_2 ^ {7}}}+8\,\frac {\beta\,{c_1}^{2} \Upsilon_1}{\Upsilon_2} \Bigg)\Bigg\}\Bigg[6 {\pi} \left( \frac {\Upsilon_1 }{28c_0\, \Upsilon_2 } -c_1\right) ^{2}\Bigg]^{-1} \,, \hspace*{2cm} {\color{blue} ({\mathbf D}_2)}
 \end{eqnarray*}
where $\Upsilon=\sqrt {c_0\, \left(480\,\pi \,\rho\,\beta\,c_0 -5120\,{\beta}^{2}{c_0}^{3}+1024\,\beta\,{c_0}^{2}-23\,c_0 -16\,\pi \,\rho \right) }$, $\Upsilon_1=\left( -80\,\beta\,{c_0}^{2}+5\,c_0+\Upsilon \right)$, $\Upsilon_2=\left( -1+30\,\beta\,c_0 \right)$.
\newpage
{\centerline{Appendix E:  The different forces of on the
model.}}\vspace{0.3cm} The forces that act on the model presented
in Appendix A have the following forms:
\renewcommand{\eqref}{}
\begin{eqnarray*}
&&F_g=\frac {{c_0}^{2}r}{16 \left(c_0\,{r}^{2} -c_1 \right) ^{6}\pi } \Bigg(21\,{c_0}^{5}{r}^{10}c_1-9\,c_0\,{r}^{2}-2\,\beta \,{c_0}^{15}{r}^{28}c_1-168\,\beta\,{c_0}^{2}{r}^{2}c_1-12870\,\beta\,{c_0}^{8}{r}^{14}c_1-10010\,
\beta\,{ c_0}^{9}{r}^{16}c_1\nonumber\\
&&-840\,\beta\,{c_0}^{3}{r}^{4}c_1-2660\,\beta\,{c_0}^{4}{r}^{6}c_1-5964\,\beta\,{c_0}^{5}{r}^{8}c_1-9996\,\beta\,{c_0}^{6}{r}^{10}c_1- 12868\,\beta\,{c_0}^{7}{r}^{12}c_1-6006\,\beta\,{c_0} ^{10}{r}^{18}c_1-2730\,\beta\,{c_0}^{11}{r}^{20}c_1\nonumber\\
&&- 910\,\beta\,{c_0}^{12}{r}^{22}c_1-210\,\beta\,{c_0}^{ 13}{r}^{24}c_1-30\,\beta\,{c_0}^{14}{r}^{26}c_1+c_1+1176\,\beta\,{c_0}^{3}{r}^{4}+4340\,\beta\,{c_0}^{4}{r }^{6}+11284\,\beta\,{c_0}^{5}{r}^{8}+21924\,\beta\,{c_0}^{ 6}{r}^{10}\nonumber\\
&&+32860\,\beta\,{c_0}^{7}{r}^{12}+38606\,\beta\,{c_0}^{8}{r}^{14}+35750\,\beta\,{c_0}^{9}{r}^{16}+26026\,\beta\, {c_0}^{10}{r}^{18}+14742\,\beta\,{c_0}^{11}{r}^{20}+6370\, \beta\,{c_0}^{12}{r}^{22}+2030\,\beta\,{c_0}^{13}{r}^{24}\nonumber\\
&&+ 450\,\beta\,{c_0}^{14}{r}^{26}+62\,\beta\,{c_0}^{15}{r}^{ 28}+200\,\beta\,{c_0}^{2}{r}^{2}+{c_0}^{7}{r}^{14}c_1+ 7\,{c_0}^{6}{r}^{12}c_1+7\,c_0\,{r}^{2}c_1+21\,{c_0}^{2}{r}^{4}c_1+35\,{c_0}^{3}{r}^{6}c_1+35\,{c_0}^{4}{r}^{8}c_1\nonumber\\
&&-16\, \beta\,c_0\,c_1+4\,\beta\,{c_0}^{16}{r}^{30}+16\, \beta\,c_0-35\,{c_0}^{2}{r}^{4}-77\,{c_0}^{3}{r}^{6}- 105\,{c_0}^{4}{r}^{8}-91\,{c_0}^{5}{r}^{10}-49\,{c_0}^{ 6}{r}^{12}-15\,{c_0}^{7}{r}^{14}-2\,{c_0}^{8}{r}^{16}-1 \Bigg) \,,\nonumber\\
&&F_a=\frac {{c_0}^{2} }{4r\pi \, \left(c_0\,{r}^{2} -c_1 \right) ^{2 }}\Bigg\{ 80\,{r}^{2}-11648\,\beta\,{c_0}^{14}{r}^{28}c_1-8128\,\beta\,c_0\,{r}^{2}c_1- 4864600\,\beta\,{c_0}^{7}{r}^{14}c_1-4896320\,\beta\,{c_0}^{8}{r}^{16}c_1\nonumber\\
&&-67200\,\beta\,{c_0}^{2}{r}^{4}c_1-315840\,\beta\,{c_0}^{3}{r}^{6}c_1-987392\,\beta\,{c_0}^{4}{r}^{8}
c_1-2220960\,\beta\,{c_0}^{5}{r}^{10}c_1-3750528\,\beta\,{c_0}^{6}{r}^{12}c_1-3827824\, \beta\,{c_0}^{9}{r}^{18}c_1\nonumber\\
&&-2306304\,\beta\,{c_0}^{10 }{r}^{20}c_1-1051960\,\beta\,{c_0}^{11}{r}^{22}c_1- 351680\,\beta\,{c_0}^{12}{r}^{24}c_1-81360\,\beta\,{c_0}^{13}{r}^{26}c_1+1680\,\beta\,{c_0}^{2}{c_1}^{2}{r }^{4}+8512\,\beta\,{c_0}^{3}{c_1}^{2}{r}^{6}\nonumber\\
&&+23856\,\beta \,{c_0}^{4}{c_1}^{2}{r}^{8}+45696\,\beta\,{c_0}^{5}{c_1}^{2}{r}^{10}+64340\,\beta\,{c_0}^{6}{c_1}^{2}{r}^{ 12}+68640\,\beta\,{c_0}^{7}{c_1}^{2}{r}^{14}+56056\,\beta \,{c_0}^{8}{r}^{16}{c_1}^{2}+34944\,\beta\,{c_0}^{9}{r }^{18}{c_1}^{2}\nonumber\\
&&+16380\,\beta\,{c_0}^{10}{r}^{20}{c_1}^ {2}+5600\,\beta\,{c_0}^{11}{r}^{22}{c_1}^{2}+1320\,\beta\, {c_0}^{12}{r}^{24}{c_1}^{2}+192\,\beta\,{c_0}^{13}{r}^ {26}{c_1}^{2}+13\,\beta\,{c_0}^{14}{r}^{28}{c_1}^{2}- 778\,\beta\,{c_0}^{15}{r}^{30}c_1\nonumber\\
&&+1209208\,\beta\,{c_0}^{10}{r}^{20}+1166880\,\beta\,{c_0}^{9}{r}^{18}+682180\,\beta \,{c_0}^{8}{r}^{16}+46592\,\beta\,{c_0}^{7}{r}^{14}-400960 \,\beta\,{c_0}^{5}{r}^{10}-356496\,\beta\,{c_0}^{6}{r}^{12 }\nonumber\\
&&-94080\,\beta\,{c_0}^{3}{r}^{6}-245840\,\beta\,{c_0}^{4}{ r}^{8}+352\,{c_0}^{6}{r}^{14}c_1-2176\,\beta\,c_0\,{r} ^{2}+392\,c_0\,{r}^{4}c_1+1064\,{c_0}^{5}{r}^{12}c_1+873600\,\beta\,{c_0}^{11}{r}^{22}\nonumber\\
&&+453180\,\beta\,{c_0} ^{12}{r}^{24}+166880\,\beta\,{c_0}^{13}{r}^{26}+41640\,\beta\,{ c_0}^{14}{r}^{28}+6336\,\beta\,{c_0}^{15}{r}^{30}+1792\,{c_0}^{4}{r}^{10}c_1+1820\,{c_0}^{3}{r}^{8}c_1+1120 \,{c_0}^{2}{r}^{6}c_1\nonumber\\
&&+112\,{c_1}^{2}c_0\,{r}^{4}+ 210\,{c_1}^{2}{c_0}^{2}{r}^{6}+50\,{c_0}^{7}{r}^{16}c_1+445\,\beta\,{c_0}^{16}{r}^{32}+140\,{c_1}^{2}{c_0}^{4}{r}^{10}+224\,{c_1}^{2}{c_0}^{3}{r}^{8}+7\,{c_1 }^{2}{c_0}^{6}{r}^{14}+48\,{c_1}^{2}{c_0}^{5}{r}^{12}\nonumber\\
&&- 21248\,\beta\,{c_0}^{2}{r}^{4}+64\,{r}^{2}c_1-384\,\beta\, c_1-64\,\beta\,{c_1}^{2}+28\,{r}^{2}{c_1}^{2}+576\,c_0\,{r}^{4}+1820\,{c_0}^{2}{r}^{6}+3248\,{c_0}^{3}{r}^{ 8}+3570\,{c_0}^{4}{r}^{10}\nonumber\\
&&+2464\,{c_0}^{5}{r}^{12}+1036\,{c_0}^{6}{r}^{14}+240\,{c_0}^{7}{r}^{16}+23\,{c_0}^{8}{r} ^{18} \Bigg\}
\,,\nonumber\\
&&
F_h=\frac {c_0 }{8\pi \, \left( c_1-c_0\,{r}^{2} \right)}\, \Bigg\{ -28\,\beta\,{c_0}^{15}{r}^{27}c_1-896\,\beta\,{c_0}^{2}rc_1-160160\,\beta\,{c_0 }^{8}{r}^{13}c_1-128128\,\beta\,{c_0}^{9}{r}^{15}c_1- 6720\,\beta\,{c_0}^{3}{r}^{3}c_1\nonumber\\
&&-25536\,\beta\,{c_0}^ {4}{r}^{5}c_1-63616\,\beta\,{c_0}^{5}{r}^{7}c_1-114240 \,\beta\,{c_0}^{6}{r}^{9}c_1-154416\,\beta\,{c_0}^{7} {r}^{11}c_1-78624\,\beta\,{c_0}^{10}{r}^{17}c_1-36400 \,\beta\,{c_0}^{11}{r}^{19}c_1\nonumber\\
&&-12320\,\beta\,{c_0}^{ 12}{r}^{21}c_1-2880\,\beta\,{c_0}^{13}{r}^{23}c_1-416 \,\beta\,{c_0}^{14}{r}^{25}c_1-240\,{c_0}^{8}{r}^{15}- 240\,c_0\,r-1680\,{c_0}^{2}{r}^{3}-5040\,{c_0}^{3}{r}^{ 5}-8400\,{c_0}^{4}{r}^{7}\nonumber\\
&&-8400\,{c_0}^{5}{r}^{9}-5040\,{c_0}^{6}{r}^{11}-1680\,{c_0}^{7}{r}^{13}+66304\,\beta\,{c_0}^{3}{r}^{3}+346080\,
\beta\,{c_0}^{4}{r}^{5}+1137920\, \beta\,{c_0}^{5}{r}^{7}+2633120\,\beta\,{c_0}^{6}{r}^{9}\nonumber\\
&&+ 4526976\,\beta\,{c_0}^{7}{r}^{11}+5945464\,\beta\,{c_0}^{8 }{r}^{13}+6040320\,\beta\,{c_0}^{9}{r}^{15}+4756752\,\beta\,{c_0}^{10}{r}^{17}+2882880\,\beta\,{c_0}^{11}{r}^{19}+1321320 \,\beta\,{c_0}^{12}{r}^{21}\nonumber\\
&&+443520\,\beta\,{c_0}^{13}{r}^{ 23}+102960\,\beta\,{c_0}^{14}{r}^{25}+14784\,\beta\,{c_0}^ {15}{r}^{27}+6016\,\beta\,{c_0}^{2}r-14\,{c_0}^{7}{r}^{13}c_1-96\,{c_0}^{6}{r}^{11}c_1-56\,c_0\,rc_1-224 \,{c_0}^{2}{r}^{3}c_1\nonumber\\
&&-420\,{c_0}^{3}{r}^{5}c_1-448 \,{c_0}^{4}{r}^{7}c_1-280\,{c_0}^{5}{r}^{9}c_1+990 \,\beta\,{c_0}^{16}{r}^{29} \Bigg\}\,,\hspace*{1cm} {\color{blue} ({\mathbf E})}
 \end{eqnarray*}
 \newpage
 {\centerline{Appendix F:  The form of the mass in terms of the central density at the surface of the star}}\vspace{0.3cm}
The form of the mass in terms of the central density at the surface of the star takes the form:
\renewcommand{\eqref}{}
\begin{eqnarray*}
&&M(\rho_0)=\frac {1}{{2836619246897071128576\beta}^{15}\pi } \Bigg\{\left( 3+\Upsilon_3 \right) {l}^{3} \left( 3\,{l}^{2}+{l}^{2}\Upsilon_3 +96\,\beta \right)\Bigg( 3\,{ l}^{4}+{l}^{4}\Upsilon_3 +16\,{l}^{4}\beta \,\rho_0\,\pi +768\,{\beta}^{2}+48\,{l}^{2}\beta \nonumber\\
&& +16\,{l}^{2} \beta\,\Upsilon_3 \Bigg)\times\Bigg( 9\,{l}^ {8}+3\,{l}^{8}\Upsilon_3 +96\,{l}^{8}\beta \,\rho_0\,\pi +16\,{l}^{8}\Upsilon_3 \beta\,\rho_0\,\pi +128\,{l}^{8}{\beta}^{2}{\rho_0}^{2}{\pi }^{2}+288\,{l}^{6}\beta+96\,{l}^{6}\beta\,\Upsilon_3 +2304\,{l}^{6}{\beta}^{2}\rho_0\,\pi \nonumber\\
&& +256\,{l}^{6}{ \beta}^{2}\Upsilon_3 \rho_0\,\pi +3456 \,{l}^{4}{\beta}^{2}+1152\,{l}^{4}{\beta}^{2}\Upsilon_3 +18432\,{l}^{4}{\beta}^{3}\rho_0\,\pi +18432\,{l} ^{2}{\beta}^{3}+6144\,{l}^{2}{\beta}^{3}\Upsilon_3 +147456\,{\beta}^{4} \Bigg)  \Bigg( -1358954496\,{\beta }^{7} \nonumber\\
&&+81\,{l}^{14}+6193152\,{l}^{8}{\beta}^{4}\Upsilon_3 \rho_0\,\pi +5505024\,{l}^{8}{\beta}^{5}\Upsilon_3 {\rho_0}^{2}{\pi }^{2}+41287680\,{l}^ {6}{\beta}^{5}\sqrt {9+96\,\beta\,\rho_0\,\pi }\rho_0\,\pi + 132120576\,{l}^{4}{\beta}^{6}\Upsilon_3 \rho_0\,\pi \nonumber\\
&& +23040\,{l}^{12}{\beta}^{2}\Upsilon_3 \rho_0\,\pi +73728\,{l}^{12}{\beta}^{3}\Upsilon_3{\rho_0}^{2}{\pi }^{2}+32768\,{l}^{12}{ \beta}^{4}\Upsilon_3{\rho_0}^{3}{\pi }^{3}+145152\,{l}^{10}{\beta}^{2}+23224320\,{l}^{6}{\beta}^{4}+ 148635648\,{l}^{4}{\beta}^{5} \nonumber\\
&&+5184\,{l}^{12}\beta+2322432\,{l}^{8}{ \beta}^{3}+594542592\,{l}^{2}{\beta}^{6}+516096\,{l}^{10}{\beta}^{3 }\Upsilon_3\rho_0\,\pi +1032192\,{l}^{ 10}{\beta}^{4}\Upsilon_3{\rho_0}^{2}{ \pi }^{2}+432\,{l}^{14}\Upsilon_3\beta\,\rho_0\,\pi \nonumber\\
&& +1920\,{l}^{14}\Upsilon_3{ \beta}^{2}{\rho_0}^{2}{\pi }^{2}+2048\,{l}^{14}\Upsilon_3{\beta}^{3}{\rho_0}^{3}{\pi }^{3}+96768 \,{l}^{12}{\beta}^{2}\rho_0\,\pi +516096\,{l}^{12}{\beta}^{3}{\rho_0}^{2}{\pi }^{2}+688128\,{l}^{12}{\beta}^{4}{\rho_0}^{3}{ \pi }^{3}+2322432\,{l}^{10}{\beta}^{3}\rho_0\,\pi \nonumber\\
&& +9289728\,{l}^{ 10}{\beta}^{4}{\rho_0}^{2}{\pi }^{2}+5505024\,{l}^{10}{\beta}^{5 }{\rho_0}^{3}{\pi }^{3}+30965760\,{l}^{8}{\beta}^{4}\rho_0\, \pi +82575360\,{l}^{8}{\beta}^{5}{\rho_0}^{2}{\pi }^{2}+247726080 \,{l}^{6}{\beta}^{5}\rho_0\,\pi \nonumber\\
&& +330301440\,{l}^{6}{\beta}^{6}{\rho_0}^{2}{\pi }^{2}+1189085184\,{l}^{4}{\beta}^{6}\rho_0\, \pi +3170893824\,{l}^{2}{\beta}^{7}\rho_0\,\pi +1728\,{l}^{14} \beta\,\rho_0\,\pi +11520\,{l}^{14}{\beta}^{2}{\rho_0}^{2}{ \pi }^{2} \nonumber\\
&&+24576\,{l}^{14}{\beta}^{3}{\rho_0}^{3}{\pi }^{3}+8192\, {l}^{14}{\beta}^{4}{\rho_0}^{4}{\pi }^{4}+452984832\,\Upsilon_3 {\beta}^{7}+27\,{l}^{14}\Upsilon_3+1728\,{l}^{12}\beta\,\Upsilon_3+48384\,{l}^{10}{\beta}^{2}\Upsilon_3+774144\,{l}^{8}{\beta}^{3}
\Upsilon_3 \nonumber\\
&&+49545216\,{l}^{4}{\beta}^{5}\Upsilon_3 +7741440\,{l}^{6}{\beta}^{4}\Upsilon_3 +198180864\,{l}^{2}{\beta}^{6}\Upsilon_3 \Bigg)\Bigg\}\,,\hspace*{7cm} {\color{blue} ({\mathbf F}_1)}
\end{eqnarray*}
$\Upsilon_3=\sqrt {9+96\,\beta\,\rho_0\,\pi }$.  The form of the derivative of $M(\rho_0)$ w.r.t. the central density takes the form
\newpage
\begin{eqnarray*}
&&\frac{\partial M(\rho_0)}{\partial \rho_0}=\frac {{l}^{3} }{461689330549653504{\beta}^{14}\Upsilon_3} \Bigg\{ 50497270528868352\,{\alpha}^{12}{l}^{4}+215205838848\,{l}^{18}{\alpha}^{5}\Upsilon_3+509483520\,{l}^{22}{\beta}^{3}\Upsilon_3\nonumber\\
&&+262440\,{l}^{26}\beta\,\Upsilon_3 +12227604480\,{l}^{20}{\beta}^{4}\Upsilon_3+6673792914948096\,{\beta}^{10}{l}^{8}\Upsilon_3+22355562473717760\,{\beta}^{11}{l}^{6}\Upsilon\nonumber\\
&&+1442426916962304\,{\beta}^{9}{l}^{10}\Upsilon_3+2869411184640\,{l}^{16}{\beta}^{6} \Upsilon_3+67329694038491136\,{\beta}^{13 }{l}^{2}\Upsilon_3+14696640\,{l}^{24}{ \beta}^{2}\Upsilon_3\nonumber\\
&&+29495597727744\,{ \beta}^{7}{l}^{14}\Upsilon_3+ 235047487537152\,{\beta}^{8}{l}^{12}\Upsilon_3+88486793183232\,{\beta}^{7}{l}^{14}+4327280750886912\,{\beta}^ {9}{l}^{10}\nonumber\\
&& +67066687421153280\,{\beta}^{11}{l}^{6}+151491811586605056 \,{\beta}^{12}{l}^{4}+705142462611456\,{\beta}^{8}{l}^{12}+ 201989082115473408\,{\beta}^{13}{l}^{2}\nonumber\\
&&+44089920\,{l}^{24}{\beta}^{2 }+36682813440\,{l}^{20}{\beta}^{4}+645617516544\,{l}^{18}{\beta}^{5} +787320\,{l}^{26}\beta+1528450560\,{l}^{22}{\beta}^{3}+8608233553920 \,{l}^{16}{\beta}^{6}\nonumber\\
&&+20021378744844288\,{\beta}^{10}{l}^{8}+ 76948221758275584\,{\beta}^{14}\Upsilon_3 +2187\,{l}^{28}\Upsilon_3+81618807029760\, {l}^{18}{\beta}^{8}\Upsilon_3{\rho_0} ^{3}{\pi }^{3}\nonumber\\
&&+8115506380800\,{l}^{20}{\beta}^{7}\Upsilon_3 {\rho_0}^{3}{\pi }^{3}+1672151040\,{l}^{26}{ \beta}^{5}\Upsilon_3{\rho_0}^{4}{\pi }^{4}+901722931200\,{l}^{22}{\beta}^{7}\Upsilon_3 {\rho_0}^{4}{\pi }^{4}+23224320\,{l}^{28}{\beta}^{4}{\rho_0}^{4}{\pi }^{4}\Upsilon_3\nonumber\\
&&+ 1981808640\,{l}^{26}{\beta}^{6}{\rho_0}^{5}{\pi }^{5}\Upsilon_3 +52022476800\,{l}^{24}{\beta}^{6}{\rho_0}^{4}{\pi }^{4}\Upsilon_3+37158912\,{ l}^{28}{\beta}^{5}{\rho_0}^{5}{\pi }^{5}\Upsilon_3 +660602880\,{l}^{26}{\beta}^{7}{\rho_0}^{6}{\pi } ^{6}\Upsilon_3\nonumber\\
&&+41617981440\,{l}^{24}{ \beta}^{7}{\rho_0}^{5}{\pi }^{5}\Upsilon_3 +412216197120\,{l}^{22}{\beta}^{8}{\rho_0}^{5}{\pi }^{5} \Upsilon_3+9274864435200\,{l}^{20}{\beta} ^{8}{\rho_0}^{4}{\pi }^{4}\Upsilon_3+ 54412538019840\,{l}^{18}{\beta}^{9}{\rho_0}^{4}{\pi }^{4}\Upsilon_3\nonumber\\
&& +22020096\,{l}^{28}{\beta}^{6}{\rho_0}^{6}{\pi }^{6}\Upsilon_3+5284823040 \,{l}^{24}{\beta}^{8}{\rho_0}^{6}{\pi }^{6}\Upsilon_3 +1648864788480\,{l}^{20}{\beta}^{9}{\rho_0}^{5}{ \pi }^{5}\Upsilon_3+2097152\,{l}^{28} \Upsilon_3{\beta}^{7}{\rho_0}^{7}{ \pi }^{7}\nonumber\\
&&+53562342113280\,{l}^{16}{\beta}^{7}\Upsilon_3 \rho_0\,\pi +306070526361600\,{l}^{16}{\beta}^{8} \Upsilon_3{\rho_0}^{2}{\pi }^{2}+ 544125380198400\,{l}^{16}{\beta}^{9}\Upsilon_3 {\rho_0}^{3}{\pi }^{3}\nonumber\\
&&+145100101386240\,{l}^{16}{\beta}^{10} \Upsilon_3{\rho_0}^{4}{\pi }^{4}+ 471978486005760\,{\beta}^{8}{l}^{14}\Upsilon_3 \rho_0\,\pi +2097986565832704\,{\beta}^{9}{l}^{14}\Upsilon_3 {\rho_0}^{2}{\pi }^{2}\nonumber\\
&&+ 2238339386179584\,{\beta}^{10}{l}^{14}\Upsilon_3{\rho_0}^{3}{\pi }^{3}+3136706152759296\,{\beta}^{9}{l}^{ 12}\Upsilon_3\rho_0\,\pi + 10050610019696640\,{\beta}^{10}{l}^{12}\Upsilon_3 {\rho_0}^{2}{\pi }^{2}\nonumber\\
&&+4476678772359168\,{\beta}^{11}{l}^ {12}\Upsilon_3{\rho_0}^{3}{\pi }^{3}+ 15451638768599040\,{\beta}^{10}{l}^{10}\Upsilon_3 \rho_0\,\pi +31122449718312960\,{\beta}^{11}{l}^{10} \Upsilon_3{\rho_0}^{2}{\pi }^{2}\nonumber\\
&&+ 54267032044044288\,{\beta}^{11}{l}^{8}\Upsilon_3\rho_0\,\pi +49795919549300736\,{\beta}^{12}{l}^{8}\Upsilon_3 {\rho_0}^{2}{\pi }^{2}+ 126243176322170880\,{\beta}^{12}{l}^{6}\Upsilon_3 \rho_0\,\pi\nonumber\\
&& +168324235096227840\,{\beta}^{13}{l}^{4} \Upsilon_3\rho_0\,\pi +13586227200\,{l }^{22}{\beta}^{4}\Upsilon_3\rho_0\, \pi +8398080\,{l}^{26}{\beta}^{2}\Upsilon_3 \rho_0\,\pi +102643200\,{l}^{26}{\beta}^{3}\Upsilon_3 {\rho_0}^{2}{\pi }^{2}\nonumber\\
&&+597196800\,{l}^{26 }{\beta}^{4}\Upsilon_3{\rho_0}^{3}{ \pi }^{3}+431101440\,{l}^{24}{\beta}^{3}\Upsilon_3\rho_0\,\pi +4702924800\,{l}^{24}{\beta}^{4}\Upsilon_3 {\rho_0}^{2}{\pi }^{2}+130427781120\, {l}^{22}{\beta}^{5}\Upsilon_3{\rho_0} ^{2}{\pi }^{2}\nonumber\\
&&+293462507520\,{l}^{20}{\beta}^{5}\Upsilon_3 \rho_0\,\pi +4591057895424\,{l}^{18}{\beta}^{6} \Upsilon_3\rho_0\,\pi +75816\,{l}^{28} \Upsilon_3\beta\,\rho_0\,\pi +1026432 \,{l}^{28}\Upsilon_3{\beta}^{2}{\rho_0}^{2}{\pi }^{2}\nonumber\\
&&+6842880\,{l}^{28}\Upsilon_3 {\beta}^{3}{\rho_0}^{3}{\pi }^{3}+541033758720\,{l}^{22}{ \beta}^{6}\Upsilon_3{\rho_0}^{3}{\pi }^{3}+2434651914240\,{l}^{20}{\beta}^{6}\Upsilon_3{\rho_0}^{2}{\pi }^{2}+32137405267968\,{l}^{18}{\beta }^{7}\Upsilon_3{\rho_0}^{2}{\pi }^{2}\nonumber\\
&&+ 23410114560\,{l}^{24}{\beta}^{5}\Upsilon_3{\rho_0}^{3}{\pi }^{3}+6561\,{l}^{28}+1887865021661184\,{\beta}^ {8}{l}^{14}\rho_0\,\pi +12587136637206528\,{\beta}^{9}{l}^{14}{\rho_0}^{2}{\pi }^{2}\nonumber\\
&&+26856941602996224\,{\beta}^{10}{l}^{14}{\rho_0}^{3}{\pi }^{3}+8954749114122240\,{\beta}^{11}{l}^{14}{\rho_0}^{4}{\pi }^{4}+13170878258872320\,{\beta}^{9}{l}^{12}\rho_0 \,\pi\nonumber\\
&& +70310435701653504\,{\beta}^{10}{l}^{12}{\rho_0}^{2}{\pi }^ {2}+93893362389614592\,{\beta}^{11}{l}^{12}{\rho_0}^{3}{\pi }^{3} +69433746977193984\,{\beta}^{10}{l}^{10}\rho_0\,\pi \nonumber\\
&&+ 279050020995465216\,{\beta}^{11}{l}^{10}{\rho_0}^{2}{\pi }^{2}+ 166921533137092608\,{\beta}^{12}{l}^{10}{\rho_0}^{3}{\pi }^{3}+ 269581782771302400\,{\beta}^{11}{l}^{8}\rho_0\,\pi\nonumber\\
&& + 732911773648158720\,{\beta}^{12}{l}^{8}{\rho_0}^{2}{\pi }^{2}+ 736418528545996800\,{\beta}^{12}{l}^{6}\rho_0\,\pi + 1066053488942776320\,{\beta}^{13}{l}^{6}{\rho_0}^{2}{\pi }^{2}\nonumber\\
&&+ 1312929033750577152\,{\beta}^{13}{l}^{4}\rho_0\,\pi + 1615912656923787264\,{\beta}^{14}{l}^{2}\rho_0\,\pi +262440\,{l}^ {28}\beta\,\rho_0\,\pi +4199040\,{l}^{28}{\beta}^{2}{\rho_0} ^{2}{\pi }^{2}\nonumber\\
&&+34214400\,{l}^{28}{\beta}^{3}{\rho_0}^{3}{\pi }^{3 }+149299200\,{l}^{28}{\beta}^{4}{\rho_0}^{4}{\pi }^{4}+29393280\, {l}^{26}{\beta}^{2}\rho_0\,\pi +431101440\,{l}^{26}{\beta}^{3}{\rho_0}^{2}{\pi }^{2}+3135283200\,{l}^{26}{\beta}^{4}{\rho_0}^ {3}{\pi }^{3}\nonumber\\
&&+1528450560\,{l}^{24}{\beta}^{3}\rho_0\,\pi + 20379340800\,{l}^{24}{\beta}^{4}{\rho_0}^{2}{\pi }^{2}+ 130427781120\,{l}^{24}{\beta}^{5}{\rho_0}^{3}{\pi }^{3}+ 48910417920\,{l}^{22}{\beta}^{4}\rho_0\,\pi\nonumber\\
&& +586925015040\,{l}^{ 22}{\beta}^{5}{\rho_0}^{2}{\pi }^{2}+1076029194240\,{l}^{20}{ \beta}^{5}\rho_0\,\pi +11477644738560\,{l}^{20}{\beta}^{6}{\rho_0}^{2}{\pi }^{2}+17216467107840\,{l}^{18}{\beta}^{6}\rho_0\, \pi\nonumber\\
&& +206597605294080\,{l}^{16}{\beta}^{7}\rho_0\,\pi + 405775319040\,{l}^{24}{\beta}^{6}{\rho_0}^{4}{\pi }^{4}+ 3246202552320\,{l}^{22}{\beta}^{6}{\rho_0}^{3}{\pi }^{3}+ 53562342113280\,{l}^{20}{\beta}^{7}{\rho_0}^{3}{\pi }^{3}\nonumber\\
&&+ 160687026339840\,{l}^{18}{\beta}^{7}{\rho_0}^{2}{\pi }^{2}+ 1652780842352640\,{l}^{16}{\beta}^{8}{\rho_0}^{2}{\pi }^{2}+ 102023508787200\,{l}^{20}{\beta}^{8}{\rho_0}^{4}{\pi }^{4}\nonumber\\
&&+ 612141052723200\,{l}^{18}{\beta}^{8}{\rho_0}^{3}{\pi }^{3}+ 4897128421785600\,{l}^{16}{\beta}^{9}{\rho_0}^{3}{\pi }^{3}+ 816188070297600\,{l}^{18}{\beta}^{9}{\rho_0}^{4}{\pi }^{4}+ 11705057280\,{l}^{26}{\beta}^{5}{\rho_0}^{4}{\pi }^{4}\nonumber\\
&&+ 8115506380800\,{l}^{22}{\beta}^{7}{\rho_0}^{4}{\pi }^{4}+ 3917702737428480\,{l}^{16}{\beta}^{10}{\rho_0}^{4}{\pi }^{4}+ 334430208\,{l}^{28}{\beta}^{5}{\rho_0}^{5}{\pi }^{5}+330301440\,{ l}^{28}{\beta}^{6}{\rho_0}^{6}{\pi }^{6}\nonumber\\
&&+20808990720\,{l}^{26}{ \beta}^{6}{\rho_0}^{5}{\pi }^{5}+541033758720\,{l}^{24}{\beta}^{ 7}{\rho_0}^{5}{\pi }^{5}+94371840\,{l}^{28}{\beta}^{7}{\rho_0 }^{7}{\pi }^{7}+13872660480\,{l}^{26}{\beta}^{7}{\rho_0}^{6}{\pi }^{6}\nonumber\\
&&+206108098560\,{l}^{24}{\beta}^{8}{\rho_0}^{6}{\pi }^{6}+ 7419891548160\,{l}^{22}{\beta}^{8}{\rho_0}^{5}{\pi }^{5}+ 54412538019840\,{l}^{20}{\beta}^{9}{\rho_0}^{5}{\pi }^{5}+ 1509949440\,{l}^{26}{\beta}^{8}{\rho_0}^{7}{\pi }^{7}\nonumber\\
&&+ 1099243192320\,{l}^{22}{\beta}^{9}{\rho_0}^{6}{\pi }^{6}+ 174120121663488\,{l}^{18}{\beta}^{10}{\rho_0}^{5}{\pi }^{5} \Bigg\}\,. \hspace*{7cm} {\color{blue} ({\mathbf F}_2)}
\end{eqnarray*}

\begin{thebibliography}{129}%
\makeatletter
\providecommand \@ifxundefined [1]{%
 \@ifx{#1\undefined}
}%
\providecommand \@ifnum [1]{%
 \ifnum #1\expandafter \@firstoftwo
 \else \expandafter \@secondoftwo
 \fi
}%
\providecommand \@ifx [1]{%
 \ifx #1\expandafter \@firstoftwo
 \else \expandafter \@secondoftwo
 \fi
}%
\providecommand \natexlab [1]{#1}%
\providecommand \enquote  [1]{``#1''}%
\providecommand \bibnamefont  [1]{#1}%
\providecommand \bibfnamefont [1]{#1}%
\providecommand \citenamefont [1]{#1}%
\providecommand \href@noop [0]{\@secondoftwo}%
\providecommand \href [0]{\begingroup \@sanitize@url \@href}%
\providecommand \@href[1]{\@@startlink{#1}\@@href}%
\providecommand \@@href[1]{\endgroup#1\@@endlink}%
\providecommand \@sanitize@url [0]{\catcode `\\12\catcode `\$12\catcode
  `\&12\catcode `\#12\catcode `\^12\catcode `\_12\catcode `\%12\relax}%
\providecommand \@@startlink[1]{}%
\providecommand \@@endlink[0]{}%
\providecommand \url  [0]{\begingroup\@sanitize@url \@url }%
\providecommand \@url [1]{\endgroup\@href {#1}{\urlprefix }}%
\providecommand \urlprefix  [0]{URL }%
\providecommand \Eprint [0]{\href }%
\providecommand \doibase [0]{http://dx.doi.org/}%
\providecommand \selectlanguage [0]{\@gobble}%
\providecommand \bibinfo  [0]{\@secondoftwo}%
\providecommand \bibfield  [0]{\@secondoftwo}%
\providecommand \translation [1]{[#1]}%
\providecommand \BibitemOpen [0]{}%
\providecommand \bibitemStop [0]{}%
\providecommand \bibitemNoStop [0]{.\EOS\space}%
\providecommand \EOS [0]{\spacefactor3000\relax}%
\providecommand \BibitemShut  [1]{\csname bibitem#1\endcsname}%
\let\auto@bib@innerbib\@empty
\bibitem [{\citenamefont {Banerjee}\ \emph
  {et~al.}(2021{\natexlab{a}})\citenamefont {Banerjee}, \citenamefont
  {Tangphati},\ and\ \citenamefont {Channuie}}]{Banerjee:2020yhu}%
  \BibitemOpen
  \bibfield  {author} {\bibinfo {author} {\bibfnamefont {A.}~\bibnamefont
  {Banerjee}}, \bibinfo {author} {\bibfnamefont {T.}~\bibnamefont {Tangphati}},
  \ and\ \bibinfo {author} {\bibfnamefont {P.}~\bibnamefont {Channuie}},\
  }\href {\doibase 10.3847/1538-4357/abd094} {\bibfield  {journal} {\bibinfo
  {journal} {Astrophys. J.}\ }\textbf {\bibinfo {volume} {909}},\ \bibinfo
  {pages} {14} (\bibinfo {year} {2021}{\natexlab{a}})},\ \Eprint
  {http://arxiv.org/abs/2006.00479} {arXiv:2006.00479 [gr-qc]} \BibitemShut
  {NoStop}%
\bibitem [{\citenamefont {Hennigar}\ \emph {et~al.}(2020)\citenamefont
  {Hennigar}, \citenamefont {Kubiz\v{n}\'ak}, \citenamefont {Mann},\ and\
  \citenamefont {Pollack}}]{Hennigar:2020lsl}%
  \BibitemOpen
  \bibfield  {author} {\bibinfo {author} {\bibfnamefont {R.~A.}\ \bibnamefont
  {Hennigar}}, \bibinfo {author} {\bibfnamefont {D.}~\bibnamefont
  {Kubiz\v{n}\'ak}}, \bibinfo {author} {\bibfnamefont {R.~B.}\ \bibnamefont
  {Mann}}, \ and\ \bibinfo {author} {\bibfnamefont {C.}~\bibnamefont
  {Pollack}},\ }\href {\doibase 10.1007/JHEP07(2020)027} {\bibfield  {journal}
  {\bibinfo  {journal} {JHEP}\ }\textbf {\bibinfo {volume} {07}},\ \bibinfo
  {pages} {027} (\bibinfo {year} {2020})},\ \Eprint
  {http://arxiv.org/abs/2004.09472} {arXiv:2004.09472 [gr-qc]} \BibitemShut
  {NoStop}%
\bibitem [{\citenamefont {Casalino}\ \emph {et~al.}(2021)\citenamefont
  {Casalino}, \citenamefont {Colleaux}, \citenamefont {Rinaldi},\ and\
  \citenamefont {Vicentini}}]{Casalino:2020kbt}%
  \BibitemOpen
  \bibfield  {author} {\bibinfo {author} {\bibfnamefont {A.}~\bibnamefont
  {Casalino}}, \bibinfo {author} {\bibfnamefont {A.}~\bibnamefont {Colleaux}},
  \bibinfo {author} {\bibfnamefont {M.}~\bibnamefont {Rinaldi}}, \ and\
  \bibinfo {author} {\bibfnamefont {S.}~\bibnamefont {Vicentini}},\ }\href
  {\doibase 10.1016/j.dark.2020.100770} {\bibfield  {journal} {\bibinfo
  {journal} {Phys. Dark Univ.}\ }\textbf {\bibinfo {volume} {31}},\ \bibinfo
  {pages} {100770} (\bibinfo {year} {2021})},\ \Eprint
  {http://arxiv.org/abs/2003.07068} {arXiv:2003.07068 [gr-qc]} \BibitemShut
  {NoStop}%
\bibitem [{\citenamefont {Aoki}\ \emph {et~al.}(2020)\citenamefont {Aoki},
  \citenamefont {Gorji},\ and\ \citenamefont {Mukohyama}}]{Aoki:2020lig}%
  \BibitemOpen
  \bibfield  {author} {\bibinfo {author} {\bibfnamefont {K.}~\bibnamefont
  {Aoki}}, \bibinfo {author} {\bibfnamefont {M.~A.}\ \bibnamefont {Gorji}}, \
  and\ \bibinfo {author} {\bibfnamefont {S.}~\bibnamefont {Mukohyama}},\ }\href
  {\doibase 10.1016/j.physletb.2020.135843} {\bibfield  {journal} {\bibinfo
  {journal} {Phys. Lett. B}\ }\textbf {\bibinfo {volume} {810}},\ \bibinfo
  {pages} {135843} (\bibinfo {year} {2020})},\ \Eprint
  {http://arxiv.org/abs/2005.03859} {arXiv:2005.03859 [gr-qc]} \BibitemShut
  {NoStop}%
\bibitem [{\citenamefont {Aasi}\ \emph {et~al.}(2015)\citenamefont {Aasi} \emph
  {et~al.}}]{LIGOScientific:2014pky}%
  \BibitemOpen
  \bibfield  {author} {\bibinfo {author} {\bibfnamefont {J.}~\bibnamefont
  {Aasi}} \emph {et~al.} (\bibinfo {collaboration} {LIGO Scientific}),\ }\href
  {\doibase 10.1088/0264-9381/32/7/074001} {\bibfield  {journal} {\bibinfo
  {journal} {Class. Quant. Grav.}\ }\textbf {\bibinfo {volume} {32}},\ \bibinfo
  {pages} {074001} (\bibinfo {year} {2015})},\ \Eprint
  {http://arxiv.org/abs/1411.4547} {arXiv:1411.4547 [gr-qc]} \BibitemShut
  {NoStop}%
\bibitem [{\citenamefont {Somiya}(2012)}]{Somiya:2011np}%
  \BibitemOpen
  \bibfield  {author} {\bibinfo {author} {\bibfnamefont {K.}~\bibnamefont
  {Somiya}} (\bibinfo {collaboration} {KAGRA}),\ }\href {\doibase
  10.1088/0264-9381/29/12/124007} {\bibfield  {journal} {\bibinfo  {journal}
  {Class. Quant. Grav.}\ }\textbf {\bibinfo {volume} {29}},\ \bibinfo {pages}
  {124007} (\bibinfo {year} {2012})},\ \Eprint {http://arxiv.org/abs/1111.7185}
  {arXiv:1111.7185 [gr-qc]} \BibitemShut {NoStop}%
\bibitem [{\citenamefont {Aso}\ \emph {et~al.}(2013)\citenamefont {Aso},
  \citenamefont {Michimura}, \citenamefont {Somiya}, \citenamefont {Ando},
  \citenamefont {Miyakawa}, \citenamefont {Sekiguchi}, \citenamefont
  {Tatsumi},\ and\ \citenamefont {Yamamoto}}]{Aso:2013eba}%
  \BibitemOpen
  \bibfield  {author} {\bibinfo {author} {\bibfnamefont {Y.}~\bibnamefont
  {Aso}}, \bibinfo {author} {\bibfnamefont {Y.}~\bibnamefont {Michimura}},
  \bibinfo {author} {\bibfnamefont {K.}~\bibnamefont {Somiya}}, \bibinfo
  {author} {\bibfnamefont {M.}~\bibnamefont {Ando}}, \bibinfo {author}
  {\bibfnamefont {O.}~\bibnamefont {Miyakawa}}, \bibinfo {author}
  {\bibfnamefont {T.}~\bibnamefont {Sekiguchi}}, \bibinfo {author}
  {\bibfnamefont {D.}~\bibnamefont {Tatsumi}}, \ and\ \bibinfo {author}
  {\bibfnamefont {H.}~\bibnamefont {Yamamoto}} (\bibinfo {collaboration}
  {KAGRA}),\ }\href {\doibase 10.1103/PhysRevD.88.043007} {\bibfield  {journal}
  {\bibinfo  {journal} {Phys. Rev. D}\ }\textbf {\bibinfo {volume} {88}},\
  \bibinfo {pages} {043007} (\bibinfo {year} {2013})},\ \Eprint
  {http://arxiv.org/abs/1306.6747} {arXiv:1306.6747 [gr-qc]} \BibitemShut
  {NoStop}%
\bibitem [{\citenamefont {Acernese}\ \emph {et~al.}(2015)\citenamefont
  {Acernese} \emph {et~al.}}]{VIRGO:2014yos}%
  \BibitemOpen
  \bibfield  {author} {\bibinfo {author} {\bibfnamefont {F.}~\bibnamefont
  {Acernese}} \emph {et~al.} (\bibinfo {collaboration} {VIRGO}),\ }\href
  {\doibase 10.1088/0264-9381/32/2/024001} {\bibfield  {journal} {\bibinfo
  {journal} {Class. Quant. Grav.}\ }\textbf {\bibinfo {volume} {32}},\ \bibinfo
  {pages} {024001} (\bibinfo {year} {2015})},\ \Eprint
  {http://arxiv.org/abs/1408.3978} {arXiv:1408.3978 [gr-qc]} \BibitemShut
  {NoStop}%
\bibitem [{\citenamefont {Psaltis}\ \emph {et~al.}(2020)\citenamefont {Psaltis}
  \emph {et~al.}}]{EventHorizonTelescope:2020qrl}%
  \BibitemOpen
  \bibfield  {author} {\bibinfo {author} {\bibfnamefont {D.}~\bibnamefont
  {Psaltis}} \emph {et~al.} (\bibinfo {collaboration} {Event Horizon
  Telescope}),\ }\href {\doibase 10.1103/PhysRevLett.125.141104} {\bibfield
  {journal} {\bibinfo  {journal} {Phys. Rev. Lett.}\ }\textbf {\bibinfo
  {volume} {125}},\ \bibinfo {pages} {141104} (\bibinfo {year} {2020})},\
  \Eprint {http://arxiv.org/abs/2010.01055} {arXiv:2010.01055 [gr-qc]}
  \BibitemShut {NoStop}%
\bibitem [{\citenamefont {Raaijmakers}\ \emph {et~al.}(2020)\citenamefont
  {Raaijmakers} \emph {et~al.}}]{Raaijmakers:2019dks}%
  \BibitemOpen
  \bibfield  {author} {\bibinfo {author} {\bibfnamefont {G.}~\bibnamefont
  {Raaijmakers}} \emph {et~al.},\ }\href {\doibase 10.3847/2041-8213/ab822f}
  {\bibfield  {journal} {\bibinfo  {journal} {Astrophys. J. Lett.}\ }\textbf
  {\bibinfo {volume} {893}},\ \bibinfo {pages} {L21} (\bibinfo {year}
  {2020})},\ \Eprint {http://arxiv.org/abs/1912.11031} {arXiv:1912.11031
  [astro-ph.HE]} \BibitemShut {NoStop}%
\bibitem [{\citenamefont {Astashenok}\ \emph {et~al.}(2013)\citenamefont
  {Astashenok}, \citenamefont {Capozziello},\ and\ \citenamefont
  {Odintsov}}]{Astashenok:2013vza}%
  \BibitemOpen
  \bibfield  {author} {\bibinfo {author} {\bibfnamefont {A.~V.}\ \bibnamefont
  {Astashenok}}, \bibinfo {author} {\bibfnamefont {S.}~\bibnamefont
  {Capozziello}}, \ and\ \bibinfo {author} {\bibfnamefont {S.~D.}\ \bibnamefont
  {Odintsov}},\ }\href {\doibase 10.1088/1475-7516/2013/12/040} {\bibfield
  {journal} {\bibinfo  {journal} {JCAP}\ }\textbf {\bibinfo {volume} {12}},\
  \bibinfo {pages} {040} (\bibinfo {year} {2013})},\ \Eprint
  {http://arxiv.org/abs/1309.1978} {arXiv:1309.1978 [gr-qc]} \BibitemShut
  {NoStop}%
\bibitem [{\citenamefont {Astashenok}\ \emph {et~al.}(2014)\citenamefont
  {Astashenok}, \citenamefont {Capozziello},\ and\ \citenamefont
  {Odintsov}}]{Astashenok:2014pua}%
  \BibitemOpen
  \bibfield  {author} {\bibinfo {author} {\bibfnamefont {A.~V.}\ \bibnamefont
  {Astashenok}}, \bibinfo {author} {\bibfnamefont {S.}~\bibnamefont
  {Capozziello}}, \ and\ \bibinfo {author} {\bibfnamefont {S.~D.}\ \bibnamefont
  {Odintsov}},\ }\href {\doibase 10.1103/PhysRevD.89.103509} {\bibfield
  {journal} {\bibinfo  {journal} {Phys. Rev. D}\ }\textbf {\bibinfo {volume}
  {89}},\ \bibinfo {pages} {103509} (\bibinfo {year} {2014})},\ \Eprint
  {http://arxiv.org/abs/1401.4546} {arXiv:1401.4546 [gr-qc]} \BibitemShut
  {NoStop}%
\bibitem [{\citenamefont {Abbott}\ \emph
  {et~al.}(2020{\natexlab{a}})\citenamefont {Abbott} \emph
  {et~al.}}]{LIGOScientific:2020zkf}%
  \BibitemOpen
  \bibfield  {author} {\bibinfo {author} {\bibfnamefont {R.}~\bibnamefont
  {Abbott}} \emph {et~al.} (\bibinfo {collaboration} {LIGO Scientific,
  Virgo}),\ }\href {\doibase 10.3847/2041-8213/ab960f} {\bibfield  {journal}
  {\bibinfo  {journal} {Astrophys. J. Lett.}\ }\textbf {\bibinfo {volume}
  {896}},\ \bibinfo {pages} {L44} (\bibinfo {year} {2020}{\natexlab{a}})},\
  \Eprint {http://arxiv.org/abs/2006.12611} {arXiv:2006.12611 [astro-ph.HE]}
  \BibitemShut {NoStop}%
\bibitem [{\citenamefont {Huang}\ \emph {et~al.}(2020)\citenamefont {Huang},
  \citenamefont {Hu}, \citenamefont {Zhang},\ and\ \citenamefont
  {Shen}}]{Huang:2020cab}%
  \BibitemOpen
  \bibfield  {author} {\bibinfo {author} {\bibfnamefont {K.}~\bibnamefont
  {Huang}}, \bibinfo {author} {\bibfnamefont {J.}~\bibnamefont {Hu}}, \bibinfo
  {author} {\bibfnamefont {Y.}~\bibnamefont {Zhang}}, \ and\ \bibinfo {author}
  {\bibfnamefont {H.}~\bibnamefont {Shen}},\ }\href {\doibase
  10.3847/1538-4357/abbb37} {\bibfield  {journal} {\bibinfo  {journal}
  {Astrophys. J.}\ }\textbf {\bibinfo {volume} {904}},\ \bibinfo {pages} {39}
  (\bibinfo {year} {2020})},\ \Eprint {http://arxiv.org/abs/2008.04491}
  {arXiv:2008.04491 [nucl-th]} \BibitemShut {NoStop}%
\bibitem [{\citenamefont {Bombaci}\ \emph {et~al.}(2021)\citenamefont
  {Bombaci}, \citenamefont {Drago}, \citenamefont {Logoteta}, \citenamefont
  {Pagliara},\ and\ \citenamefont {Vida\~na}}]{Bombaci:2020vgw}%
  \BibitemOpen
  \bibfield  {author} {\bibinfo {author} {\bibfnamefont {I.}~\bibnamefont
  {Bombaci}}, \bibinfo {author} {\bibfnamefont {A.}~\bibnamefont {Drago}},
  \bibinfo {author} {\bibfnamefont {D.}~\bibnamefont {Logoteta}}, \bibinfo
  {author} {\bibfnamefont {G.}~\bibnamefont {Pagliara}}, \ and\ \bibinfo
  {author} {\bibfnamefont {I.}~\bibnamefont {Vida\~na}},\ }\href {\doibase
  10.1103/PhysRevLett.126.162702} {\bibfield  {journal} {\bibinfo  {journal}
  {Phys. Rev. Lett.}\ }\textbf {\bibinfo {volume} {126}},\ \bibinfo {pages}
  {162702} (\bibinfo {year} {2021})},\ \Eprint
  {http://arxiv.org/abs/2010.01509} {arXiv:2010.01509 [nucl-th]} \BibitemShut
  {NoStop}%
\bibitem [{\citenamefont {Roupas}\ \emph {et~al.}(2021)\citenamefont {Roupas},
  \citenamefont {Panotopoulos},\ and\ \citenamefont {Lopes}}]{Roupas:2020nua}%
  \BibitemOpen
  \bibfield  {author} {\bibinfo {author} {\bibfnamefont {Z.}~\bibnamefont
  {Roupas}}, \bibinfo {author} {\bibfnamefont {G.}~\bibnamefont
  {Panotopoulos}}, \ and\ \bibinfo {author} {\bibfnamefont {I.}~\bibnamefont
  {Lopes}},\ }\href {\doibase 10.1103/PhysRevD.103.083015} {\bibfield
  {journal} {\bibinfo  {journal} {Phys. Rev. D}\ }\textbf {\bibinfo {volume}
  {103}},\ \bibinfo {pages} {083015} (\bibinfo {year} {2021})},\ \Eprint
  {http://arxiv.org/abs/2010.11020} {arXiv:2010.11020 [astro-ph.HE]}
  \BibitemShut {NoStop}%
\bibitem [{\citenamefont {Zhou}\ \emph {et~al.}(2021)\citenamefont {Zhou},
  \citenamefont {Li},\ and\ \citenamefont {Li}}]{Zhou:2020xan}%
  \BibitemOpen
  \bibfield  {author} {\bibinfo {author} {\bibfnamefont {X.}~\bibnamefont
  {Zhou}}, \bibinfo {author} {\bibfnamefont {A.}~\bibnamefont {Li}}, \ and\
  \bibinfo {author} {\bibfnamefont {B.-A.}\ \bibnamefont {Li}},\ }\href
  {\doibase 10.3847/1538-4357/abe538} {\bibfield  {journal} {\bibinfo
  {journal} {Astrophys. J.}\ }\textbf {\bibinfo {volume} {910}},\ \bibinfo
  {pages} {62} (\bibinfo {year} {2021})},\ \Eprint
  {http://arxiv.org/abs/2011.11934} {arXiv:2011.11934 [astro-ph.HE]}
  \BibitemShut {NoStop}%
\bibitem [{\citenamefont {Awad}\ and\ \citenamefont
  {Nashed}(2017)}]{Awad:2017sau}%
  \BibitemOpen
  \bibfield  {author} {\bibinfo {author} {\bibfnamefont {A.}~\bibnamefont
  {Awad}}\ and\ \bibinfo {author} {\bibfnamefont {G.}~\bibnamefont {Nashed}},\
  }\href {\doibase 10.1088/1475-7516/2017/02/046} {\bibfield  {journal}
  {\bibinfo  {journal} {JCAP}\ }\textbf {\bibinfo {volume} {1702}},\ \bibinfo
  {pages} {046} (\bibinfo {year} {2017})},\ \Eprint
  {http://arxiv.org/abs/1701.06899} {arXiv:1701.06899 [gr-qc]} \BibitemShut
  {NoStop}%
\bibitem [{\citenamefont {Most}\ \emph {et~al.}(2020)\citenamefont {Most},
  \citenamefont {Papenfort}, \citenamefont {Weih},\ and\ \citenamefont
  {Rezzolla}}]{Most:2020bba}%
  \BibitemOpen
  \bibfield  {author} {\bibinfo {author} {\bibfnamefont {E.~R.}\ \bibnamefont
  {Most}}, \bibinfo {author} {\bibfnamefont {L.~J.}\ \bibnamefont {Papenfort}},
  \bibinfo {author} {\bibfnamefont {L.~R.}\ \bibnamefont {Weih}}, \ and\
  \bibinfo {author} {\bibfnamefont {L.}~\bibnamefont {Rezzolla}},\ }\href
  {\doibase 10.1093/mnrasl/slaa168} {\bibfield  {journal} {\bibinfo  {journal}
  {Mon. Not. Roy. Astron. Soc.}\ }\textbf {\bibinfo {volume} {499}},\ \bibinfo
  {pages} {L82} (\bibinfo {year} {2020})},\ \Eprint
  {http://arxiv.org/abs/2006.14601} {arXiv:2006.14601 [astro-ph.HE]}
  \BibitemShut {NoStop}%
\bibitem [{\citenamefont {Nashed}(2014)}]{Nashed:2014sea}%
  \BibitemOpen
  \bibfield  {author} {\bibinfo {author} {\bibfnamefont {G.~G.~L.}\
  \bibnamefont {Nashed}},\ }\href {\doibase 10.1209/0295-5075/105/10001}
  {\bibfield  {journal} {\bibinfo  {journal} {EPL}\ }\textbf {\bibinfo {volume}
  {105}},\ \bibinfo {pages} {10001} (\bibinfo {year} {2014})},\ \Eprint
  {http://arxiv.org/abs/1501.00974} {arXiv:1501.00974 [gr-qc]} \BibitemShut
  {NoStop}%
\bibitem [{\citenamefont {Nashed}(2006)}]{Nashed:2005kn}%
  \BibitemOpen
  \bibfield  {author} {\bibinfo {author} {\bibfnamefont {G.~G.~L.}\
  \bibnamefont {Nashed}},\ }\href {\doibase 10.1142/S0217751X06031478}
  {\bibfield  {journal} {\bibinfo  {journal} {Int. J. Mod. Phys. A}\ }\textbf
  {\bibinfo {volume} {21}},\ \bibinfo {pages} {3181} (\bibinfo {year}
  {2006})},\ \Eprint {http://arxiv.org/abs/gr-qc/0501002} {arXiv:gr-qc/0501002}
  \BibitemShut {NoStop}%
\bibitem [{\citenamefont {Tan}\ \emph {et~al.}(2020)\citenamefont {Tan},
  \citenamefont {Noronha-Hostler},\ and\ \citenamefont {Yunes}}]{Tan:2020ics}%
  \BibitemOpen
  \bibfield  {author} {\bibinfo {author} {\bibfnamefont {H.}~\bibnamefont
  {Tan}}, \bibinfo {author} {\bibfnamefont {J.}~\bibnamefont
  {Noronha-Hostler}}, \ and\ \bibinfo {author} {\bibfnamefont {N.}~\bibnamefont
  {Yunes}},\ }\href {\doibase 10.1103/PhysRevLett.125.261104} {\bibfield
  {journal} {\bibinfo  {journal} {Phys. Rev. Lett.}\ }\textbf {\bibinfo
  {volume} {125}},\ \bibinfo {pages} {261104} (\bibinfo {year} {2020})},\
  \Eprint {http://arxiv.org/abs/2006.16296} {arXiv:2006.16296 [astro-ph.HE]}
  \BibitemShut {NoStop}%
\bibitem [{\citenamefont {Vattis}\ \emph {et~al.}(2020)\citenamefont {Vattis},
  \citenamefont {Goldstein},\ and\ \citenamefont
  {Koushiappas}}]{Vattis:2020iuz}%
  \BibitemOpen
  \bibfield  {author} {\bibinfo {author} {\bibfnamefont {K.}~\bibnamefont
  {Vattis}}, \bibinfo {author} {\bibfnamefont {I.~S.}\ \bibnamefont
  {Goldstein}}, \ and\ \bibinfo {author} {\bibfnamefont {S.~M.}\ \bibnamefont
  {Koushiappas}},\ }\href {\doibase 10.1103/PhysRevD.102.061301} {\bibfield
  {journal} {\bibinfo  {journal} {Phys. Rev. D}\ }\textbf {\bibinfo {volume}
  {102}},\ \bibinfo {pages} {061301} (\bibinfo {year} {2020})},\ \Eprint
  {http://arxiv.org/abs/2006.15675} {arXiv:2006.15675 [astro-ph.HE]}
  \BibitemShut {NoStop}%
\bibitem [{\citenamefont {Zhang}\ and\ \citenamefont
  {Li}(2020)}]{Zhang:2020zsc}%
  \BibitemOpen
  \bibfield  {author} {\bibinfo {author} {\bibfnamefont {N.-B.}\ \bibnamefont
  {Zhang}}\ and\ \bibinfo {author} {\bibfnamefont {B.-A.}\ \bibnamefont {Li}},\
  }\href {\doibase 10.3847/1538-4357/abb470} {\bibfield  {journal} {\bibinfo
  {journal} {Astrophys. J.}\ }\textbf {\bibinfo {volume} {902}},\ \bibinfo
  {pages} {38} (\bibinfo {year} {2020})},\ \Eprint
  {http://arxiv.org/abs/2007.02513} {arXiv:2007.02513 [astro-ph.HE]}
  \BibitemShut {NoStop}%
\bibitem [{\citenamefont {Fattoyev}\ \emph {et~al.}(2020)\citenamefont
  {Fattoyev}, \citenamefont {Horowitz}, \citenamefont {Piekarewicz},\ and\
  \citenamefont {Reed}}]{Fattoyev:2020cws}%
  \BibitemOpen
  \bibfield  {author} {\bibinfo {author} {\bibfnamefont {F.~J.}\ \bibnamefont
  {Fattoyev}}, \bibinfo {author} {\bibfnamefont {C.~J.}\ \bibnamefont
  {Horowitz}}, \bibinfo {author} {\bibfnamefont {J.}~\bibnamefont
  {Piekarewicz}}, \ and\ \bibinfo {author} {\bibfnamefont {B.}~\bibnamefont
  {Reed}},\ }\href {\doibase 10.1103/PhysRevC.102.065805} {\bibfield  {journal}
  {\bibinfo  {journal} {Phys. Rev. C}\ }\textbf {\bibinfo {volume} {102}},\
  \bibinfo {pages} {065805} (\bibinfo {year} {2020})},\ \Eprint
  {http://arxiv.org/abs/2007.03799} {arXiv:2007.03799 [nucl-th]} \BibitemShut
  {NoStop}%
\bibitem [{\citenamefont {Shirafuji}\ \emph {et~al.}(1996)\citenamefont
  {Shirafuji}, \citenamefont {Nashed},\ and\ \citenamefont
  {Kobayashi}}]{Shirafuji:1996im}%
  \BibitemOpen
  \bibfield  {author} {\bibinfo {author} {\bibfnamefont {T.}~\bibnamefont
  {Shirafuji}}, \bibinfo {author} {\bibfnamefont {G.~G.~L.}\ \bibnamefont
  {Nashed}}, \ and\ \bibinfo {author} {\bibfnamefont {Y.}~\bibnamefont
  {Kobayashi}},\ }\href {\doibase 10.1143/PTP.96.933} {\bibfield  {journal}
  {\bibinfo  {journal} {Prog. Theor. Phys.}\ }\textbf {\bibinfo {volume}
  {96}},\ \bibinfo {pages} {933} (\bibinfo {year} {1996})},\ \Eprint
  {http://arxiv.org/abs/gr-qc/9609060} {arXiv:gr-qc/9609060} \BibitemShut
  {NoStop}%
\bibitem [{\citenamefont {Tsokaros}\ \emph {et~al.}(2020)\citenamefont
  {Tsokaros}, \citenamefont {Ruiz},\ and\ \citenamefont
  {Shapiro}}]{Tsokaros:2020hli}%
  \BibitemOpen
  \bibfield  {author} {\bibinfo {author} {\bibfnamefont {A.}~\bibnamefont
  {Tsokaros}}, \bibinfo {author} {\bibfnamefont {M.}~\bibnamefont {Ruiz}}, \
  and\ \bibinfo {author} {\bibfnamefont {S.~L.}\ \bibnamefont {Shapiro}},\
  }\href {\doibase 10.3847/1538-4357/abc421} {\bibfield  {journal} {\bibinfo
  {journal} {Astrophys. J.}\ }\textbf {\bibinfo {volume} {905}},\ \bibinfo
  {pages} {48} (\bibinfo {year} {2020})},\ \Eprint
  {http://arxiv.org/abs/2007.05526} {arXiv:2007.05526 [astro-ph.HE]}
  \BibitemShut {NoStop}%
\bibitem [{\citenamefont {Tews}\ \emph {et~al.}(2021)\citenamefont {Tews},
  \citenamefont {Pang}, \citenamefont {Dietrich}, \citenamefont {Coughlin},
  \citenamefont {Antier}, \citenamefont {Bulla}, \citenamefont {Heinzel},\ and\
  \citenamefont {Issa}}]{Tews:2020ylw}%
  \BibitemOpen
  \bibfield  {author} {\bibinfo {author} {\bibfnamefont {I.}~\bibnamefont
  {Tews}}, \bibinfo {author} {\bibfnamefont {P.~T.~H.}\ \bibnamefont {Pang}},
  \bibinfo {author} {\bibfnamefont {T.}~\bibnamefont {Dietrich}}, \bibinfo
  {author} {\bibfnamefont {M.~W.}\ \bibnamefont {Coughlin}}, \bibinfo {author}
  {\bibfnamefont {S.}~\bibnamefont {Antier}}, \bibinfo {author} {\bibfnamefont
  {M.}~\bibnamefont {Bulla}}, \bibinfo {author} {\bibfnamefont
  {J.}~\bibnamefont {Heinzel}}, \ and\ \bibinfo {author} {\bibfnamefont
  {L.}~\bibnamefont {Issa}},\ }\href {\doibase 10.3847/2041-8213/abdaae}
  {\bibfield  {journal} {\bibinfo  {journal} {Astrophys. J. Lett.}\ }\textbf
  {\bibinfo {volume} {908}},\ \bibinfo {pages} {L1} (\bibinfo {year} {2021})},\
  \Eprint {http://arxiv.org/abs/2007.06057} {arXiv:2007.06057 [astro-ph.HE]}
  \BibitemShut {NoStop}%
\bibitem [{\citenamefont {Dexheimer}\ \emph {et~al.}(2021)\citenamefont
  {Dexheimer}, \citenamefont {Gomes}, \citenamefont {Kl\"ahn}, \citenamefont
  {Han},\ and\ \citenamefont {Salinas}}]{Dexheimer:2020rlp}%
  \BibitemOpen
  \bibfield  {author} {\bibinfo {author} {\bibfnamefont {V.}~\bibnamefont
  {Dexheimer}}, \bibinfo {author} {\bibfnamefont {R.~O.}\ \bibnamefont
  {Gomes}}, \bibinfo {author} {\bibfnamefont {T.}~\bibnamefont {Kl\"ahn}},
  \bibinfo {author} {\bibfnamefont {S.}~\bibnamefont {Han}}, \ and\ \bibinfo
  {author} {\bibfnamefont {M.}~\bibnamefont {Salinas}},\ }\href {\doibase
  10.1103/PhysRevC.103.025808} {\bibfield  {journal} {\bibinfo  {journal}
  {Phys. Rev. C}\ }\textbf {\bibinfo {volume} {103}},\ \bibinfo {pages}
  {025808} (\bibinfo {year} {2021})},\ \Eprint
  {http://arxiv.org/abs/2007.08493} {arXiv:2007.08493 [astro-ph.HE]}
  \BibitemShut {NoStop}%
\bibitem [{\citenamefont {Godzieba}\ \emph {et~al.}(2021)\citenamefont
  {Godzieba}, \citenamefont {Radice},\ and\ \citenamefont
  {Bernuzzi}}]{Godzieba:2020tjn}%
  \BibitemOpen
  \bibfield  {author} {\bibinfo {author} {\bibfnamefont {D.~A.}\ \bibnamefont
  {Godzieba}}, \bibinfo {author} {\bibfnamefont {D.}~\bibnamefont {Radice}}, \
  and\ \bibinfo {author} {\bibfnamefont {S.}~\bibnamefont {Bernuzzi}},\ }\href
  {\doibase 10.3847/1538-4357/abd4dd} {\bibfield  {journal} {\bibinfo
  {journal} {Astrophys. J.}\ }\textbf {\bibinfo {volume} {908}},\ \bibinfo
  {pages} {122} (\bibinfo {year} {2021})},\ \Eprint
  {http://arxiv.org/abs/2007.10999} {arXiv:2007.10999 [astro-ph.HE]}
  \BibitemShut {NoStop}%
\bibitem [{\citenamefont {Kanakis-Pegios}\ \emph {et~al.}(2021)\citenamefont
  {Kanakis-Pegios}, \citenamefont {Koliogiannis},\ and\ \citenamefont
  {Moustakidis}}]{Kanakis-Pegios:2020kzp}%
  \BibitemOpen
  \bibfield  {author} {\bibinfo {author} {\bibfnamefont {A.}~\bibnamefont
  {Kanakis-Pegios}}, \bibinfo {author} {\bibfnamefont {P.~S.}\ \bibnamefont
  {Koliogiannis}}, \ and\ \bibinfo {author} {\bibfnamefont {C.~C.}\
  \bibnamefont {Moustakidis}},\ }\href {\doibase 10.3390/sym13020183}
  {\bibfield  {journal} {\bibinfo  {journal} {Symmetry}\ }\textbf {\bibinfo
  {volume} {13}},\ \bibinfo {pages} {183} (\bibinfo {year} {2021})},\ \Eprint
  {http://arxiv.org/abs/2012.09580} {arXiv:2012.09580 [astro-ph.HE]}
  \BibitemShut {NoStop}%
\bibitem [{\citenamefont {Nathanail}\ \emph {et~al.}(2021)\citenamefont
  {Nathanail}, \citenamefont {Most},\ and\ \citenamefont
  {Rezzolla}}]{Nathanail:2021tay}%
  \BibitemOpen
  \bibfield  {author} {\bibinfo {author} {\bibfnamefont {A.}~\bibnamefont
  {Nathanail}}, \bibinfo {author} {\bibfnamefont {E.~R.}\ \bibnamefont {Most}},
  \ and\ \bibinfo {author} {\bibfnamefont {L.}~\bibnamefont {Rezzolla}},\
  }\href {\doibase 10.3847/2041-8213/abdfc6} {\bibfield  {journal} {\bibinfo
  {journal} {Astrophys. J. Lett.}\ }\textbf {\bibinfo {volume} {908}},\
  \bibinfo {pages} {L28} (\bibinfo {year} {2021})},\ \Eprint
  {http://arxiv.org/abs/2101.01735} {arXiv:2101.01735 [astro-ph.HE]}
  \BibitemShut {NoStop}%
\bibitem [{\citenamefont {Roupas}(2021)}]{Roupas:2020jyv}%
  \BibitemOpen
  \bibfield  {author} {\bibinfo {author} {\bibfnamefont {Z.}~\bibnamefont
  {Roupas}},\ }\href {\doibase 10.1007/s10509-021-03919-5} {\bibfield
  {journal} {\bibinfo  {journal} {Astrophys. Space Sci.}\ }\textbf {\bibinfo
  {volume} {366}},\ \bibinfo {pages} {9} (\bibinfo {year} {2021})},\ \Eprint
  {http://arxiv.org/abs/2007.10679} {arXiv:2007.10679 [gr-qc]} \BibitemShut
  {NoStop}%
\bibitem [{\citenamefont {{Biswas}}\ \emph {et~al.}(2021)\citenamefont
  {{Biswas}}, \citenamefont {{Nandi}}, \citenamefont {{Char}}, \citenamefont
  {{Bose}},\ and\ \citenamefont {{Stergioulas}}}]{2021MNRAS.505.1600B}%
  \BibitemOpen
  \bibfield  {author} {\bibinfo {author} {\bibfnamefont {B.}~\bibnamefont
  {{Biswas}}}, \bibinfo {author} {\bibfnamefont {R.}~\bibnamefont {{Nandi}}},
  \bibinfo {author} {\bibfnamefont {P.}~\bibnamefont {{Char}}}, \bibinfo
  {author} {\bibfnamefont {S.}~\bibnamefont {{Bose}}}, \ and\ \bibinfo {author}
  {\bibfnamefont {N.}~\bibnamefont {{Stergioulas}}},\ }\href {\doibase
  10.1093/mnras/stab1383} {\bibfield  {journal} {\bibinfo  {journal} {mnras}\
  }\textbf {\bibinfo {volume} {505}},\ \bibinfo {pages} {1600} (\bibinfo {year}
  {2021})},\ \Eprint {http://arxiv.org/abs/2010.02090} {arXiv:2010.02090
  [astro-ph.HE]} \BibitemShut {NoStop}%
\bibitem [{\citenamefont {Nunes}\ \emph {et~al.}(2020)\citenamefont {Nunes},
  \citenamefont {Coelho},\ and\ \citenamefont {de~Araujo}}]{Nunes:2020cuz}%
  \BibitemOpen
  \bibfield  {author} {\bibinfo {author} {\bibfnamefont {R.~C.}\ \bibnamefont
  {Nunes}}, \bibinfo {author} {\bibfnamefont {J.~G.}\ \bibnamefont {Coelho}}, \
  and\ \bibinfo {author} {\bibfnamefont {J.~C.~N.}\ \bibnamefont {de~Araujo}},\
  }\href {\doibase 10.1140/epjc/s10052-020-08695-0} {\bibfield  {journal}
  {\bibinfo  {journal} {Eur. Phys. J. C}\ }\textbf {\bibinfo {volume} {80}},\
  \bibinfo {pages} {1115} (\bibinfo {year} {2020})},\ \Eprint
  {http://arxiv.org/abs/2008.10395} {arXiv:2008.10395 [astro-ph.HE]}
  \BibitemShut {NoStop}%
\bibitem [{\citenamefont {Astashenok}\ \emph {et~al.}(2020)\citenamefont
  {Astashenok}, \citenamefont {Capozziello}, \citenamefont {Odintsov},\ and\
  \citenamefont {Oikonomou}}]{Astashenok:2020qds}%
  \BibitemOpen
  \bibfield  {author} {\bibinfo {author} {\bibfnamefont {A.~V.}\ \bibnamefont
  {Astashenok}}, \bibinfo {author} {\bibfnamefont {S.}~\bibnamefont
  {Capozziello}}, \bibinfo {author} {\bibfnamefont {S.~D.}\ \bibnamefont
  {Odintsov}}, \ and\ \bibinfo {author} {\bibfnamefont {V.~K.}\ \bibnamefont
  {Oikonomou}},\ }\href {\doibase 10.1016/j.physletb.2020.135910} {\bibfield
  {journal} {\bibinfo  {journal} {Phys. Lett. B}\ }\textbf {\bibinfo {volume}
  {811}},\ \bibinfo {pages} {135910} (\bibinfo {year} {2020})},\ \Eprint
  {http://arxiv.org/abs/2008.10884} {arXiv:2008.10884 [gr-qc]} \BibitemShut
  {NoStop}%
\bibitem [{\citenamefont {Astashenok}\ \emph
  {et~al.}(2021{\natexlab{a}})\citenamefont {Astashenok}, \citenamefont
  {Capozziello}, \citenamefont {Odintsov},\ and\ \citenamefont
  {Oikonomou}}]{Astashenok:2021peo}%
  \BibitemOpen
  \bibfield  {author} {\bibinfo {author} {\bibfnamefont {A.~V.}\ \bibnamefont
  {Astashenok}}, \bibinfo {author} {\bibfnamefont {S.}~\bibnamefont
  {Capozziello}}, \bibinfo {author} {\bibfnamefont {S.~D.}\ \bibnamefont
  {Odintsov}}, \ and\ \bibinfo {author} {\bibfnamefont {V.~K.}\ \bibnamefont
  {Oikonomou}},\ }\href {\doibase 10.1016/j.physletb.2021.136222} {\bibfield
  {journal} {\bibinfo  {journal} {Phys. Lett. B}\ }\textbf {\bibinfo {volume}
  {816}},\ \bibinfo {pages} {136222} (\bibinfo {year} {2021}{\natexlab{a}})},\
  \Eprint {http://arxiv.org/abs/2103.04144} {arXiv:2103.04144 [gr-qc]}
  \BibitemShut {NoStop}%
\bibitem [{\citenamefont {Mustafa}\ \emph {et~al.}(2020)\citenamefont
  {Mustafa}, \citenamefont {Shamir},\ and\ \citenamefont
  {Tie-Cheng}}]{Mustafa:2020jln}%
  \BibitemOpen
  \bibfield  {author} {\bibinfo {author} {\bibfnamefont {G.}~\bibnamefont
  {Mustafa}}, \bibinfo {author} {\bibfnamefont {M.~F.}\ \bibnamefont {Shamir}},
  \ and\ \bibinfo {author} {\bibfnamefont {X.}~\bibnamefont {Tie-Cheng}},\
  }\href {\doibase 10.1103/PhysRevD.101.104013} {\bibfield  {journal} {\bibinfo
   {journal} {Phys. Rev. D}\ }\textbf {\bibinfo {volume} {101}},\ \bibinfo
  {pages} {104013} (\bibinfo {year} {2020})},\ \Eprint
  {http://arxiv.org/abs/2005.03997} {arXiv:2005.03997 [gr-qc]} \BibitemShut
  {NoStop}%
\bibitem [{\citenamefont {Nashed}(2021)}]{Nashed:2021pkc}%
  \BibitemOpen
  \bibfield  {author} {\bibinfo {author} {\bibfnamefont {G.~G.~L.}\
  \bibnamefont {Nashed}},\ }\href {\doibase 10.3847/1538-4357/ac19bb}
  {\bibfield  {journal} {\bibinfo  {journal} {Astrophys. J.}\ }\textbf
  {\bibinfo {volume} {919}},\ \bibinfo {pages} {113} (\bibinfo {year}
  {2021})},\ \Eprint {http://arxiv.org/abs/2108.04060} {arXiv:2108.04060
  [gr-qc]} \BibitemShut {NoStop}%
\bibitem [{\citenamefont {Nashed}\ \emph {et~al.}(2021)\citenamefont {Nashed},
  \citenamefont {Odintsov},\ and\ \citenamefont {Oikonomou}}]{Nashed:2021gkp}%
  \BibitemOpen
  \bibfield  {author} {\bibinfo {author} {\bibfnamefont {G.~G.~L.}\
  \bibnamefont {Nashed}}, \bibinfo {author} {\bibfnamefont {S.~D.}\
  \bibnamefont {Odintsov}}, \ and\ \bibinfo {author} {\bibfnamefont {V.~K.}\
  \bibnamefont {Oikonomou}},\ }\href {\doibase 10.1140/epjc/s10052-021-09321-3}
  {\bibfield  {journal} {\bibinfo  {journal} {Eur. Phys. J. C}\ }\textbf
  {\bibinfo {volume} {81}},\ \bibinfo {pages} {528} (\bibinfo {year} {2021})},\
  \Eprint {http://arxiv.org/abs/2106.13607} {arXiv:2106.13607 [gr-qc]}
  \BibitemShut {NoStop}%
\bibitem [{\citenamefont {Nashed}\ and\ \citenamefont
  {Capozziello}(2021)}]{Nashed:2021sji}%
  \BibitemOpen
  \bibfield  {author} {\bibinfo {author} {\bibfnamefont {G.~G.~L.}\
  \bibnamefont {Nashed}}\ and\ \bibinfo {author} {\bibfnamefont
  {S.}~\bibnamefont {Capozziello}},\ }\href {\doibase
  10.1140/epjc/s10052-021-09273-8} {\bibfield  {journal} {\bibinfo  {journal}
  {Eur. Phys. J. C}\ }\textbf {\bibinfo {volume} {81}},\ \bibinfo {pages} {481}
  (\bibinfo {year} {2021})},\ \Eprint {http://arxiv.org/abs/2105.11975}
  {arXiv:2105.11975 [gr-qc]} \BibitemShut {NoStop}%
\bibitem [{\citenamefont {Astashenok}\ \emph
  {et~al.}(2021{\natexlab{b}})\citenamefont {Astashenok}, \citenamefont
  {Capozziello}, \citenamefont {Odintsov},\ and\ \citenamefont
  {Oikonomou}}]{Astashenok:2021btj}%
  \BibitemOpen
  \bibfield  {author} {\bibinfo {author} {\bibfnamefont {A.~V.}\ \bibnamefont
  {Astashenok}}, \bibinfo {author} {\bibfnamefont {S.}~\bibnamefont
  {Capozziello}}, \bibinfo {author} {\bibfnamefont {S.~D.}\ \bibnamefont
  {Odintsov}}, \ and\ \bibinfo {author} {\bibfnamefont {V.~K.}\ \bibnamefont
  {Oikonomou}},\ }\href {\doibase 10.1209/0295-5075/ac3d6c} {\  (\bibinfo
  {year} {2021}{\natexlab{b}}),\ 10.1209/0295-5075/ac3d6c},\ \Eprint
  {http://arxiv.org/abs/2111.14179} {arXiv:2111.14179 [gr-qc]} \BibitemShut
  {NoStop}%
\bibitem [{\citenamefont {Astashenok}\ \emph
  {et~al.}(2021{\natexlab{c}})\citenamefont {Astashenok}, \citenamefont
  {Capozziello}, \citenamefont {Odintsov},\ and\ \citenamefont
  {Oikonomou}}]{Astashenok:2021xpm}%
  \BibitemOpen
  \bibfield  {author} {\bibinfo {author} {\bibfnamefont {A.~V.}\ \bibnamefont
  {Astashenok}}, \bibinfo {author} {\bibfnamefont {S.}~\bibnamefont
  {Capozziello}}, \bibinfo {author} {\bibfnamefont {S.~D.}\ \bibnamefont
  {Odintsov}}, \ and\ \bibinfo {author} {\bibfnamefont {V.~K.}\ \bibnamefont
  {Oikonomou}},\ }\href {\doibase 10.1209/0295-5075/134/59001} {\bibfield
  {journal} {\bibinfo  {journal} {EPL}\ }\textbf {\bibinfo {volume} {134}},\
  \bibinfo {pages} {59001} (\bibinfo {year} {2021}{\natexlab{c}})},\ \Eprint
  {http://arxiv.org/abs/2106.01234} {arXiv:2106.01234 [gr-qc]} \BibitemShut
  {NoStop}%
\bibitem [{\citenamefont {Nashed}\ \emph {et~al.}(2020)\citenamefont {Nashed},
  \citenamefont {Abebe},\ and\ \citenamefont {Bamba}}]{Nashed:2020buf}%
  \BibitemOpen
  \bibfield  {author} {\bibinfo {author} {\bibfnamefont {G.~G.~L.}\
  \bibnamefont {Nashed}}, \bibinfo {author} {\bibfnamefont {A.}~\bibnamefont
  {Abebe}}, \ and\ \bibinfo {author} {\bibfnamefont {K.}~\bibnamefont
  {Bamba}},\ }\href {\doibase 10.1140/epjc/s10052-020-08671-8} {\bibfield
  {journal} {\bibinfo  {journal} {Eur. Phys. J. C}\ }\textbf {\bibinfo {volume}
  {80}},\ \bibinfo {pages} {1109} (\bibinfo {year} {2020})}\BibitemShut
  {NoStop}%
\bibitem [{\citenamefont {Nashed}\ and\ \citenamefont
  {Capozziello}(2020)}]{Nashed:2020kjh}%
  \BibitemOpen
  \bibfield  {author} {\bibinfo {author} {\bibfnamefont {G.~G.~L.}\
  \bibnamefont {Nashed}}\ and\ \bibinfo {author} {\bibfnamefont
  {S.}~\bibnamefont {Capozziello}},\ }\href {\doibase
  10.1140/epjc/s10052-020-08551-1} {\bibfield  {journal} {\bibinfo  {journal}
  {Eur. Phys. J. C}\ }\textbf {\bibinfo {volume} {80}},\ \bibinfo {pages} {969}
  (\bibinfo {year} {2020})},\ \Eprint {http://arxiv.org/abs/2010.06355}
  {arXiv:2010.06355 [gr-qc]} \BibitemShut {NoStop}%
\bibitem [{\citenamefont {\"Ovg\"un}(2021)}]{Ovgun:2021ttv}%
  \BibitemOpen
  \bibfield  {author} {\bibinfo {author} {\bibfnamefont {A.}~\bibnamefont
  {\"Ovg\"un}},\ }\href {\doibase 10.1016/j.physletb.2021.136517} {\bibfield
  {journal} {\bibinfo  {journal} {Phys. Lett. B}\ }\textbf {\bibinfo {volume}
  {820}},\ \bibinfo {pages} {136517} (\bibinfo {year} {2021})},\ \Eprint
  {http://arxiv.org/abs/2105.05035} {arXiv:2105.05035 [gr-qc]} \BibitemShut
  {NoStop}%
\bibitem [{\citenamefont {Abbott}\ \emph
  {et~al.}(2017{\natexlab{a}})\citenamefont {Abbott} \emph
  {et~al.}}]{LIGOScientific:2017vwq}%
  \BibitemOpen
  \bibfield  {author} {\bibinfo {author} {\bibfnamefont {B.~P.}\ \bibnamefont
  {Abbott}} \emph {et~al.} (\bibinfo {collaboration} {LIGO Scientific,
  Virgo}),\ }\href {\doibase 10.1103/PhysRevLett.119.161101} {\bibfield
  {journal} {\bibinfo  {journal} {Phys. Rev. Lett.}\ }\textbf {\bibinfo
  {volume} {119}},\ \bibinfo {pages} {161101} (\bibinfo {year}
  {2017}{\natexlab{a}})},\ \Eprint {http://arxiv.org/abs/1710.05832}
  {arXiv:1710.05832 [gr-qc]} \BibitemShut {NoStop}%
\bibitem [{\citenamefont {Abbott}\ \emph
  {et~al.}(2020{\natexlab{b}})\citenamefont {Abbott} \emph
  {et~al.}}]{LIGOScientific:2020aai}%
  \BibitemOpen
  \bibfield  {author} {\bibinfo {author} {\bibfnamefont {B.~P.}\ \bibnamefont
  {Abbott}} \emph {et~al.} (\bibinfo {collaboration} {LIGO Scientific,
  Virgo}),\ }\href {\doibase 10.3847/2041-8213/ab75f5} {\bibfield  {journal}
  {\bibinfo  {journal} {Astrophys. J. Lett.}\ }\textbf {\bibinfo {volume}
  {892}},\ \bibinfo {pages} {L3} (\bibinfo {year} {2020}{\natexlab{b}})},\
  \Eprint {http://arxiv.org/abs/2001.01761} {arXiv:2001.01761 [astro-ph.HE]}
  \BibitemShut {NoStop}%
\bibitem [{\citenamefont {Abbott}\ \emph
  {et~al.}(2018{\natexlab{a}})\citenamefont {Abbott} \emph
  {et~al.}}]{KAGRA:2013rdx}%
  \BibitemOpen
  \bibfield  {author} {\bibinfo {author} {\bibfnamefont {B.~P.}\ \bibnamefont
  {Abbott}} \emph {et~al.} (\bibinfo {collaboration} {KAGRA, LIGO Scientific,
  Virgo, VIRGO}),\ }\href {\doibase 10.1007/s41114-020-00026-9} {\bibfield
  {journal} {\bibinfo  {journal} {Living Rev. Rel.}\ }\textbf {\bibinfo
  {volume} {21}},\ \bibinfo {pages} {3} (\bibinfo {year}
  {2018}{\natexlab{a}})},\ \Eprint {http://arxiv.org/abs/1304.0670}
  {arXiv:1304.0670 [gr-qc]} \BibitemShut {NoStop}%
\bibitem [{\citenamefont {Abbott}\ \emph
  {et~al.}(2017{\natexlab{b}})\citenamefont {Abbott} \emph
  {et~al.}}]{LIGOScientific:2017ync}%
  \BibitemOpen
  \bibfield  {author} {\bibinfo {author} {\bibfnamefont {B.~P.}\ \bibnamefont
  {Abbott}} \emph {et~al.} (\bibinfo {collaboration} {LIGO Scientific, Virgo,
  Fermi GBM, INTEGRAL, IceCube, AstroSat Cadmium Zinc Telluride Imager Team,
  IPN, Insight-Hxmt, ANTARES, Swift, AGILE Team, 1M2H Team, Dark Energy Camera
  GW-EM, DES, DLT40, GRAWITA, Fermi-LAT, ATCA, ASKAP, Las Cumbres Observatory
  Group, OzGrav, DWF (Deeper Wider Faster Program), AST3, CAASTRO, VINROUGE,
  MASTER, J-GEM, GROWTH, JAGWAR, CaltechNRAO, TTU-NRAO, NuSTAR, Pan-STARRS,
  MAXI Team, TZAC Consortium, KU, Nordic Optical Telescope, ePESSTO, GROND,
  Texas Tech University, SALT Group, TOROS, BOOTES, MWA, CALET, IKI-GW
  Follow-up, H.E.S.S., LOFAR, LWA, HAWC, Pierre Auger, ALMA, Euro VLBI Team, Pi
  of Sky, Chandra Team at McGill University, DFN, ATLAS Telescopes, High Time
  Resolution Universe Survey, RIMAS, RATIR, SKA South Africa/MeerKAT}),\ }\href
  {\doibase 10.3847/2041-8213/aa91c9} {\bibfield  {journal} {\bibinfo
  {journal} {Astrophys. J. Lett.}\ }\textbf {\bibinfo {volume} {848}},\
  \bibinfo {pages} {L12} (\bibinfo {year} {2017}{\natexlab{b}})},\ \Eprint
  {http://arxiv.org/abs/1710.05833} {arXiv:1710.05833 [astro-ph.HE]}
  \BibitemShut {NoStop}%
\bibitem [{\citenamefont {Abbott}\ \emph
  {et~al.}(2017{\natexlab{c}})\citenamefont {Abbott} \emph
  {et~al.}}]{LIGOScientific:2017zic}%
  \BibitemOpen
  \bibfield  {author} {\bibinfo {author} {\bibfnamefont {B.~P.}\ \bibnamefont
  {Abbott}} \emph {et~al.} (\bibinfo {collaboration} {LIGO Scientific, Virgo,
  Fermi-GBM, INTEGRAL}),\ }\href {\doibase 10.3847/2041-8213/aa920c} {\bibfield
   {journal} {\bibinfo  {journal} {Astrophys. J. Lett.}\ }\textbf {\bibinfo
  {volume} {848}},\ \bibinfo {pages} {L13} (\bibinfo {year}
  {2017}{\natexlab{c}})},\ \Eprint {http://arxiv.org/abs/1710.05834}
  {arXiv:1710.05834 [astro-ph.HE]} \BibitemShut {NoStop}%
\bibitem [{\citenamefont {Goldstein}\ \emph {et~al.}(2017)\citenamefont
  {Goldstein} \emph {et~al.}}]{Goldstein:2017mmi}%
  \BibitemOpen
  \bibfield  {author} {\bibinfo {author} {\bibfnamefont {A.}~\bibnamefont
  {Goldstein}} \emph {et~al.},\ }\href {\doibase 10.3847/2041-8213/aa8f41}
  {\bibfield  {journal} {\bibinfo  {journal} {Astrophys. J. Lett.}\ }\textbf
  {\bibinfo {volume} {848}},\ \bibinfo {pages} {L14} (\bibinfo {year}
  {2017})},\ \Eprint {http://arxiv.org/abs/1710.05446} {arXiv:1710.05446
  [astro-ph.HE]} \BibitemShut {NoStop}%
\bibitem [{\citenamefont {Bauswein}\ \emph {et~al.}(2017)\citenamefont
  {Bauswein}, \citenamefont {Just}, \citenamefont {Janka},\ and\ \citenamefont
  {Stergioulas}}]{Bauswein:2017vtn}%
  \BibitemOpen
  \bibfield  {author} {\bibinfo {author} {\bibfnamefont {A.}~\bibnamefont
  {Bauswein}}, \bibinfo {author} {\bibfnamefont {O.}~\bibnamefont {Just}},
  \bibinfo {author} {\bibfnamefont {H.-T.}\ \bibnamefont {Janka}}, \ and\
  \bibinfo {author} {\bibfnamefont {N.}~\bibnamefont {Stergioulas}},\ }\href
  {\doibase 10.3847/2041-8213/aa9994} {\bibfield  {journal} {\bibinfo
  {journal} {Astrophys. J. Lett.}\ }\textbf {\bibinfo {volume} {850}},\
  \bibinfo {pages} {L34} (\bibinfo {year} {2017})},\ \Eprint
  {http://arxiv.org/abs/1710.06843} {arXiv:1710.06843 [astro-ph.HE]}
  \BibitemShut {NoStop}%
\bibitem [{\citenamefont {Abbott}\ \emph
  {et~al.}(2018{\natexlab{b}})\citenamefont {Abbott} \emph
  {et~al.}}]{LIGOScientific:2018cki}%
  \BibitemOpen
  \bibfield  {author} {\bibinfo {author} {\bibfnamefont {B.~P.}\ \bibnamefont
  {Abbott}} \emph {et~al.} (\bibinfo {collaboration} {LIGO Scientific,
  Virgo}),\ }\href {\doibase 10.1103/PhysRevLett.121.161101} {\bibfield
  {journal} {\bibinfo  {journal} {Phys. Rev. Lett.}\ }\textbf {\bibinfo
  {volume} {121}},\ \bibinfo {pages} {161101} (\bibinfo {year}
  {2018}{\natexlab{b}})},\ \Eprint {http://arxiv.org/abs/1805.11581}
  {arXiv:1805.11581 [gr-qc]} \BibitemShut {NoStop}%
\bibitem [{\citenamefont {Capano}\ \emph {et~al.}(2020)\citenamefont {Capano},
  \citenamefont {Tews}, \citenamefont {Brown}, \citenamefont {Margalit},
  \citenamefont {De}, \citenamefont {Kumar}, \citenamefont {Brown},
  \citenamefont {Krishnan},\ and\ \citenamefont {Reddy}}]{Capano:2019eae}%
  \BibitemOpen
  \bibfield  {author} {\bibinfo {author} {\bibfnamefont {C.~D.}\ \bibnamefont
  {Capano}}, \bibinfo {author} {\bibfnamefont {I.}~\bibnamefont {Tews}},
  \bibinfo {author} {\bibfnamefont {S.~M.}\ \bibnamefont {Brown}}, \bibinfo
  {author} {\bibfnamefont {B.}~\bibnamefont {Margalit}}, \bibinfo {author}
  {\bibfnamefont {S.}~\bibnamefont {De}}, \bibinfo {author} {\bibfnamefont
  {S.}~\bibnamefont {Kumar}}, \bibinfo {author} {\bibfnamefont {D.~A.}\
  \bibnamefont {Brown}}, \bibinfo {author} {\bibfnamefont {B.}~\bibnamefont
  {Krishnan}}, \ and\ \bibinfo {author} {\bibfnamefont {S.}~\bibnamefont
  {Reddy}},\ }\href {\doibase 10.1038/s41550-020-1014-6} {\bibfield  {journal}
  {\bibinfo  {journal} {Nature Astron.}\ }\textbf {\bibinfo {volume} {4}},\
  \bibinfo {pages} {625} (\bibinfo {year} {2020})},\ \Eprint
  {http://arxiv.org/abs/1908.10352} {arXiv:1908.10352 [astro-ph.HE]}
  \BibitemShut {NoStop}%
\bibitem [{\citenamefont {Thapa}\ \emph {et~al.}(2021)\citenamefont {Thapa},
  \citenamefont {Kumar},\ and\ \citenamefont {Sinha}}]{Thapa:2021ifv}%
  \BibitemOpen
  \bibfield  {author} {\bibinfo {author} {\bibfnamefont {V.~B.}\ \bibnamefont
  {Thapa}}, \bibinfo {author} {\bibfnamefont {A.}~\bibnamefont {Kumar}}, \ and\
  \bibinfo {author} {\bibfnamefont {M.}~\bibnamefont {Sinha}},\ }\href
  {\doibase 10.1093/mnras/stab2327} {\bibfield  {journal} {\bibinfo  {journal}
  {Mon. Not. Roy. Astron. Soc.}\ }\textbf {\bibinfo {volume} {507}},\ \bibinfo
  {pages} {2} (\bibinfo {year} {2021})},\ \Eprint
  {http://arxiv.org/abs/2108.04318} {arXiv:2108.04318 [astro-ph.HE]}
  \BibitemShut {NoStop}%
\bibitem [{\citenamefont {Dietrich}\ \emph {et~al.}(2020)\citenamefont
  {Dietrich}, \citenamefont {Coughlin}, \citenamefont {Pang}, \citenamefont
  {Bulla}, \citenamefont {Heinzel}, \citenamefont {Issa}, \citenamefont
  {Tews},\ and\ \citenamefont {Antier}}]{Dietrich:2020efo}%
  \BibitemOpen
  \bibfield  {author} {\bibinfo {author} {\bibfnamefont {T.}~\bibnamefont
  {Dietrich}}, \bibinfo {author} {\bibfnamefont {M.~W.}\ \bibnamefont
  {Coughlin}}, \bibinfo {author} {\bibfnamefont {P.~T.~H.}\ \bibnamefont
  {Pang}}, \bibinfo {author} {\bibfnamefont {M.}~\bibnamefont {Bulla}},
  \bibinfo {author} {\bibfnamefont {J.}~\bibnamefont {Heinzel}}, \bibinfo
  {author} {\bibfnamefont {L.}~\bibnamefont {Issa}}, \bibinfo {author}
  {\bibfnamefont {I.}~\bibnamefont {Tews}}, \ and\ \bibinfo {author}
  {\bibfnamefont {S.}~\bibnamefont {Antier}},\ }\href {\doibase
  10.1126/science.abb4317} {\bibfield  {journal} {\bibinfo  {journal}
  {Science}\ }\textbf {\bibinfo {volume} {370}},\ \bibinfo {pages} {1450}
  (\bibinfo {year} {2020})},\ \Eprint {http://arxiv.org/abs/2002.11355}
  {arXiv:2002.11355 [astro-ph.HE]} \BibitemShut {NoStop}%
\bibitem [{\citenamefont {Breschi}\ \emph {et~al.}(2021)\citenamefont
  {Breschi}, \citenamefont {Perego}, \citenamefont {Bernuzzi}, \citenamefont
  {Del~Pozzo}, \citenamefont {Nedora}, \citenamefont {Radice},\ and\
  \citenamefont {Vescovi}}]{Breschi:2021tbm}%
  \BibitemOpen
  \bibfield  {author} {\bibinfo {author} {\bibfnamefont {M.}~\bibnamefont
  {Breschi}}, \bibinfo {author} {\bibfnamefont {A.}~\bibnamefont {Perego}},
  \bibinfo {author} {\bibfnamefont {S.}~\bibnamefont {Bernuzzi}}, \bibinfo
  {author} {\bibfnamefont {W.}~\bibnamefont {Del~Pozzo}}, \bibinfo {author}
  {\bibfnamefont {V.}~\bibnamefont {Nedora}}, \bibinfo {author} {\bibfnamefont
  {D.}~\bibnamefont {Radice}}, \ and\ \bibinfo {author} {\bibfnamefont
  {D.}~\bibnamefont {Vescovi}},\ }\href {\doibase 10.1093/mnras/stab1287}
  {\bibfield  {journal} {\bibinfo  {journal} {Mon. Not. Roy. Astron. Soc.}\
  }\textbf {\bibinfo {volume} {505}},\ \bibinfo {pages} {1661} (\bibinfo {year}
  {2021})},\ \Eprint {http://arxiv.org/abs/2101.01201} {arXiv:2101.01201
  [astro-ph.HE]} \BibitemShut {NoStop}%
\bibitem [{\citenamefont {Chatziioannou}(2020)}]{Chatziioannou:2020pqz}%
  \BibitemOpen
  \bibfield  {author} {\bibinfo {author} {\bibfnamefont {K.}~\bibnamefont
  {Chatziioannou}},\ }\href {\doibase 10.1007/s10714-020-02754-3} {\bibfield
  {journal} {\bibinfo  {journal} {Gen. Rel. Grav.}\ }\textbf {\bibinfo {volume}
  {52}},\ \bibinfo {pages} {109} (\bibinfo {year} {2020})},\ \Eprint
  {http://arxiv.org/abs/2006.03168} {arXiv:2006.03168 [gr-qc]} \BibitemShut
  {NoStop}%
\bibitem [{\citenamefont {Del~Pozzo}\ \emph {et~al.}(2013)\citenamefont
  {Del~Pozzo}, \citenamefont {Li}, \citenamefont {Agathos}, \citenamefont {Van
  Den~Broeck},\ and\ \citenamefont {Vitale}}]{DelPozzo:2013ala}%
  \BibitemOpen
  \bibfield  {author} {\bibinfo {author} {\bibfnamefont {W.}~\bibnamefont
  {Del~Pozzo}}, \bibinfo {author} {\bibfnamefont {T.~G.~F.}\ \bibnamefont
  {Li}}, \bibinfo {author} {\bibfnamefont {M.}~\bibnamefont {Agathos}},
  \bibinfo {author} {\bibfnamefont {C.}~\bibnamefont {Van Den~Broeck}}, \ and\
  \bibinfo {author} {\bibfnamefont {S.}~\bibnamefont {Vitale}},\ }\href
  {\doibase 10.1103/PhysRevLett.111.071101} {\bibfield  {journal} {\bibinfo
  {journal} {Phys. Rev. Lett.}\ }\textbf {\bibinfo {volume} {111}},\ \bibinfo
  {pages} {071101} (\bibinfo {year} {2013})},\ \Eprint
  {http://arxiv.org/abs/1307.8338} {arXiv:1307.8338 [gr-qc]} \BibitemShut
  {NoStop}%
\bibitem [{\citenamefont {Chatziioannou}\ \emph {et~al.}(2015)\citenamefont
  {Chatziioannou}, \citenamefont {Yagi}, \citenamefont {Klein}, \citenamefont
  {Cornish},\ and\ \citenamefont {Yunes}}]{Chatziioannou:2015uea}%
  \BibitemOpen
  \bibfield  {author} {\bibinfo {author} {\bibfnamefont {K.}~\bibnamefont
  {Chatziioannou}}, \bibinfo {author} {\bibfnamefont {K.}~\bibnamefont {Yagi}},
  \bibinfo {author} {\bibfnamefont {A.}~\bibnamefont {Klein}}, \bibinfo
  {author} {\bibfnamefont {N.}~\bibnamefont {Cornish}}, \ and\ \bibinfo
  {author} {\bibfnamefont {N.}~\bibnamefont {Yunes}},\ }\href {\doibase
  10.1103/PhysRevD.92.104008} {\bibfield  {journal} {\bibinfo  {journal} {Phys.
  Rev. D}\ }\textbf {\bibinfo {volume} {92}},\ \bibinfo {pages} {104008}
  (\bibinfo {year} {2015})},\ \Eprint {http://arxiv.org/abs/1508.02062}
  {arXiv:1508.02062 [gr-qc]} \BibitemShut {NoStop}%
\bibitem [{\citenamefont {Lackey}\ and\ \citenamefont
  {Wade}(2015)}]{Lackey:2014fwa}%
  \BibitemOpen
  \bibfield  {author} {\bibinfo {author} {\bibfnamefont {B.~D.}\ \bibnamefont
  {Lackey}}\ and\ \bibinfo {author} {\bibfnamefont {L.}~\bibnamefont {Wade}},\
  }\href {\doibase 10.1103/PhysRevD.91.043002} {\bibfield  {journal} {\bibinfo
  {journal} {Phys. Rev. D}\ }\textbf {\bibinfo {volume} {91}},\ \bibinfo
  {pages} {043002} (\bibinfo {year} {2015})},\ \Eprint
  {http://arxiv.org/abs/1410.8866} {arXiv:1410.8866 [gr-qc]} \BibitemShut
  {NoStop}%
\bibitem [{\citenamefont {Hernandez~Vivanco}\ \emph {et~al.}(2019)\citenamefont
  {Hernandez~Vivanco}, \citenamefont {Smith}, \citenamefont {Thrane},
  \citenamefont {Lasky}, \citenamefont {Talbot},\ and\ \citenamefont
  {Raymond}}]{HernandezVivanco:2019vvk}%
  \BibitemOpen
  \bibfield  {author} {\bibinfo {author} {\bibfnamefont {F.}~\bibnamefont
  {Hernandez~Vivanco}}, \bibinfo {author} {\bibfnamefont {R.}~\bibnamefont
  {Smith}}, \bibinfo {author} {\bibfnamefont {E.}~\bibnamefont {Thrane}},
  \bibinfo {author} {\bibfnamefont {P.~D.}\ \bibnamefont {Lasky}}, \bibinfo
  {author} {\bibfnamefont {C.}~\bibnamefont {Talbot}}, \ and\ \bibinfo {author}
  {\bibfnamefont {V.}~\bibnamefont {Raymond}},\ }\href {\doibase
  10.1103/PhysRevD.100.103009} {\bibfield  {journal} {\bibinfo  {journal}
  {Phys. Rev. D}\ }\textbf {\bibinfo {volume} {100}},\ \bibinfo {pages}
  {103009} (\bibinfo {year} {2019})},\ \Eprint
  {http://arxiv.org/abs/1909.02698} {arXiv:1909.02698 [gr-qc]} \BibitemShut
  {NoStop}%
\bibitem [{\citenamefont {Chatziioannou}\ and\ \citenamefont
  {Han}(2020)}]{Chatziioannou:2019yko}%
  \BibitemOpen
  \bibfield  {author} {\bibinfo {author} {\bibfnamefont {K.}~\bibnamefont
  {Chatziioannou}}\ and\ \bibinfo {author} {\bibfnamefont {S.}~\bibnamefont
  {Han}},\ }\href {\doibase 10.1103/PhysRevD.101.044019} {\bibfield  {journal}
  {\bibinfo  {journal} {Phys. Rev. D}\ }\textbf {\bibinfo {volume} {101}},\
  \bibinfo {pages} {044019} (\bibinfo {year} {2020})},\ \Eprint
  {http://arxiv.org/abs/1911.07091} {arXiv:1911.07091 [gr-qc]} \BibitemShut
  {NoStop}%
\bibitem [{\citenamefont {Abbott}\ \emph
  {et~al.}(2017{\natexlab{d}})\citenamefont {Abbott} \emph
  {et~al.}}]{LIGOScientific:2017fdd}%
  \BibitemOpen
  \bibfield  {author} {\bibinfo {author} {\bibfnamefont {B.~P.}\ \bibnamefont
  {Abbott}} \emph {et~al.} (\bibinfo {collaboration} {LIGO Scientific,
  Virgo}),\ }\href {\doibase 10.3847/2041-8213/aa9a35} {\bibfield  {journal}
  {\bibinfo  {journal} {Astrophys. J. Lett.}\ }\textbf {\bibinfo {volume}
  {851}},\ \bibinfo {pages} {L16} (\bibinfo {year} {2017}{\natexlab{d}})},\
  \Eprint {http://arxiv.org/abs/1710.09320} {arXiv:1710.09320 [astro-ph.HE]}
  \BibitemShut {NoStop}%
\bibitem [{\citenamefont {Abbott}\ \emph
  {et~al.}(2017{\natexlab{e}})\citenamefont {Abbott} \emph
  {et~al.}}]{LIGOScientific:2016wof}%
  \BibitemOpen
  \bibfield  {author} {\bibinfo {author} {\bibfnamefont {B.~P.}\ \bibnamefont
  {Abbott}} \emph {et~al.} (\bibinfo {collaboration} {LIGO Scientific}),\
  }\href {\doibase 10.1088/1361-6382/aa51f4} {\bibfield  {journal} {\bibinfo
  {journal} {Class. Quant. Grav.}\ }\textbf {\bibinfo {volume} {34}},\ \bibinfo
  {pages} {044001} (\bibinfo {year} {2017}{\natexlab{e}})},\ \Eprint
  {http://arxiv.org/abs/1607.08697} {arXiv:1607.08697 [astro-ph.IM]}
  \BibitemShut {NoStop}%
\bibitem [{\citenamefont {Maggiore}\ \emph {et~al.}(2020)\citenamefont
  {Maggiore} \emph {et~al.}}]{Maggiore:2019uih}%
  \BibitemOpen
  \bibfield  {author} {\bibinfo {author} {\bibfnamefont {M.}~\bibnamefont
  {Maggiore}} \emph {et~al.},\ }\href {\doibase 10.1088/1475-7516/2020/03/050}
  {\bibfield  {journal} {\bibinfo  {journal} {JCAP}\ }\textbf {\bibinfo
  {volume} {03}},\ \bibinfo {pages} {050} (\bibinfo {year} {2020})},\ \Eprint
  {http://arxiv.org/abs/1912.02622} {arXiv:1912.02622 [astro-ph.CO]}
  \BibitemShut {NoStop}%
\bibitem [{\citenamefont {Ganapathy}\ \emph {et~al.}(2021)\citenamefont
  {Ganapathy}, \citenamefont {McCuller}, \citenamefont {Graef~Rollins},
  \citenamefont {Hall}, \citenamefont {Barsotti},\ and\ \citenamefont
  {Evans}}]{Ganapathy:2020thy}%
  \BibitemOpen
  \bibfield  {author} {\bibinfo {author} {\bibfnamefont {D.}~\bibnamefont
  {Ganapathy}}, \bibinfo {author} {\bibfnamefont {L.}~\bibnamefont {McCuller}},
  \bibinfo {author} {\bibfnamefont {J.}~\bibnamefont {Graef~Rollins}}, \bibinfo
  {author} {\bibfnamefont {E.~D.}\ \bibnamefont {Hall}}, \bibinfo {author}
  {\bibfnamefont {L.}~\bibnamefont {Barsotti}}, \ and\ \bibinfo {author}
  {\bibfnamefont {M.}~\bibnamefont {Evans}},\ }\href {\doibase
  10.1103/PhysRevD.103.022002} {\bibfield  {journal} {\bibinfo  {journal}
  {Phys. Rev. D}\ }\textbf {\bibinfo {volume} {103}},\ \bibinfo {pages}
  {022002} (\bibinfo {year} {2021})},\ \Eprint
  {http://arxiv.org/abs/2010.15735} {arXiv:2010.15735 [astro-ph.IM]}
  \BibitemShut {NoStop}%
\bibitem [{\citenamefont {Page}\ \emph {et~al.}(2020)\citenamefont {Page} \emph
  {et~al.}}]{Page:2020zbr}%
  \BibitemOpen
  \bibfield  {author} {\bibinfo {author} {\bibfnamefont {M.~A.}\ \bibnamefont
  {Page}} \emph {et~al.},\ }\href@noop {} {\  (\bibinfo {year} {2020})},\
  \Eprint {http://arxiv.org/abs/2007.08766} {arXiv:2007.08766 [physics.optics]}
  \BibitemShut {NoStop}%
\bibitem [{\citenamefont {{Rasio}}\ and\ \citenamefont
  {{Shapiro}}(1992)}]{1992ApJ...401..226R}%
  \BibitemOpen
  \bibfield  {author} {\bibinfo {author} {\bibfnamefont {F.~A.}\ \bibnamefont
  {{Rasio}}}\ and\ \bibinfo {author} {\bibfnamefont {S.~L.}\ \bibnamefont
  {{Shapiro}}},\ }\href {\doibase 10.1086/172055} {\bibfield  {journal}
  {\bibinfo  {journal} {\apj}\ }\textbf {\bibinfo {volume} {401}},\ \bibinfo
  {pages} {226} (\bibinfo {year} {1992})}\BibitemShut {NoStop}%
\bibitem [{\citenamefont {Shibata}(2005)}]{Shibata:2005xz}%
  \BibitemOpen
  \bibfield  {author} {\bibinfo {author} {\bibfnamefont {M.}~\bibnamefont
  {Shibata}},\ }\href {\doibase 10.1103/PhysRevLett.94.201101} {\bibfield
  {journal} {\bibinfo  {journal} {Phys. Rev. Lett.}\ }\textbf {\bibinfo
  {volume} {94}},\ \bibinfo {pages} {201101} (\bibinfo {year} {2005})},\
  \Eprint {http://arxiv.org/abs/gr-qc/0504082} {arXiv:gr-qc/0504082}
  \BibitemShut {NoStop}%
\bibitem [{\citenamefont {Elizalde}\ \emph {et~al.}(2020)\citenamefont
  {Elizalde}, \citenamefont {Nashed}, \citenamefont {Nojiri},\ and\
  \citenamefont {Odintsov}}]{Elizalde:2020icc}%
  \BibitemOpen
  \bibfield  {author} {\bibinfo {author} {\bibfnamefont {E.}~\bibnamefont
  {Elizalde}}, \bibinfo {author} {\bibfnamefont {G.~G.~L.}\ \bibnamefont
  {Nashed}}, \bibinfo {author} {\bibfnamefont {S.}~\bibnamefont {Nojiri}}, \
  and\ \bibinfo {author} {\bibfnamefont {S.~D.}\ \bibnamefont {Odintsov}},\
  }\href {\doibase 10.1140/epjc/s10052-020-7686-3} {\bibfield  {journal}
  {\bibinfo  {journal} {Eur. Phys. J. C}\ }\textbf {\bibinfo {volume} {80}},\
  \bibinfo {pages} {109} (\bibinfo {year} {2020})},\ \Eprint
  {http://arxiv.org/abs/2001.11357} {arXiv:2001.11357 [gr-qc]} \BibitemShut
  {NoStop}%
\bibitem [{\citenamefont {Bauswein}\ and\ \citenamefont
  {Janka}(2012)}]{Bauswein:2011tp}%
  \BibitemOpen
  \bibfield  {author} {\bibinfo {author} {\bibfnamefont {A.}~\bibnamefont
  {Bauswein}}\ and\ \bibinfo {author} {\bibfnamefont {H.~T.}\ \bibnamefont
  {Janka}},\ }\href {\doibase 10.1103/PhysRevLett.108.011101} {\bibfield
  {journal} {\bibinfo  {journal} {Phys. Rev. Lett.}\ }\textbf {\bibinfo
  {volume} {108}},\ \bibinfo {pages} {011101} (\bibinfo {year} {2012})},\
  \Eprint {http://arxiv.org/abs/1106.1616} {arXiv:1106.1616 [astro-ph.SR]}
  \BibitemShut {NoStop}%
\bibitem [{\citenamefont {Clark}\ \emph {et~al.}(2014)\citenamefont {Clark},
  \citenamefont {Bauswein}, \citenamefont {Cadonati}, \citenamefont {Janka},
  \citenamefont {Pankow},\ and\ \citenamefont {Stergioulas}}]{Clark:2014wua}%
  \BibitemOpen
  \bibfield  {author} {\bibinfo {author} {\bibfnamefont {J.}~\bibnamefont
  {Clark}}, \bibinfo {author} {\bibfnamefont {A.}~\bibnamefont {Bauswein}},
  \bibinfo {author} {\bibfnamefont {L.}~\bibnamefont {Cadonati}}, \bibinfo
  {author} {\bibfnamefont {H.~T.}\ \bibnamefont {Janka}}, \bibinfo {author}
  {\bibfnamefont {C.}~\bibnamefont {Pankow}}, \ and\ \bibinfo {author}
  {\bibfnamefont {N.}~\bibnamefont {Stergioulas}},\ }\href {\doibase
  10.1103/PhysRevD.90.062004} {\bibfield  {journal} {\bibinfo  {journal} {Phys.
  Rev. D}\ }\textbf {\bibinfo {volume} {90}},\ \bibinfo {pages} {062004}
  (\bibinfo {year} {2014})},\ \Eprint {http://arxiv.org/abs/1406.5444}
  {arXiv:1406.5444 [astro-ph.HE]} \BibitemShut {NoStop}%
\bibitem [{\citenamefont {Rezzolla}\ and\ \citenamefont
  {Takami}(2016)}]{Rezzolla:2016nxn}%
  \BibitemOpen
  \bibfield  {author} {\bibinfo {author} {\bibfnamefont {L.}~\bibnamefont
  {Rezzolla}}\ and\ \bibinfo {author} {\bibfnamefont {K.}~\bibnamefont
  {Takami}},\ }\href {\doibase 10.1103/PhysRevD.93.124051} {\bibfield
  {journal} {\bibinfo  {journal} {Phys. Rev. D}\ }\textbf {\bibinfo {volume}
  {93}},\ \bibinfo {pages} {124051} (\bibinfo {year} {2016})},\ \Eprint
  {http://arxiv.org/abs/1604.00246} {arXiv:1604.00246 [gr-qc]} \BibitemShut
  {NoStop}%
\bibitem [{\citenamefont {Bauswein}\ and\ \citenamefont
  {Stergioulas}(2019)}]{Bauswein:2019ybt}%
  \BibitemOpen
  \bibfield  {author} {\bibinfo {author} {\bibfnamefont {A.}~\bibnamefont
  {Bauswein}}\ and\ \bibinfo {author} {\bibfnamefont {N.}~\bibnamefont
  {Stergioulas}},\ }\href {\doibase 10.1088/1361-6471/ab2b90} {\bibfield
  {journal} {\bibinfo  {journal} {J. Phys. G}\ }\textbf {\bibinfo {volume}
  {46}},\ \bibinfo {pages} {113002} (\bibinfo {year} {2019})},\ \Eprint
  {http://arxiv.org/abs/1901.06969} {arXiv:1901.06969 [gr-qc]} \BibitemShut
  {NoStop}%
\bibitem [{\citenamefont {Breschi}\ \emph {et~al.}(2019)\citenamefont
  {Breschi}, \citenamefont {Bernuzzi}, \citenamefont {Zappa}, \citenamefont
  {Agathos}, \citenamefont {Perego}, \citenamefont {Radice},\ and\
  \citenamefont {Nagar}}]{Breschi:2019srl}%
  \BibitemOpen
  \bibfield  {author} {\bibinfo {author} {\bibfnamefont {M.}~\bibnamefont
  {Breschi}}, \bibinfo {author} {\bibfnamefont {S.}~\bibnamefont {Bernuzzi}},
  \bibinfo {author} {\bibfnamefont {F.}~\bibnamefont {Zappa}}, \bibinfo
  {author} {\bibfnamefont {M.}~\bibnamefont {Agathos}}, \bibinfo {author}
  {\bibfnamefont {A.}~\bibnamefont {Perego}}, \bibinfo {author} {\bibfnamefont
  {D.}~\bibnamefont {Radice}}, \ and\ \bibinfo {author} {\bibfnamefont
  {A.}~\bibnamefont {Nagar}},\ }\href {\doibase 10.1103/PhysRevD.100.104029}
  {\bibfield  {journal} {\bibinfo  {journal} {Phys. Rev. D}\ }\textbf {\bibinfo
  {volume} {100}},\ \bibinfo {pages} {104029} (\bibinfo {year} {2019})},\
  \Eprint {http://arxiv.org/abs/1908.11418} {arXiv:1908.11418 [gr-qc]}
  \BibitemShut {NoStop}%
\bibitem [{\citenamefont {Tsang}\ \emph {et~al.}(2019)\citenamefont {Tsang},
  \citenamefont {Dietrich},\ and\ \citenamefont {Van
  Den~Broeck}}]{Tsang:2019esi}%
  \BibitemOpen
  \bibfield  {author} {\bibinfo {author} {\bibfnamefont {K.~W.}\ \bibnamefont
  {Tsang}}, \bibinfo {author} {\bibfnamefont {T.}~\bibnamefont {Dietrich}}, \
  and\ \bibinfo {author} {\bibfnamefont {C.}~\bibnamefont {Van Den~Broeck}},\
  }\href {\doibase 10.1103/PhysRevD.100.044047} {\bibfield  {journal} {\bibinfo
   {journal} {Phys. Rev. D}\ }\textbf {\bibinfo {volume} {100}},\ \bibinfo
  {pages} {044047} (\bibinfo {year} {2019})},\ \Eprint
  {http://arxiv.org/abs/1907.02424} {arXiv:1907.02424 [gr-qc]} \BibitemShut
  {NoStop}%
\bibitem [{\citenamefont {Vretinaris}\ \emph {et~al.}(2020)\citenamefont
  {Vretinaris}, \citenamefont {Stergioulas},\ and\ \citenamefont
  {Bauswein}}]{Vretinaris:2019spn}%
  \BibitemOpen
  \bibfield  {author} {\bibinfo {author} {\bibfnamefont {S.}~\bibnamefont
  {Vretinaris}}, \bibinfo {author} {\bibfnamefont {N.}~\bibnamefont
  {Stergioulas}}, \ and\ \bibinfo {author} {\bibfnamefont {A.}~\bibnamefont
  {Bauswein}},\ }\href {\doibase 10.1103/PhysRevD.101.084039} {\bibfield
  {journal} {\bibinfo  {journal} {Phys. Rev. D}\ }\textbf {\bibinfo {volume}
  {101}},\ \bibinfo {pages} {084039} (\bibinfo {year} {2020})},\ \Eprint
  {http://arxiv.org/abs/1910.10856} {arXiv:1910.10856 [gr-qc]} \BibitemShut
  {NoStop}%
\bibitem [{\citenamefont {Easter}\ \emph {et~al.}(2020)\citenamefont {Easter},
  \citenamefont {Ghonge}, \citenamefont {Lasky}, \citenamefont {Casey},
  \citenamefont {Clark}, \citenamefont {Vivanco},\ and\ \citenamefont
  {Chatziioannou}}]{Easter:2020ifj}%
  \BibitemOpen
  \bibfield  {author} {\bibinfo {author} {\bibfnamefont {P.~J.}\ \bibnamefont
  {Easter}}, \bibinfo {author} {\bibfnamefont {S.}~\bibnamefont {Ghonge}},
  \bibinfo {author} {\bibfnamefont {P.~D.}\ \bibnamefont {Lasky}}, \bibinfo
  {author} {\bibfnamefont {A.~R.}\ \bibnamefont {Casey}}, \bibinfo {author}
  {\bibfnamefont {J.~A.}\ \bibnamefont {Clark}}, \bibinfo {author}
  {\bibfnamefont {F.~H.}\ \bibnamefont {Vivanco}}, \ and\ \bibinfo {author}
  {\bibfnamefont {K.}~\bibnamefont {Chatziioannou}},\ }\href {\doibase
  10.1103/PhysRevD.102.043011} {\bibfield  {journal} {\bibinfo  {journal}
  {Phys. Rev. D}\ }\textbf {\bibinfo {volume} {102}},\ \bibinfo {pages}
  {043011} (\bibinfo {year} {2020})},\ \Eprint
  {http://arxiv.org/abs/2006.04396} {arXiv:2006.04396 [astro-ph.HE]}
  \BibitemShut {NoStop}%
\bibitem [{\citenamefont {Friedman}\ and\ \citenamefont
  {Stergioulas}(2020)}]{Friedman:2020xac}%
  \BibitemOpen
  \bibfield  {author} {\bibinfo {author} {\bibfnamefont {J.~L.}\ \bibnamefont
  {Friedman}}\ and\ \bibinfo {author} {\bibfnamefont {N.}~\bibnamefont
  {Stergioulas}},\ }\href {\doibase 10.1142/S0218271820410151} {\bibfield
  {journal} {\bibinfo  {journal} {Int. J. Mod. Phys. D}\ }\textbf {\bibinfo
  {volume} {29}},\ \bibinfo {pages} {2041015} (\bibinfo {year} {2020})},\
  \Eprint {http://arxiv.org/abs/2005.14135} {arXiv:2005.14135 [astro-ph.HE]}
  \BibitemShut {NoStop}%
\bibitem [{\citenamefont {Zwiebach}(1985)}]{zwiebach1985curvature}%
  \BibitemOpen
  \bibfield  {author} {\bibinfo {author} {\bibfnamefont {B.}~\bibnamefont
  {Zwiebach}},\ }\href@noop {} {\bibfield  {journal} {\bibinfo  {journal}
  {Physics Letters B}\ }\textbf {\bibinfo {volume} {156}},\ \bibinfo {pages}
  {315} (\bibinfo {year} {1985})}\BibitemShut {NoStop}%
\bibitem [{\citenamefont {Gross}\ and\ \citenamefont
  {Sloan}(1987)}]{gross1987quartic}%
  \BibitemOpen
  \bibfield  {author} {\bibinfo {author} {\bibfnamefont {D.~J.}\ \bibnamefont
  {Gross}}\ and\ \bibinfo {author} {\bibfnamefont {J.~H.}\ \bibnamefont
  {Sloan}},\ }\href@noop {} {\bibfield  {journal} {\bibinfo  {journal} {Nuclear
  Physics B}\ }\textbf {\bibinfo {volume} {291}},\ \bibinfo {pages} {41}
  (\bibinfo {year} {1987})}\BibitemShut {NoStop}%
\bibitem [{\citenamefont {Glavan}\ and\ \citenamefont
  {Lin}(2020)}]{Glavan:2019inb}%
  \BibitemOpen
  \bibfield  {author} {\bibinfo {author} {\bibfnamefont {D.}~\bibnamefont
  {Glavan}}\ and\ \bibinfo {author} {\bibfnamefont {C.}~\bibnamefont {Lin}},\
  }\href {\doibase 10.1103/PhysRevLett.124.081301} {\bibfield  {journal}
  {\bibinfo  {journal} {Phys. Rev. Lett.}\ }\textbf {\bibinfo {volume} {124}},\
  \bibinfo {pages} {081301} (\bibinfo {year} {2020})},\ \Eprint
  {http://arxiv.org/abs/1905.03601} {arXiv:1905.03601 [gr-qc]} \BibitemShut
  {NoStop}%
\bibitem [{\citenamefont {Mann}\ and\ \citenamefont
  {Ross}(1993)}]{Mann:1992ar}%
  \BibitemOpen
  \bibfield  {author} {\bibinfo {author} {\bibfnamefont {R.~B.}\ \bibnamefont
  {Mann}}\ and\ \bibinfo {author} {\bibfnamefont {S.~F.}\ \bibnamefont
  {Ross}},\ }\href {\doibase 10.1088/0264-9381/10/7/015} {\bibfield  {journal}
  {\bibinfo  {journal} {Class. Quant. Grav.}\ }\textbf {\bibinfo {volume}
  {10}},\ \bibinfo {pages} {1405} (\bibinfo {year} {1993})},\ \Eprint
  {http://arxiv.org/abs/gr-qc/9208004} {arXiv:gr-qc/9208004} \BibitemShut
  {NoStop}%
\bibitem [{\citenamefont {Nojiri}\ and\ \citenamefont
  {Odintsov}(2020)}]{Nojiri:2020tph}%
  \BibitemOpen
  \bibfield  {author} {\bibinfo {author} {\bibfnamefont {S.}~\bibnamefont
  {Nojiri}}\ and\ \bibinfo {author} {\bibfnamefont {S.~D.}\ \bibnamefont
  {Odintsov}},\ }\href {\doibase 10.1209/0295-5075/130/10004} {\bibfield
  {journal} {\bibinfo  {journal} {EPL}\ }\textbf {\bibinfo {volume} {130}},\
  \bibinfo {pages} {10004} (\bibinfo {year} {2020})},\ \Eprint
  {http://arxiv.org/abs/2004.01404} {arXiv:2004.01404 [hep-th]} \BibitemShut
  {NoStop}%
\bibitem [{\citenamefont {Torii}\ and\ \citenamefont
  {Shinkai}(2008)}]{Torii:2008ru}%
  \BibitemOpen
  \bibfield  {author} {\bibinfo {author} {\bibfnamefont {T.}~\bibnamefont
  {Torii}}\ and\ \bibinfo {author} {\bibfnamefont {H.-a.}\ \bibnamefont
  {Shinkai}},\ }\href {\doibase 10.1103/PhysRevD.78.084037} {\bibfield
  {journal} {\bibinfo  {journal} {Phys. Rev. D}\ }\textbf {\bibinfo {volume}
  {78}},\ \bibinfo {pages} {084037} (\bibinfo {year} {2008})},\ \Eprint
  {http://arxiv.org/abs/0810.1790} {arXiv:0810.1790 [gr-qc]} \BibitemShut
  {NoStop}%
\bibitem [{\citenamefont {Mardones}\ and\ \citenamefont
  {Zanelli}(1991)}]{mardones1991lovelock}%
  \BibitemOpen
  \bibfield  {author} {\bibinfo {author} {\bibfnamefont {A.}~\bibnamefont
  {Mardones}}\ and\ \bibinfo {author} {\bibfnamefont {J.}~\bibnamefont
  {Zanelli}},\ }\href@noop {} {\bibfield  {journal} {\bibinfo  {journal}
  {Classical and Quantum Gravity}\ }\textbf {\bibinfo {volume} {8}},\ \bibinfo
  {pages} {1545} (\bibinfo {year} {1991})}\BibitemShut {NoStop}%
\bibitem [{\citenamefont {Woodard}(2015)}]{Woodard:2015zca}%
  \BibitemOpen
  \bibfield  {author} {\bibinfo {author} {\bibfnamefont {R.~P.}\ \bibnamefont
  {Woodard}},\ }\href {\doibase 10.4249/scholarpedia.32243} {\bibfield
  {journal} {\bibinfo  {journal} {Scholarpedia}\ }\textbf {\bibinfo {volume}
  {10}},\ \bibinfo {pages} {32243} (\bibinfo {year} {2015})},\ \Eprint
  {http://arxiv.org/abs/1506.02210} {arXiv:1506.02210 [hep-th]} \BibitemShut
  {NoStop}%
\bibitem [{\citenamefont {Tomozawa}(2011)}]{Tomozawa:2011gp}%
  \BibitemOpen
  \bibfield  {author} {\bibinfo {author} {\bibfnamefont {Y.}~\bibnamefont
  {Tomozawa}},\ }\href@noop {} {\  (\bibinfo {year} {2011})},\ \Eprint
  {http://arxiv.org/abs/1107.1424} {arXiv:1107.1424 [gr-qc]} \BibitemShut
  {NoStop}%
\bibitem [{\citenamefont {Aoki}\ \emph {et~al.}(2021)\citenamefont {Aoki},
  \citenamefont {Gorji}, \citenamefont {Mizuno},\ and\ \citenamefont
  {Mukohyama}}]{Aoki:2020ila}%
  \BibitemOpen
  \bibfield  {author} {\bibinfo {author} {\bibfnamefont {K.}~\bibnamefont
  {Aoki}}, \bibinfo {author} {\bibfnamefont {M.~A.}\ \bibnamefont {Gorji}},
  \bibinfo {author} {\bibfnamefont {S.}~\bibnamefont {Mizuno}}, \ and\ \bibinfo
  {author} {\bibfnamefont {S.}~\bibnamefont {Mukohyama}},\ }\href {\doibase
  10.1088/1475-7516/2021/01/054} {\bibfield  {journal} {\bibinfo  {journal}
  {JCAP}\ }\textbf {\bibinfo {volume} {01}},\ \bibinfo {pages} {054} (\bibinfo
  {year} {2021})},\ \Eprint {http://arxiv.org/abs/2010.03973} {arXiv:2010.03973
  [gr-qc]} \BibitemShut {NoStop}%
\bibitem [{\citenamefont {Bonifacio}\ \emph {et~al.}(2020)\citenamefont
  {Bonifacio}, \citenamefont {Hinterbichler},\ and\ \citenamefont
  {Johnson}}]{Bonifacio:2020vbk}%
  \BibitemOpen
  \bibfield  {author} {\bibinfo {author} {\bibfnamefont {J.}~\bibnamefont
  {Bonifacio}}, \bibinfo {author} {\bibfnamefont {K.}~\bibnamefont
  {Hinterbichler}}, \ and\ \bibinfo {author} {\bibfnamefont {L.~A.}\
  \bibnamefont {Johnson}},\ }\href {\doibase 10.1103/PhysRevD.102.024029}
  {\bibfield  {journal} {\bibinfo  {journal} {Phys. Rev. D}\ }\textbf {\bibinfo
  {volume} {102}},\ \bibinfo {pages} {024029} (\bibinfo {year} {2020})},\
  \Eprint {http://arxiv.org/abs/2004.10716} {arXiv:2004.10716 [hep-th]}
  \BibitemShut {NoStop}%
\bibitem [{\citenamefont {Ai}(2020)}]{Ai:2020peo}%
  \BibitemOpen
  \bibfield  {author} {\bibinfo {author} {\bibfnamefont {W.-Y.}\ \bibnamefont
  {Ai}},\ }\href {\doibase 10.1088/1572-9494/aba242} {\bibfield  {journal}
  {\bibinfo  {journal} {Commun. Theor. Phys.}\ }\textbf {\bibinfo {volume}
  {72}},\ \bibinfo {pages} {095402} (\bibinfo {year} {2020})},\ \Eprint
  {http://arxiv.org/abs/2004.02858} {arXiv:2004.02858 [gr-qc]} \BibitemShut
  {NoStop}%
\bibitem [{\citenamefont {Mahapatra}(2020)}]{Mahapatra:2020rds}%
  \BibitemOpen
  \bibfield  {author} {\bibinfo {author} {\bibfnamefont {S.}~\bibnamefont
  {Mahapatra}},\ }\href {\doibase 10.1140/epjc/s10052-020-08568-6} {\bibfield
  {journal} {\bibinfo  {journal} {Eur. Phys. J. C}\ }\textbf {\bibinfo {volume}
  {80}},\ \bibinfo {pages} {992} (\bibinfo {year} {2020})},\ \Eprint
  {http://arxiv.org/abs/2004.09214} {arXiv:2004.09214 [gr-qc]} \BibitemShut
  {NoStop}%
\bibitem [{\citenamefont {G\"urses}\ \emph {et~al.}(2020)\citenamefont
  {G\"urses}, \citenamefont {\c{S}i\c{s}man},\ and\ \citenamefont
  {Tekin}}]{Gurses:2020ofy}%
  \BibitemOpen
  \bibfield  {author} {\bibinfo {author} {\bibfnamefont {M.}~\bibnamefont
  {G\"urses}}, \bibinfo {author} {\bibfnamefont {T.~c.}\ \bibnamefont
  {\c{S}i\c{s}man}}, \ and\ \bibinfo {author} {\bibfnamefont {B.}~\bibnamefont
  {Tekin}},\ }\href {\doibase 10.1140/epjc/s10052-020-8200-7} {\bibfield
  {journal} {\bibinfo  {journal} {Eur. Phys. J. C}\ }\textbf {\bibinfo {volume}
  {80}},\ \bibinfo {pages} {647} (\bibinfo {year} {2020})},\ \Eprint
  {http://arxiv.org/abs/2004.03390} {arXiv:2004.03390 [gr-qc]} \BibitemShut
  {NoStop}%
\bibitem [{\citenamefont {Hohmann}\ \emph {et~al.}(2021)\citenamefont
  {Hohmann}, \citenamefont {Pfeifer},\ and\ \citenamefont
  {Voicu}}]{hohmann2021canonical}%
  \BibitemOpen
  \bibfield  {author} {\bibinfo {author} {\bibfnamefont {M.}~\bibnamefont
  {Hohmann}}, \bibinfo {author} {\bibfnamefont {C.}~\bibnamefont {Pfeifer}}, \
  and\ \bibinfo {author} {\bibfnamefont {N.}~\bibnamefont {Voicu}},\
  }\href@noop {} {\bibfield  {journal} {\bibinfo  {journal} {The European
  Physical Journal Plus}\ }\textbf {\bibinfo {volume} {136}},\ \bibinfo {pages}
  {1} (\bibinfo {year} {2021})}\BibitemShut {NoStop}%
\bibitem [{\citenamefont {Ma}\ and\ \citenamefont {Lu}(2020)}]{Ma:2020ufk}%
  \BibitemOpen
  \bibfield  {author} {\bibinfo {author} {\bibfnamefont {L.}~\bibnamefont
  {Ma}}\ and\ \bibinfo {author} {\bibfnamefont {H.}~\bibnamefont {Lu}},\ }\href
  {\doibase 10.1140/epjc/s10052-020-08780-4} {\bibfield  {journal} {\bibinfo
  {journal} {Eur. Phys. J. C}\ }\textbf {\bibinfo {volume} {80}},\ \bibinfo
  {pages} {1209} (\bibinfo {year} {2020})},\ \Eprint
  {http://arxiv.org/abs/2004.14738} {arXiv:2004.14738 [gr-qc]} \BibitemShut
  {NoStop}%
\bibitem [{\citenamefont {Lu}\ and\ \citenamefont {Pang}(2020)}]{Lu:2020iav}%
  \BibitemOpen
  \bibfield  {author} {\bibinfo {author} {\bibfnamefont {H.}~\bibnamefont
  {Lu}}\ and\ \bibinfo {author} {\bibfnamefont {Y.}~\bibnamefont {Pang}},\
  }\href {\doibase 10.1016/j.physletb.2020.135717} {\bibfield  {journal}
  {\bibinfo  {journal} {Phys. Lett. B}\ }\textbf {\bibinfo {volume} {809}},\
  \bibinfo {pages} {135717} (\bibinfo {year} {2020})},\ \Eprint
  {http://arxiv.org/abs/2003.11552} {arXiv:2003.11552 [gr-qc]} \BibitemShut
  {NoStop}%
\bibitem [{\citenamefont {Fernandes}\ \emph {et~al.}(2020)\citenamefont
  {Fernandes}, \citenamefont {Carrilho}, \citenamefont {Clifton},\ and\
  \citenamefont {Mulryne}}]{Fernandes:2020nbq}%
  \BibitemOpen
  \bibfield  {author} {\bibinfo {author} {\bibfnamefont {P.~G.~S.}\
  \bibnamefont {Fernandes}}, \bibinfo {author} {\bibfnamefont {P.}~\bibnamefont
  {Carrilho}}, \bibinfo {author} {\bibfnamefont {T.}~\bibnamefont {Clifton}}, \
  and\ \bibinfo {author} {\bibfnamefont {D.~J.}\ \bibnamefont {Mulryne}},\
  }\href {\doibase 10.1103/PhysRevD.102.024025} {\bibfield  {journal} {\bibinfo
   {journal} {Phys. Rev. D}\ }\textbf {\bibinfo {volume} {102}},\ \bibinfo
  {pages} {024025} (\bibinfo {year} {2020})},\ \Eprint
  {http://arxiv.org/abs/2004.08362} {arXiv:2004.08362 [gr-qc]} \BibitemShut
  {NoStop}%
\bibitem [{\citenamefont {Banerjee}\ \emph
  {et~al.}(2021{\natexlab{b}})\citenamefont {Banerjee}, \citenamefont
  {Hansraj},\ and\ \citenamefont {Moodly}}]{Banerjee:2021bmv}%
  \BibitemOpen
  \bibfield  {author} {\bibinfo {author} {\bibfnamefont {A.}~\bibnamefont
  {Banerjee}}, \bibinfo {author} {\bibfnamefont {S.}~\bibnamefont {Hansraj}}, \
  and\ \bibinfo {author} {\bibfnamefont {L.}~\bibnamefont {Moodly}},\ }\href
  {\doibase 10.1140/epjc/s10052-021-09585-9} {\bibfield  {journal} {\bibinfo
  {journal} {Eur. Phys. J. C}\ }\textbf {\bibinfo {volume} {81}},\ \bibinfo
  {pages} {790} (\bibinfo {year} {2021}{\natexlab{b}})}\BibitemShut {NoStop}%
\bibitem [{\citenamefont {Ghosh}\ and\ \citenamefont
  {Kumar}(2020)}]{Ghosh:2020syx}%
  \BibitemOpen
  \bibfield  {author} {\bibinfo {author} {\bibfnamefont {S.~G.}\ \bibnamefont
  {Ghosh}}\ and\ \bibinfo {author} {\bibfnamefont {R.}~\bibnamefont {Kumar}},\
  }\href {\doibase 10.1088/1361-6382/abc134} {\bibfield  {journal} {\bibinfo
  {journal} {Class. Quant. Grav.}\ }\textbf {\bibinfo {volume} {37}},\ \bibinfo
  {pages} {245008} (\bibinfo {year} {2020})},\ \Eprint
  {http://arxiv.org/abs/2003.12291} {arXiv:2003.12291 [gr-qc]} \BibitemShut
  {NoStop}%
\bibitem [{\citenamefont {Kumar}\ and\ \citenamefont
  {Ghosh}(2020)}]{Kumar:2020xvu}%
  \BibitemOpen
  \bibfield  {author} {\bibinfo {author} {\bibfnamefont {A.}~\bibnamefont
  {Kumar}}\ and\ \bibinfo {author} {\bibfnamefont {S.~G.}\ \bibnamefont
  {Ghosh}},\ }\href@noop {} {\  (\bibinfo {year} {2020})},\ \Eprint
  {http://arxiv.org/abs/2004.01131} {arXiv:2004.01131 [gr-qc]} \BibitemShut
  {NoStop}%
\bibitem [{\citenamefont {Kumar}\ and\ \citenamefont
  {Kumar}(2020)}]{Kumar:2020uyz}%
  \BibitemOpen
  \bibfield  {author} {\bibinfo {author} {\bibfnamefont {A.}~\bibnamefont
  {Kumar}}\ and\ \bibinfo {author} {\bibfnamefont {R.}~\bibnamefont {Kumar}},\
  }\href@noop {} {\  (\bibinfo {year} {2020})},\ \Eprint
  {http://arxiv.org/abs/2003.13104} {arXiv:2003.13104 [gr-qc]} \BibitemShut
  {NoStop}%
\bibitem [{\citenamefont {Zhang}\ \emph
  {et~al.}(2020{\natexlab{a}})\citenamefont {Zhang}, \citenamefont {Zhang},
  \citenamefont {Li},\ and\ \citenamefont {Guo}}]{Zhang:2020sjh}%
  \BibitemOpen
  \bibfield  {author} {\bibinfo {author} {\bibfnamefont {C.-Y.}\ \bibnamefont
  {Zhang}}, \bibinfo {author} {\bibfnamefont {S.-J.}\ \bibnamefont {Zhang}},
  \bibinfo {author} {\bibfnamefont {P.-C.}\ \bibnamefont {Li}}, \ and\ \bibinfo
  {author} {\bibfnamefont {M.}~\bibnamefont {Guo}},\ }\href {\doibase
  10.1007/JHEP08(2020)105} {\bibfield  {journal} {\bibinfo  {journal} {JHEP}\
  }\textbf {\bibinfo {volume} {08}},\ \bibinfo {pages} {105} (\bibinfo {year}
  {2020}{\natexlab{a}})},\ \Eprint {http://arxiv.org/abs/2004.03141}
  {arXiv:2004.03141 [gr-qc]} \BibitemShut {NoStop}%
\bibitem [{\citenamefont {Hosseini~Mansoori}(2021)}]{HosseiniMansoori:2020yfj}%
  \BibitemOpen
  \bibfield  {author} {\bibinfo {author} {\bibfnamefont {S.~A.}\ \bibnamefont
  {Hosseini~Mansoori}},\ }\href {\doibase 10.1016/j.dark.2021.100776}
  {\bibfield  {journal} {\bibinfo  {journal} {Phys. Dark Univ.}\ }\textbf
  {\bibinfo {volume} {31}},\ \bibinfo {pages} {100776} (\bibinfo {year}
  {2021})},\ \Eprint {http://arxiv.org/abs/2003.13382} {arXiv:2003.13382
  [gr-qc]} \BibitemShut {NoStop}%
\bibitem [{\citenamefont {Mishra}(2020)}]{Mishra:2020gce}%
  \BibitemOpen
  \bibfield  {author} {\bibinfo {author} {\bibfnamefont {A.~K.}\ \bibnamefont
  {Mishra}},\ }\href {\doibase 10.1007/s10714-020-02763-2} {\bibfield
  {journal} {\bibinfo  {journal} {Gen. Rel. Grav.}\ }\textbf {\bibinfo {volume}
  {52}},\ \bibinfo {pages} {106} (\bibinfo {year} {2020})},\ \Eprint
  {http://arxiv.org/abs/2004.01243} {arXiv:2004.01243 [gr-qc]} \BibitemShut
  {NoStop}%
\bibitem [{\citenamefont {Zhang}\ \emph
  {et~al.}(2020{\natexlab{b}})\citenamefont {Zhang}, \citenamefont {Li},\ and\
  \citenamefont {Guo}}]{Zhang:2020qam}%
  \BibitemOpen
  \bibfield  {author} {\bibinfo {author} {\bibfnamefont {C.-Y.}\ \bibnamefont
  {Zhang}}, \bibinfo {author} {\bibfnamefont {P.-C.}\ \bibnamefont {Li}}, \
  and\ \bibinfo {author} {\bibfnamefont {M.}~\bibnamefont {Guo}},\ }\href
  {\doibase 10.1140/epjc/s10052-020-08448-z} {\bibfield  {journal} {\bibinfo
  {journal} {Eur. Phys. J. C}\ }\textbf {\bibinfo {volume} {80}},\ \bibinfo
  {pages} {874} (\bibinfo {year} {2020}{\natexlab{b}})},\ \Eprint
  {http://arxiv.org/abs/2003.13068} {arXiv:2003.13068 [hep-th]} \BibitemShut
  {NoStop}%
\bibitem [{\citenamefont {Zhang}\ \emph
  {et~al.}(2020{\natexlab{c}})\citenamefont {Zhang}, \citenamefont {Wei},\ and\
  \citenamefont {Liu}}]{Zhang:2020qew}%
  \BibitemOpen
  \bibfield  {author} {\bibinfo {author} {\bibfnamefont {Y.-P.}\ \bibnamefont
  {Zhang}}, \bibinfo {author} {\bibfnamefont {S.-W.}\ \bibnamefont {Wei}}, \
  and\ \bibinfo {author} {\bibfnamefont {Y.-X.}\ \bibnamefont {Liu}},\ }\href
  {\doibase 10.3390/universe6080103} {\bibfield  {journal} {\bibinfo  {journal}
  {Universe}\ }\textbf {\bibinfo {volume} {6}},\ \bibinfo {pages} {103}
  (\bibinfo {year} {2020}{\natexlab{c}})},\ \Eprint
  {http://arxiv.org/abs/2003.10960} {arXiv:2003.10960 [gr-qc]} \BibitemShut
  {NoStop}%
\bibitem [{\citenamefont {Jusufi}\ \emph {et~al.}(2020)\citenamefont {Jusufi},
  \citenamefont {Banerjee},\ and\ \citenamefont {Ghosh}}]{Jusufi:2020yus}%
  \BibitemOpen
  \bibfield  {author} {\bibinfo {author} {\bibfnamefont {K.}~\bibnamefont
  {Jusufi}}, \bibinfo {author} {\bibfnamefont {A.}~\bibnamefont {Banerjee}}, \
  and\ \bibinfo {author} {\bibfnamefont {S.~G.}\ \bibnamefont {Ghosh}},\ }\href
  {\doibase 10.1140/epjc/s10052-020-8287-x} {\bibfield  {journal} {\bibinfo
  {journal} {Eur. Phys. J. C}\ }\textbf {\bibinfo {volume} {80}},\ \bibinfo
  {pages} {698} (\bibinfo {year} {2020})},\ \Eprint
  {http://arxiv.org/abs/2004.10750} {arXiv:2004.10750 [gr-qc]} \BibitemShut
  {NoStop}%
\bibitem [{\citenamefont {Eslam~Panah}\ \emph {et~al.}(2020)\citenamefont
  {Eslam~Panah}, \citenamefont {Jafarzade},\ and\ \citenamefont
  {Hendi}}]{EslamPanah:2020hoj}%
  \BibitemOpen
  \bibfield  {author} {\bibinfo {author} {\bibfnamefont {B.}~\bibnamefont
  {Eslam~Panah}}, \bibinfo {author} {\bibfnamefont {K.}~\bibnamefont
  {Jafarzade}}, \ and\ \bibinfo {author} {\bibfnamefont {S.~H.}\ \bibnamefont
  {Hendi}},\ }\href {\doibase 10.1016/j.nuclphysb.2020.115269} {\bibfield
  {journal} {\bibinfo  {journal} {Nucl. Phys. B}\ }\textbf {\bibinfo {volume}
  {961}},\ \bibinfo {pages} {115269} (\bibinfo {year} {2020})},\ \Eprint
  {http://arxiv.org/abs/2004.04058} {arXiv:2004.04058 [hep-th]} \BibitemShut
  {NoStop}%
\bibitem [{\citenamefont {Antoniadis}\ \emph {et~al.}(2013)\citenamefont
  {Antoniadis} \emph {et~al.}}]{Antoniadis:2013pzd}%
  \BibitemOpen
  \bibfield  {author} {\bibinfo {author} {\bibfnamefont {J.}~\bibnamefont
  {Antoniadis}} \emph {et~al.},\ }\href {\doibase 10.1126/science.1233232}
  {\bibfield  {journal} {\bibinfo  {journal} {Science}\ }\textbf {\bibinfo
  {volume} {340}},\ \bibinfo {pages} {6131} (\bibinfo {year} {2013})},\ \Eprint
  {http://arxiv.org/abs/1304.6875} {arXiv:1304.6875 [astro-ph.HE]} \BibitemShut
  {NoStop}%
\bibitem [{\citenamefont {Bodmer}(1971)}]{bodmer1971collapsed}%
  \BibitemOpen
  \bibfield  {author} {\bibinfo {author} {\bibfnamefont {A.}~\bibnamefont
  {Bodmer}},\ }\href@noop {} {\bibfield  {journal} {\bibinfo  {journal}
  {Physical Review D}\ }\textbf {\bibinfo {volume} {4}},\ \bibinfo {pages}
  {1601} (\bibinfo {year} {1971})}\BibitemShut {NoStop}%
\bibitem [{\citenamefont {Witten}(1984)}]{PhysRevD.30.272}%
  \BibitemOpen
  \bibfield  {author} {\bibinfo {author} {\bibfnamefont {E.}~\bibnamefont
  {Witten}},\ }\href {\doibase 10.1103/PhysRevD.30.272} {\bibfield  {journal}
  {\bibinfo  {journal} {Phys. Rev. D}\ }\textbf {\bibinfo {volume} {30}},\
  \bibinfo {pages} {272} (\bibinfo {year} {1984})}\BibitemShut {NoStop}%
\bibitem [{\citenamefont {Ghosh}\ and\ \citenamefont
  {Maharaj}(2020)}]{Ghosh:2020vpc}%
  \BibitemOpen
  \bibfield  {author} {\bibinfo {author} {\bibfnamefont {S.~G.}\ \bibnamefont
  {Ghosh}}\ and\ \bibinfo {author} {\bibfnamefont {S.~D.}\ \bibnamefont
  {Maharaj}},\ }\href {\doibase 10.1016/j.dark.2020.100687} {\bibfield
  {journal} {\bibinfo  {journal} {Phys. Dark Univ.}\ }\textbf {\bibinfo
  {volume} {30}},\ \bibinfo {pages} {100687} (\bibinfo {year} {2020})},\
  \Eprint {http://arxiv.org/abs/2003.09841} {arXiv:2003.09841 [gr-qc]}
  \BibitemShut {NoStop}%
\bibitem [{\citenamefont {{Herrera}}(1994)}]{1994PhLA..188..402H}%
  \BibitemOpen
  \bibfield  {author} {\bibinfo {author} {\bibfnamefont {L.}~\bibnamefont
  {{Herrera}}},\ }\href {\doibase 10.1016/0375-9601(94)90485-5} {\bibfield
  {journal} {\bibinfo  {journal} {Physics Letters A}\ }\textbf {\bibinfo
  {volume} {188}},\ \bibinfo {pages} {402} (\bibinfo {year}
  {1994})}\BibitemShut {NoStop}%
\bibitem [{\citenamefont {Abreu}\ \emph {et~al.}(2007)\citenamefont {Abreu},
  \citenamefont {Hernandez},\ and\ \citenamefont {Nunez}}]{Abreu:2007ew}%
  \BibitemOpen
  \bibfield  {author} {\bibinfo {author} {\bibfnamefont {H.}~\bibnamefont
  {Abreu}}, \bibinfo {author} {\bibfnamefont {H.}~\bibnamefont {Hernandez}}, \
  and\ \bibinfo {author} {\bibfnamefont {L.~A.}\ \bibnamefont {Nunez}},\ }\href
  {\doibase 10.1088/0264-9381/24/18/005} {\bibfield  {journal} {\bibinfo
  {journal} {Class. Quant. Grav.}\ }\textbf {\bibinfo {volume} {24}},\ \bibinfo
  {pages} {4631} (\bibinfo {year} {2007})},\ \Eprint
  {http://arxiv.org/abs/0706.3452} {arXiv:0706.3452 [gr-qc]} \BibitemShut
  {NoStop}%
\bibitem [{\citenamefont {Bamba}\ \emph {et~al.}(2017)\citenamefont {Bamba},
  \citenamefont {Ilyas}, \citenamefont {Bhatti},\ and\ \citenamefont
  {Yousaf}}]{Bamba:2017cjr}%
  \BibitemOpen
  \bibfield  {author} {\bibinfo {author} {\bibfnamefont {K.}~\bibnamefont
  {Bamba}}, \bibinfo {author} {\bibfnamefont {M.}~\bibnamefont {Ilyas}},
  \bibinfo {author} {\bibfnamefont {M.~Z.}\ \bibnamefont {Bhatti}}, \ and\
  \bibinfo {author} {\bibfnamefont {Z.}~\bibnamefont {Yousaf}},\ }\href
  {\doibase 10.1007/s10714-017-2276-x} {\bibfield  {journal} {\bibinfo
  {journal} {Gen. Rel. Grav.}\ }\textbf {\bibinfo {volume} {49}},\ \bibinfo
  {pages} {112} (\bibinfo {year} {2017})},\ \Eprint
  {http://arxiv.org/abs/1707.07386} {arXiv:1707.07386 [gr-qc]} \BibitemShut
  {NoStop}%
\bibitem [{\citenamefont {Das}\ \emph {et~al.}(2019)\citenamefont {Das},
  \citenamefont {Rahaman},\ and\ \citenamefont {Baskey}}]{Das:2019dkn}%
  \BibitemOpen
  \bibfield  {author} {\bibinfo {author} {\bibfnamefont {S.}~\bibnamefont
  {Das}}, \bibinfo {author} {\bibfnamefont {F.}~\bibnamefont {Rahaman}}, \ and\
  \bibinfo {author} {\bibfnamefont {L.}~\bibnamefont {Baskey}},\ }\href
  {\doibase 10.1140/epjc/s10052-019-7367-2} {\bibfield  {journal} {\bibinfo
  {journal} {Eur. Phys. J.}\ }\textbf {\bibinfo {volume} {C79}},\ \bibinfo
  {pages} {853} (\bibinfo {year} {2019})}\BibitemShut {NoStop}%
\bibitem [{\citenamefont {Tolman}(1939)}]{PhysRev.55.364}%
  \BibitemOpen
  \bibfield  {author} {\bibinfo {author} {\bibfnamefont {R.~C.}\ \bibnamefont
  {Tolman}},\ }\href {\doibase 10.1103/PhysRev.55.364} {\bibfield  {journal}
  {\bibinfo  {journal} {Phys. Rev.}\ }\textbf {\bibinfo {volume} {55}},\
  \bibinfo {pages} {364} (\bibinfo {year} {1939})}\BibitemShut {NoStop}%
\bibitem [{\citenamefont {Oppenheimer}\ and\ \citenamefont
  {Volkoff}(1939)}]{PhysRev.55.374}%
  \BibitemOpen
  \bibfield  {author} {\bibinfo {author} {\bibfnamefont {J.~R.}\ \bibnamefont
  {Oppenheimer}}\ and\ \bibinfo {author} {\bibfnamefont {G.~M.}\ \bibnamefont
  {Volkoff}},\ }\href {\doibase 10.1103/PhysRev.55.374} {\bibfield  {journal}
  {\bibinfo  {journal} {Phys. Rev.}\ }\textbf {\bibinfo {volume} {55}},\
  \bibinfo {pages} {374} (\bibinfo {year} {1939})}\BibitemShut {NoStop}%
\bibitem [{\citenamefont {Ponce~de Leon}(1993)}]{PoncedeLeon1993}%
  \BibitemOpen
  \bibfield  {author} {\bibinfo {author} {\bibfnamefont {J.}~\bibnamefont
  {Ponce~de Leon}},\ }\href {\doibase 10.1007/BF00763756} {\bibfield  {journal}
  {\bibinfo  {journal} {General Relativity and Gravitation}\ }\textbf {\bibinfo
  {volume} {25}},\ \bibinfo {pages} {1123} (\bibinfo {year}
  {1993})}\BibitemShut {NoStop}%
\bibitem [{\citenamefont {Moustakidis}(2017)}]{Moustakidis:2016ndw}%
  \BibitemOpen
  \bibfield  {author} {\bibinfo {author} {\bibfnamefont {C.~C.}\ \bibnamefont
  {Moustakidis}},\ }\href {\doibase 10.1007/s10714-017-2232-9} {\bibfield
  {journal} {\bibinfo  {journal} {Gen. Rel. Grav.}\ }\textbf {\bibinfo {volume}
  {49}},\ \bibinfo {pages} {68} (\bibinfo {year} {2017})},\ \Eprint
  {http://arxiv.org/abs/1612.01726} {arXiv:1612.01726 [gr-qc]} \BibitemShut
  {NoStop}%
\bibitem [{\citenamefont {{Heintzmann}}\ and\ \citenamefont
  {{Hillebrandt}}(1975)}]{1975A&A....38...51H}%
  \BibitemOpen
  \bibfield  {author} {\bibinfo {author} {\bibfnamefont {H.}~\bibnamefont
  {{Heintzmann}}}\ and\ \bibinfo {author} {\bibfnamefont {W.}~\bibnamefont
  {{Hillebrandt}}},\ }\href@noop {} {\bibfield  {journal} {\bibinfo  {journal}
  {aap}\ }\textbf {\bibinfo {volume} {38}},\ \bibinfo {pages} {51} (\bibinfo
  {year} {1975})}\BibitemShut {NoStop}%
\bibitem [{\citenamefont {Chan}\ \emph {et~al.}(1993)\citenamefont {Chan},
  \citenamefont {Herrera},\ and\ \citenamefont
  {Santos}}]{10.1093/mnras/265.3.533}%
  \BibitemOpen
  \bibfield  {author} {\bibinfo {author} {\bibfnamefont {R.}~\bibnamefont
  {Chan}}, \bibinfo {author} {\bibfnamefont {L.}~\bibnamefont {Herrera}}, \
  and\ \bibinfo {author} {\bibfnamefont {N.~O.}\ \bibnamefont {Santos}},\
  }\href {\doibase 10.1093/mnras/265.3.533} {\bibfield  {journal} {\bibinfo
  {journal} {Monthly Notices of the Royal Astronomical Society}\ }\textbf
  {\bibinfo {volume} {265}},\ \bibinfo {pages} {533} (\bibinfo {year}
  {1993})},\ \Eprint
  {http://arxiv.org/abs/http://oup.prod.sis.lan/mnras/article-pdf/265/3/533/3807712/mnras265-0533.pdf}
  {http://oup.prod.sis.lan/mnras/article-pdf/265/3/533/3807712/mnras265-0533.pdf}
  \BibitemShut {NoStop}%
\bibitem [{\citenamefont {{Harrison}}\ \emph {et~al.}(1965)\citenamefont
  {{Harrison}}, \citenamefont {{Thorne}}, \citenamefont {{Wakano}},\ and\
  \citenamefont {{Wheeler}}}]{1965gtgc.book.....H}%
  \BibitemOpen
  \bibfield  {author} {\bibinfo {author} {\bibfnamefont {B.~K.}\ \bibnamefont
  {{Harrison}}}, \bibinfo {author} {\bibfnamefont {K.~S.}\ \bibnamefont
  {{Thorne}}}, \bibinfo {author} {\bibfnamefont {M.}~\bibnamefont {{Wakano}}},
  \ and\ \bibinfo {author} {\bibfnamefont {J.~A.}\ \bibnamefont {{Wheeler}}},\
  }\href@noop {} {\emph {\bibinfo {title} {{Gravitation Theory and
  Gravitational Collapse}}}}\ (\bibinfo {year} {1965})\BibitemShut {NoStop}%
\bibitem [{\citenamefont {{Zeldovich}}\ and\ \citenamefont
  {{Novikov}}(1971)}]{1971reas.book.....Z}%
  \BibitemOpen
  \bibfield  {author} {\bibinfo {author} {\bibfnamefont {Y.~B.}\ \bibnamefont
  {{Zeldovich}}}\ and\ \bibinfo {author} {\bibfnamefont {I.~D.}\ \bibnamefont
  {{Novikov}}},\ }\href@noop {} {\emph {\bibinfo {title} {{Relativistic
  astrophysics. Vol.1: Stars and relativity}}}}\ (\bibinfo {year}
  {1971})\BibitemShut {NoStop}%
\bibitem [{\citenamefont {{Zeldovich}}\ and\ \citenamefont
  {{Novikov}}(1983)}]{1983reas.book.....Z}%
  \BibitemOpen
  \bibfield  {author} {\bibinfo {author} {\bibfnamefont {I.~B.}\ \bibnamefont
  {{Zeldovich}}}\ and\ \bibinfo {author} {\bibfnamefont {I.~D.}\ \bibnamefont
  {{Novikov}}},\ }\href@noop {} {\emph {\bibinfo {title} {{Relativistic
  astrophysics. Vol.2: The structure and evolution of the universe}}}}\
  (\bibinfo {year} {1983})\BibitemShut {NoStop}%
\bibitem [{\citenamefont {{{\"O}zel}}\ \emph {et~al.}(2016)\citenamefont
  {{{\"O}zel}}, \citenamefont {{Psaltis}}, \citenamefont {{G{\"u}ver}},
  \citenamefont {{Baym}}, \citenamefont {{Heinke}},\ and\ \citenamefont
  {{Guillot}}}]{2016ApJ...820...28O}%
  \BibitemOpen
  \bibfield  {author} {\bibinfo {author} {\bibfnamefont {F.}~\bibnamefont
  {{{\"O}zel}}}, \bibinfo {author} {\bibfnamefont {D.}~\bibnamefont
  {{Psaltis}}}, \bibinfo {author} {\bibfnamefont {T.}~\bibnamefont
  {{G{\"u}ver}}}, \bibinfo {author} {\bibfnamefont {G.}~\bibnamefont {{Baym}}},
  \bibinfo {author} {\bibfnamefont {C.}~\bibnamefont {{Heinke}}}, \ and\
  \bibinfo {author} {\bibfnamefont {S.}~\bibnamefont {{Guillot}}},\ }\href
  {\doibase 10.3847/0004-637X/820/1/28} {\bibfield  {journal} {\bibinfo
  {journal} {apj}\ }\textbf {\bibinfo {volume} {820}},\ \bibinfo {eid} {28}
  (\bibinfo {year} {2016})},\ \Eprint {http://arxiv.org/abs/1505.05155}
  {arXiv:1505.05155 [astro-ph.HE]} \BibitemShut {NoStop}%
\bibitem [{\citenamefont {Bhatti}\ \emph {et~al.}(2017)\citenamefont {Bhatti},
  \citenamefont {Sharif}, \citenamefont {Yousaf},\ and\ \citenamefont
  {Ilyas}}]{Bhatti:2017fov}%
  \BibitemOpen
  \bibfield  {author} {\bibinfo {author} {\bibfnamefont {M.~Z.-u.-H.}\
  \bibnamefont {Bhatti}}, \bibinfo {author} {\bibfnamefont {M.}~\bibnamefont
  {Sharif}}, \bibinfo {author} {\bibfnamefont {Z.}~\bibnamefont {Yousaf}}, \
  and\ \bibinfo {author} {\bibfnamefont {M.}~\bibnamefont {Ilyas}},\ }\href
  {\doibase 10.1142/S021827181850044X} {\bibfield  {journal} {\bibinfo
  {journal} {Int. J. Mod. Phys. D}\ }\textbf {\bibinfo {volume} {27}},\
  \bibinfo {pages} {1850044} (\bibinfo {year} {2017})}\BibitemShut {NoStop}%
\end{thebibliography}
%

\end{document}